\documentclass[aps,prx,preprintnumbers,nofootinbib,twocolumn,amsmath,amssymb,floatfix,superscriptaddress]{revtex4-2}
\usepackage[utf8]{inputenc}
\usepackage[T1]{fontenc}
\usepackage[german,english]{babel}
\usepackage{graphicx}
\usepackage{amsmath}
\usepackage{mathtools}
\usepackage{amssymb}
\usepackage{amsthm}
\usepackage{tensor}
\usepackage{microtype} 
\usepackage{quoting} 
\quotingsetup{font=small}
\usepackage{tensor}
\usepackage{graphicx}
\usepackage{braket}
\usepackage{bm}
\usepackage{comment}
\usepackage{booktabs}
\usepackage{eurosym} 
\usepackage{braket}
\usepackage{bbold} 
\usepackage{emptypage}
\usepackage{epstopdf}
\usepackage{verbatim}
\usepackage{enumitem}
\usepackage{float}
\usepackage{chemformula}

\newcommand{\beq}{\begin{equation}}
\newcommand{\eeq}{\end{equation}}

\usepackage[bookmarks=true,colorlinks,linkcolor=red,urlcolor=blue,citecolor=blue]{hyperref} 

\begin{document}
\title{Quantum reaction-limited reaction-diffusion dynamics of annihilation processes}
\author{Gabriele Perfetto}
\thanks{gabriele.perfetto@uni-tuebingen.de}
\affiliation{Institut f\"ur Theoretische Physik, Universit\"at T\"ubingen, Auf der Morgenstelle 14, 72076 T\"ubingen, Germany}
\author{Federico Carollo}
\affiliation{Institut f\"ur Theoretische Physik, Universit\"at T\"ubingen, Auf der Morgenstelle 14, 72076 T\"ubingen, Germany}
\author{Juan P. Garrahan}
\affiliation{School of Physics, Astronomy, University of Nottingham, Nottingham, NG7 2RD, UK.}
\affiliation{Centre for the Mathematics, Theoretical Physics of Quantum Non-Equilibrium Systems,
University of Nottingham, Nottingham, NG7 2RD, UK}
\author{Igor Lesanovsky}
\affiliation{Institut f\"ur Theoretische Physik, Universit\"at T\"ubingen, Auf der Morgenstelle 14, 72076 T\"ubingen, Germany}
\affiliation{School of Physics, Astronomy, University of Nottingham, Nottingham, NG7 2RD, UK.}
\affiliation{Centre for the Mathematics, Theoretical Physics of Quantum Non-Equilibrium Systems,
University of Nottingham, Nottingham, NG7 2RD, UK}

\begin{abstract}
We investigate the quantum reaction-diffusion dynamics of fermionic particles which coherently hop in a one-dimensional lattice and undergo annihilation reactions. The latter are modelled as dissipative processes which involve losses of pairs $2A \to \emptyset$, triplets $3A \to \emptyset$, and quadruplets $4A \to \emptyset$ of neighbouring particles. When considering classical particles, the corresponding decay of their density in time follows an asymptotic power-law behavior. The associated exponent in one dimension is different from the mean-field prediction whenever diffusive mixing is not too strong and spatial correlations are relevant. This specifically applies to $2A\to \emptyset$, while the mean-field power-law prediction just acquires a logarithmic correction for $3A \to \emptyset$ and is exact for $4A \to \emptyset$. A mean-field approach is also valid, for all the three processes, when the diffusive mixing is strong, i.e., in the so-called reaction-limited regime. Here, we show that the picture is different for quantum systems. We consider the quantum reaction-limited regime and we show that for all the three processes power-law behavior beyond mean field is present as a consequence of quantum coherences, which are not related to space dimensionality. The decay in $3A\to \emptyset$ is further, highly intricate, since the power-law behavior therein only appears within an intermediate time window, while at long times the density decay is not  power-law. Our results show that emergent critical behavior in quantum dynamics has a markedly different origin, based on quantum coherences, to that applying to classical critical phenomena, which is, instead, solely determined by the relevance of spatial correlations.
\end{abstract}

\maketitle

\section{Introduction}
The investigation and classification of universal behavior in nonequilibrium many-body systems is a timely and challenging research area. 
Far from equilibrium the Boltzmann-Gibbs measure does not correctly describe the system and therefore the emergence of universal behavior cannot be clearly pinpointed as in equilibrium-critical systems \cite{wilson1974renormalization,Kogut1979}. Within this perspective, reaction-diffusion (RD) systems, where classical particles hop on a lattice (diffusion in the continuum limit) at rate $\Omega$ and react at rate $\Gamma$ upon meeting, are genuine nonequilibrium systems where universal quantities can be identified and characterized \cite{henkel2008non,hinrichsen2000non,vladimir1997nonequilibrium,tauber2002dynamic,tauber2005applications,tauber2014critical,krapivsky2010kinetic}. At long times the density of particles decays as a power-law, which can be obtained, for instance, from a mean-field treatment whenever diffusive mixing is effective. This is valid when the diffusive hopping is strong, $\Gamma/\Omega \ll 1$, i.e., in the \textit{reaction-limited regime} (sometimes also called ``well stirred mixture'' regime) \cite{hinrichsen2000non,vladimir1997nonequilibrium,Redner1984,fastdiffusion1992,krapivsky2010kinetic}, or in the opposite \textit{diffusion-limited} regime of weak hopping $\Gamma/\Omega\sim 1$ in dimensions larger than the upper critical dimension $d_c$. This dimension can be identified both via exact lattice calculations \cite{toussaint1983particle,Spouge1988,Privman1994,torney1983diffusion,fluctuationseffects,redner1984scaling,kang1984universal,Racz1985} and/or renormalization group methods \cite{tauber2002dynamic,tauber2005applications,tauber2014critical,doi1976stochastic,peliti1985path,peliti1986renormalisation,QFT_RD_1998}. In the case of binary annihilation, $2A\to\emptyset$, it is $d_c=2$, so that in one dimension spatial fluctuations are relevant and they are responsible for universal power-law decay $n(t) \sim (\Omega t)^{-1/2}$ of the density $n(t)$ in time $t$ that is different from mean field \cite{toussaint1983particle,Spouge1988,Privman1994,torney1983diffusion,fluctuationseffects,redner1984scaling,tauber2002dynamic,tauber2005applications,tauber2014critical,peliti1985path,peliti1986renormalisation}. The upper critical dimension depends, however, on the number of particles involved in the reaction. For triplet annihilation $3A\to \emptyset$, it is $d_c=1$, and thus in one dimension spatial correlations are marginal and the mean-field power-law decay only acquires a logarithmic correction $n(t)\sim [\ln(\Omega t)/\Omega t]^{1/2}$ \cite{tauber2002dynamic,tauber2005applications,tauber2014critical}. For quadruplet annihilation, $4A \to \emptyset$, or higher, $d_c < 1$, so that spatial fluctuations are irrelevant and the mean-field decay $n(t)\sim (\Gamma_{4\alpha} t)^{-1/3}$ is observed in all dimensions both for the diffusion-limited and reaction-limited regimes. Note that, in the classical reaction-limited prediction, time is rescaled according to the reaction rate as $\Gamma_{k\alpha} t$ ($k=2,3,4$), differently from the diffusion-limited case, where time is rescaled as $\Omega t$ via the diffusion rate. To sum up, the classical diffusion-limited predictions for the decay of the density $n(t)$ in time in one dimensional spatially homogeneous systems are 
\begin{subequations}
\begin{align}
n(t) &\sim (\Omega t)^{-1/2}, \qquad \quad \quad 2A\to \emptyset, \\
n(t) &\sim \left[\ln (\Omega t)/\Omega t\right]^{1/2}, \quad \, \, \,  3A\to\emptyset, \\
n(t) &\sim (\Gamma_{4\alpha} t)^{-1/3}, \qquad \quad    \, 4A\to\emptyset, \\
&\, \, \mbox{(classical diffusion limited)}. \nonumber
\end{align} 
\label{eq:classical_diff_lim_intro}
\end{subequations}

Quantum RD systems, where particles move via coherent hopping while subject to dissipative reactions, represent a class of dynamical processes which are currently under intense investigation. Annihilation reactions have, indeed, a direct connection to cold-atomic experiments with two \cite{lossexp0,lossexp3,lossexp4,lossexpF1,lossexpF2,lossexpF3,lossexp5,lossexp5bis}, three \cite{lossexp2,lossexp6,lossexp7}, and four \cite{lossexp8,lossexp9} body losses, which recently received significant attention also at a theoretical level \cite{diehl2008quantum,lossth4,lossth5,lossth7,lossth8,lossth8bis,lossth9,lossth1,lossth2,lossth6,lossth3}. Analyzing their universal properties is particularly challenging since these investigations require the consideration of dissipative quantum dynamics of large systems at long times. Such dynamics, and that of similar dissipative models, such as those with kinetic constraints \cite{lesanovsky2013,olmos2014,everest2016,marcuzzi2016,buchhold2017,gutierrez2017,roscher2018,wintermantel2020,helmrich2020,nigmatullin2021,kazemi2021,carollo2022quantum}, serve as benchmark problems for numerics \cite{RDHorssen,carollo2019,gillman2019,gillman2020,jo2021} and for quantum simulators \cite{quantum_simulator_1,quantum_simulator_2,quantum_simulator_3,quantum_simulator_4}. Due to these difficulties, analytical predictions for the quantum analogue of the diffusion-limited regime \eqref{eq:classical_diff_lim_intro} are currently still missing. The quantum diffusion-limited regime $\Gamma/\Omega=1$ has been, so far, tackled only numerically for $2A\to \emptyset$ in one dimension in Ref.~\cite{RDHorssen} via exact diagonalization. Here numerical simulations found that for a completely filled initial state (unit filling fraction of the lattice) the density decays algebraically with an exponent estimated to be between $1/2$ and $1$. Such result is, however, affected by finite-size effects since a maximum of $22$ lattice sites (and particles) have been considered in Ref.~\cite{RDHorssen}.

In recent work, Ref.~\cite{QRD20222}, we presented an analytical study of quantum RD systems in their reaction-limited regime. For classical systems, the reaction-limited regime of $k$-body annihilation ($kA\to\emptyset$) is simply described by mean-field power-law decay in any spatial dimension: 
\begin{align}
\label{eq:classical_reaction_lim_intro}
n(t) &\sim  (\Gamma_{k\alpha} t)^{-1/(k-1)}, \quad \quad kA \to \emptyset, \\ 
&(\mbox{classical reaction limited}). \nonumber
\end{align}
In Ref.~\cite{QRD20222}, we showed that power-law behavior distinct from mean field \eqref{eq:classical_reaction_lim_intro} can occur for binary reactions, in contrast to the classical case. In this paper, we address the same
question for the quantum reaction-limited dynamics of three, $3A\to \emptyset$, and four-body, $4A\to \emptyset$, reactions, where spatial
fluctuations are expected to be irrelevant already in one
dimension at the classical level. We show that also for three and four-body reactions power-law behavior beyond mean field is present due to quantum coherences. In particular, we find the following results
\begin{subequations}
\begin{align}
n(t) &\sim (\Gamma_{2\alpha} t)^{-1/2}, \qquad  \,\, \,\,  2A\to \emptyset, \label{eq:q_react_lim_2}\\
n(t) &\sim (\Gamma_{3\alpha}t)^{-0.25}, \,\,\,\,\,\,\,\,\,\,\,\,\,\,\, 3A\to\emptyset, \label{eq:q_react_lim_3} \\
n(t) &\sim (\Gamma_{4\alpha} t)^{-0.1}, \quad \,\,\,\,\,\,\,\,\,\,\: 4A\to\emptyset, \label{eq:q_react_lim_4} \\
&  \mbox{(quantum reaction limited)}. \nonumber 
\end{align}    
\label{eq:quantum_reaction_lim_intro}%
\end{subequations}
Note that Eq.~\eqref{eq:q_react_lim_3} is valid only for intermediate times $\Gamma_{3\alpha}t \lesssim 10^5$.

\begin{figure}[t]
    \centering
\includegraphics[width=0.85\columnwidth]{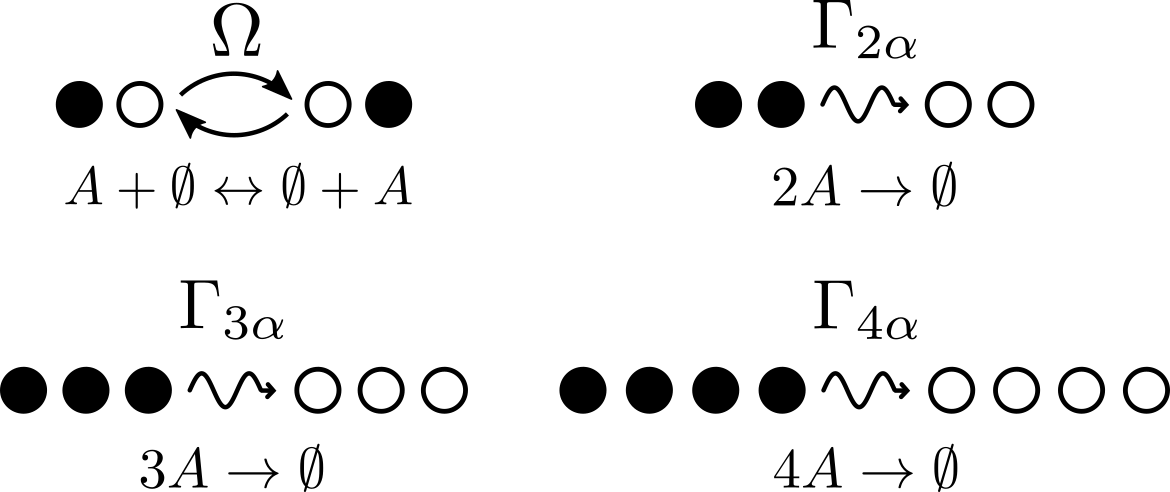}
    \caption{\textbf{Sketch of quantum RD processes}. We consider a fermionic chain where each lattice site is either empty $n_j \ket{\cdots \circ_j \cdots}=0$ or filled $n_j \ket{\cdots \bullet_{j} \cdots}=\ket{\cdots \bullet_j \cdots}$ with a fermion, with $n_j$ the number operator at site $j$. Reactions are irreversible and they are modelled through the jump operators of the Lindblad dynamics in Eqs.~\eqref{eq:master_equation} and \eqref{eq:dissipator}. In particular, we consider binary $2A\to\emptyset$ \eqref{eq:2annihilation} (rate $\Gamma_{2\alpha}$), triplet $3A\to\emptyset$ \eqref{eq:3annihilation} (rate $\Gamma_{3\alpha}$) and quadruplet $4A\to\emptyset$ \eqref{eq:4annihilation} (rate $\Gamma_{4\alpha}$) annihilation of neighbouring particles. Coherent-Hamiltonian hopping \eqref{eq:free_fermion_Hamiltonian} between neighbouring sites at rate $\Omega$ replaces classical diffusion. The particle density $\braket{n(t)}$ decays at long times towards the vacuum $\ket{\circ \circ \dots \circ \circ}$ in a universal way.  This critical decay has the root cause into quantum coherences, which can be present in any space dimensionality. Consequently, in quantum RD systems, non mean-field decay is also present for $d \geq d_c$. This is in stark contrast with classical RD systems, where critical dynamics is solely determined by spatial fluctuations in the density profile and non mean-field behavior is possible only for $d<d_c$.
    }    \label{fig:sketches_RD_processes}
\end{figure}

The results in Eq.~\eqref{eq:quantum_reaction_lim_intro} are obtained by considering the quantum RD dynamics of fermionic chains, where particles coherently hop on a lattice and are subject to dissipative annihilation reactions, as sketched in Fig.~\ref{fig:sketches_RD_processes}. The fermionic statistics naturally embodies the single occupancy constraint of each lattice site, which is often taken in the classical RD literature \cite{henkel2008non,hinrichsen2000non,vladimir1997nonequilibrium,krapivsky2010kinetic}. The dynamics is ruled by the quantum master equation \cite{gorini1976,lindblad1976,breuer2002}, where coherent hopping is given by a quadratic, number-conserving, Hamiltonian for free fermions. The latter replaces the diffusive transport of particles present in the classical RD models. It is important to stress that quantum coherent hopping gives, instead, rise to ballistic transport of particles. We will, however, refer henceforth to the open quantum systems we introduce here as quantum RD models. This terminology allows us, indeed, to connect with classical RD systems, with whom our models share some important features such as algebraic scaling in time of the density and mean-field decays (for incoherent initial conditions as we detail below). We analytically study the quantum RD dynamics in the thermodynamic limit in the reaction-limited regime by using the time-dependent generalized Gibbs ensemble approach (TGGE) \cite{tGGE1,tGGE2,tGGE3,tGGE4}. The TGGE method applies also to bosons. In the case of bosons freely hopping on the lattice, the reaction-limited asymptotic decay exponents for $k$-body onsite annihilation reactions coincide with those predicted by mean field \eqref{eq:classical_reaction_lim_intro}, see, e.g., Ref.~\cite{lossth3}. The case of fermions reveals, instead, a richer behavior, which we characterize in this paper.

In order to introduce the problem, we first review the case of binary annihilation, $2A\to \emptyset$, which had been previously studied with the TGGE method in Refs.~\cite{lossth1,lossth2,lossth6}. In this case, the exponent of the power-law decay \eqref{eq:q_react_lim_2} deviates from the mean field \eqref{eq:classical_reaction_lim_intro} ($k=2$) one as a consequence of 
quantum coherences in the initial state \cite{QRD20222}. This result is in agreement with the results of Refs.~\cite{lossth1,lossth2,QRD20222}, where the $1/2$ decay exponent has been worked out analytically from the asymptotics of the density decay. For the case of three-body annihilation, $3A\to \emptyset$, and four-body annihilation, $4A\to \emptyset$, we find that the effect of quantum coherences is even more dramatic. For $3A\to \emptyset$, we find that power-law behavior \eqref{eq:q_react_lim_3} (also distinct from mean field \eqref{eq:classical_reaction_lim_intro} with $k=3$)
is only transient for times of the order $\Gamma_{3\alpha}t\lesssim 10^5$, with an asymptotic decay of the  density for longer times, $\Gamma_{3\alpha}t \gtrsim 10^5$, which is not a power-law. The nonalgebraic long-time correction might be related to the fact that the associated classical (and quantum) diffusion-limited dynamics acquires a logarithmic correction. Our analysis, however, shows that the quantum decay is richer and the correction to the algebraic time scaling is slower than just a  logarithm. From Eq.~\eqref{eq:quantum_reaction_lim_intro}, we also find that a simple heuristic formula relating the decay exponent to the number $k$ is not possible. The case $3A\to \emptyset$, indeed, shows that in the quantum reaction-limited case the asymptotic decay does not necessarily take a power-law form. This contrasts with classical RD \eqref{eq:classical_reaction_lim_intro}, where the exponent is simply $1/(k-1)$. For $4A\to \emptyset$, we find a non mean-field power-law decay \eqref{eq:q_react_lim_4} for initial states that possess quantum coherences, similarly to $2A\to\emptyset$ case \cite{QRD20222}. For $4A\to\emptyset$, the non mean-field result \eqref{eq:q_react_lim_4} is even more surprising given that spatial fluctuations are irrelevant classically in one dimension. We emphasize that both the decay exponents $0.25$ in Eq.~\eqref{eq:q_react_lim_3} and $0.1$ in Eq.~\eqref{eq:q_react_lim_4} are approximate values, which we obtain from numerically computed effective decay exponents. In the triplet and quadruplet annihilation cases, indeed, the dynamical equation for the density following from the TGGE is more complicate than that for $2A\to\emptyset$ and the asymptotics of the density decay cannot be simply performed.

The aforementioned analysis for the results \eqref{eq:quantum_reaction_lim_intro} is exemplified by considering Fermi-sea (FS) coherent initial states with initial density of particles $n_0$. The result \eqref{eq:quantum_reaction_lim_intro} for binary, triplet and quadruplet annihilation holds for any value of $n_0 \neq 1$, i.e., whenever the FS state features quantum coherences in real space. Furthermore, our results apply, more generically, to initial states of the GGE form, which can be both pure and mixed. The necessary requirement being that the associated initial momentum occupation function is not flat in momentum. In real space, these GGE initial states are identified by a nondiagonal fermionic two-point correlation matrix and they therefore possess quantum coherences. For initial GGE states with a flat initial momentum occupation function, on the contrary, the mean-field result \eqref{eq:classical_reaction_lim_intro} is recovered. The results in Eq.~\eqref{eq:quantum_reaction_lim_intro} therefore represent a robust and universal feature of the quantum reaction-limited dynamics ensuing from coherent initial states.

Our results for $3A\to\emptyset$ and $4A\to\emptyset$ show that the universal quantum RD behavior is not solely determined by the relevance of spatial fluctuations, as in the classical case, but it also depends on quantum effects which are present regardless of space dimensionality (such as coherences in the initial state). Non mean-field universal behavior is therefore possible in quantum systems even for $d\geq d_c$ and in the absence of spatial fluctuations. This fact contrasts with the standard picture of critical phenomena, according to which the emergence of universal behavior in many-body systems is necessarily rooted into the relevance of spatial correlations and it is therefore present only in $d<d_c$ \cite{wilson1974renormalization,Kogut1979,tauber2005applications,tauber2014critical,doi1976stochastic,peliti1985path,peliti1986renormalisation,QFT_RD_1998}.. The analysis of this paper, consequently, allows us to unambiguously pinpoint a new mechanism, based on quantum coherence, underlying the emergence of the richer universal behavior of quantum many-body systems compared to that known for classical ones.

The remainder of the paper is organized as follows. In Sec.~\ref{sec:system}, we formulate quantum RD dynamics with annihilation reactions by means of a quantum master equation approach. In Sec.~\ref{sec:reaction_lim}, we first briefly discuss known results about classical RD dynamics. We then move to the reaction-limited regime of quantum RD dynamics and the associated analysis using the TGGE method. This method is used in Sec.~\ref{sec:results_annihilation_234} to address the annihilation reactions $2A\to\emptyset$, $3A\to\emptyset$ and $4A\to\emptyset$. In Sec.~\ref{sec:conclusion}, we report our conclusions. The Appenices \ref{app:3ann}, for $3A\to \emptyset$, and \ref{app:4ann}, for $4A\to\emptyset$, contain technical aspects concerning the TGGE calculations and additional results for generic fillings $n_0$ of the FS initial state.

\section{The system}
\label{sec:system}
We consider a one-dimensional lattice with $L$ sites and periodic boundary conditions. Each site $j$ can be either occupied by a fermion 
$n_j \ket{\cdots \bullet_{j} \cdots}=\ket{\cdots \bullet_j \cdots}$ or be empty $n_j \ket{\cdots \circ_j \cdots}=0$. Here, $n_j=c_j^{\dagger}c_j$ is the number operator and the operators $c_j,c_j^{\dagger}$ obey the fermionic anticommutation relations $\{c_j,c_{j'}^{\dagger}\}=\delta_{j,j'}$. Particles occupying adjacent sites may be lost into the environment through an annihilation reaction. Therefore, the ensuing dynamics is not unitary and we assume it to be governed by the quantum master equation \cite{gorini1976,lindblad1976,breuer2002} ($\hbar=1$ henceforth) 
\begin{equation}
\dot{\rho}(t)=-i[H,\rho(t)]+\mathcal{D}[\rho(t)]. 
\label{eq:master_equation}   
\end{equation}
Here, $\rho$ is the density matrix, $H$ is the quantum Hamiltonian, and the irreversible reaction processes are encoded in the dissipator $\mathcal{D}$, which we take in Lindblad form \cite{gorini1976,lindblad1976,breuer2002}
\begin{equation}
\mathcal{D}[\rho]=\sum_{j} \left[L_{j}^{\nu}\rho {L_{j}^{\nu}}^\dagger-\frac{1}{2}\left\{{L_{j}^{\nu}}^\dagger L_{j}^{\nu},\rho \right\}\right].
\label{eq:dissipator}    
\end{equation}
The $L_j^{\nu}$ are local jump operators. We consider here three different types of reaction processes, namely, binary $2A\to \emptyset$ ($\nu=2\alpha$), triplet $3A\to \emptyset$ ($\nu=3\alpha$) and quadruplet $4A \to \emptyset$ annihilation ($\nu=4\alpha$). These processes are pictorially represented in Fig.~\ref{fig:sketches_RD_processes}. In particular, for binary annihilation, $2A \to \emptyset$, of a pair of neighboring particles at rate $\Gamma_{2\alpha}$, we have
\begin{equation}
L_j^{2\alpha}=\sqrt{\Gamma_{2\alpha}}\,c_j c_{j+1}.
\label{eq:2annihilation}
\end{equation}
This process corresponds to two-body losses, which can be implemented in cold atomic gases, e.g., via inelastic scattering interactions \cite{lossexp0,lossexp3,lossexp4,lossexpF1,lossexpF2,lossexpF3}, or photoassociation into excited compounds \cite{lossexp5,lossexp5bis}. 

We also consider annihilation of triplets of neighbouring particles $3A \to \emptyset$ at rate $\Gamma_{3\alpha}$
\begin{equation}
L_j^{3\alpha}=\sqrt{\Gamma_{3\alpha}}\,c_j c_{j+1} c_{j+2}.
\label{eq:3annihilation}
\end{equation}
This triplet annihilation is also present in cold atomic gases, where it is caused by recombination of atoms into molecules \cite{lossexp2,lossexp6,lossexp7}. Finally, we consider annihilation of quadruplets of neighbouring particles $4A \to \emptyset$ at rate $\Gamma_{4\alpha}$, which is described by the jump operator
\begin{equation}
L_j^{4\alpha} = \sqrt{\Gamma_{4\alpha}}\,c_j c_{j+1} c_{j+2} c_{j+3}.
\label{eq:4annihilation}
\end{equation}    
Such four-body losses have also been experimentally detected in cold atomic gases \cite{lossexp8,lossexp9}. 

For all the three losses mechanisms in Eqs.~\eqref{eq:2annihilation}-\eqref{eq:4annihilation}, the Lindblad master equation \eqref{eq:master_equation}-\eqref{eq:dissipator} [with the Hamiltonian introduced below in Eq.~\eqref{eq:free_fermion_Hamiltonian}] we employ should be understood as an effective description of the dynamics. The derivation of the Lindblad equation from a microscopic weak system-bath coupling procedure (see, e.g., Ref.~\cite{breuer2002}) for the jump operators \eqref{eq:2annihilation}-\eqref{eq:4annihilation} is an open problem. This derivation has been so far only pursued in Ref.~\cite{lossth9} for one-body decay $L_j^{\alpha}=\sqrt{\Gamma_{\alpha}}c_j$. 

We also mention that the form of the jump operators \eqref{eq:2annihilation}-\eqref{eq:4annihilation} is dictated by the fermionic statistics, which forces annihilation reactions to occur between neighbouring particles since double (or higher) occupancy of the same lattice site is not possible. This is different from the case of bosonic systems, where annihilation reactions are typically defined between particles residing on the same site. These processes are therefore modelled \cite{lossth4,lossth5,lossth7,lossth1,lossth2,lossth3} for bosons through jump operators of the form $L_j^{k\alpha}=\sqrt{\Gamma_{k\alpha}}\, b_j^{k}$, with $k=2,3,4$ and $b_j$ a bosonic destruction operator.  

For the coherent dynamics in Eq.~\eqref{eq:master_equation}, we take quantum hopping between adjacent lattice sites at a rate $\Omega$ (cf. Fig.~\ref{fig:sketches_RD_processes}):
\begin{equation}
H=-\Omega\sum_{j=1}^{L}(c_j^{\dagger} c_{j+1}+c_{j+1}^{\dagger}c_j) \, .
\label{eq:free_fermion_Hamiltonian}
\end{equation}
This Hamiltonian replaces the diffusive motion of particles considered in classical RD dynamics \cite{henkel2008non,hinrichsen2000non,vladimir1997nonequilibrium,tauber2002dynamic,tauber2005applications,tauber2014critical} and generates, differently from the classical case, ballistic transport. The Hamiltonian $H$, however, provides a ``natural'' quantum generalization of the classical RD dynamics since it describes noninteracting motion of particles and it is number conserving, like classical diffusion. For these reasons, we will refer henceforth to the dynamics in Eqs.~\eqref{eq:master_equation}-\eqref{eq:free_fermion_Hamiltonian} as quantum RD dynamics. The Hamiltonian \eqref{eq:free_fermion_Hamiltonian} is quadratic and it can be exactly diagonalized by Fourier transform. The whole Lindblad dynamics \eqref{eq:master_equation} and \eqref{eq:dissipator}, however, is not quadratic for the jump operators in Eqs.~\eqref{eq:2annihilation}-\eqref{eq:4annihilation} and therefore it cannot be solved exactly. 

Quantum hopping introduces coherence in the dynamics, which is generically expected to affect emergent universal dynamical behavior. For the purpose of this work we quantify the latter through the long-time asymptotic decay of the particle density $\braket{n(t)}=\braket{N(t)}/L$, where $N(t)=\sum_{j}n_j(t)$ the total particle number at time $t$. The idea is as follows: While $N$ is conserved by the Hamiltonian, $[H,N]=0$, it is not conserved by the Lindblad dynamics due to the reactions \eqref{eq:2annihilation}-\eqref{eq:4annihilation}. They deplete the system, taking it towards a trivial vacuum stationary state. The nontrivial universal behavior lies in the way this stationary state is approached at long times, e.g., through a density decaying via a power law. The functional form of this asymptotic behavior is controlled by the relative strength, $\Gamma/\Omega$, of incoherent dissipation with respect to coherent Hamiltonian hopping, as we explain in the next Section. 

\section{Diffusion and reaction-limited dynamics}
\label{sec:reaction_lim}
The RD dynamics is characterized by two fundamental timescales. On the one hand, the reaction time $\sim \Gamma^{-1}$, gives the time needed for two nearby particles to react. On the other hand, the diffusion [hopping in the quantum case of Eq.~\eqref{eq:free_fermion_Hamiltonian}] time $\sim \Omega^{-1}$ yields the characteristic time needed by two separated particles to get close to each other. The dynamics is \textit{diffusion-limited} when the ratio $\Gamma/\Omega$ is at least $\Gamma/\Omega \sim 1$, 
while it is \textit{reaction-limited} when $\Gamma/\Omega \ll 1$. The universal long-time decay of the particle density $\braket{n(t)}$ is markedly different in the two regimes. In Subsec.~\ref{subsec:classical_RD} and \ref{subsec:quantum_RD}, we briefly recall some previous results concerning the RD dynamics in the diffusion and reaction-limited regimes for classical and quantum systems, respectively.

\subsection{Classical RD dynamics}
\label{subsec:classical_RD}
For classical RD systems, the reaction-limited regime is described by mean field \cite{hinrichsen2000non,vladimir1997nonequilibrium,Redner1984,fastdiffusion1992,krapivsky2010kinetic}. This is due to the fast diffusion mixing, which renders the particle density homogeneous in space. Spatial fluctuations in particle concentration are rapidly smoothed out and reactions can therefore take place everywhere with the same probability. The ensuing mean-field description, for the case of annihilation processes of $k$ particles $kA\to \emptyset$, is given by the law of mass action equation 
\begin{equation}
\frac{\mbox{d}
\braket{n}}{\mbox{d}t} = -k \, \Gamma_{k\alpha} \braket{n}^k, \, \, \braket{n} \sim (\Gamma_{k\alpha}t)^{-1/(k-1)}.
\label{eq:MF_rate_equation}
\end{equation}
In this limit the density depends therefore only on the rescaled time $\tau=\Gamma_{k\alpha}t$.

In the diffusion-limited regime, local density fluctuations are relevant for the dynamics. For this reason, the exponent of the density decay generically deviates from the mean-field prediction \eqref{eq:MF_rate_equation} \cite{toussaint1983particle,Spouge1988,Privman1994,krapivsky2010kinetic,tauber2002dynamic,tauber2005applications,tauber2014critical,doi1976stochastic,peliti1985path,peliti1986renormalisation}. This non mean-field universal behavior simply comes from the fact that at long times the dynamics relies on few remaining particles separated by large distances. In the classical case, the decay is, consequently, controlled by the large-distance properties of a random-walk, i.e., by the probability of far apart particles to meet. This is a universal quantity which depends only on the diffusion constant $\Omega$, a macroscopic quantity, and on the space dimensionality $d$, which sets the dimensionality of the underlying diffusion process. 

The random walk is recurrent in $d\le 2$ \cite{feller1957introduction}, which implies that in continuum space two particles can surely meet. This identifies the upper-critical dimension $d_c=2$ for binary annihilation $2A \to \emptyset$. The associated decay in $d=1<d_c$ is 
\begin{equation}
\braket{n(t)} \sim (\Omega t)^{-1/2}, \, \, 2A \to \emptyset, \, \, \, d=1.
\label{eq:diffusion_limited_2ann}
\end{equation}
For $d>2$, diffusive mixing is effective, as particles far apart in the continuum do not meet, universality is lost and the mean-field description \eqref{eq:MF_rate_equation} is recovered. 

For reactions $kA\to \emptyset$ involving $k>2$ particles, the upper critical dimension is $d_c=2/(k-1)$ \cite{tauber2002dynamic,tauber2005applications,tauber2014critical}, as space dimensionality must be further lowered in order to constrain a larger number of particles to meet. For triplet annihilation $3A\to\emptyset$, $d_c=1$. In $d=d_c=1$, the mean-field power law \eqref{eq:MF_rate_equation} is valid up to a logarithmic correction
\begin{equation}
\braket{n(t)} \sim \left(\frac{\ln \Omega t}{\Omega t}\right)^{1/2}, \, \, 3A \to \emptyset, \, \, \, d=1.
\label{eq:diffusion_limited_3ann}
\end{equation}
Both in Eq.~\eqref{eq:diffusion_limited_2ann} and \eqref{eq:diffusion_limited_3ann}, the density depends on the rescaled time $\tau=\Omega t$, which differs to the scaling valid in the reaction-limited regime \eqref{eq:MF_rate_equation}. 

For quadruplet annihilation $4A \to \emptyset$, instead, $d_c<1$ and the mean-field description applies also in the one-dimensional $d=1>d_c$ diffusion-limited dynamics
\begin{equation}
\braket{n(t)} \sim (\Gamma_{4\alpha} t)^{-1/3}, \, \, 4A \to \emptyset, \, \, \, d=1.
\label{eq:diffusion_limited_4ann}\end{equation}
The mean-field decay is similarly valid also for five-body, and higher-order, annihilation. 

We therefore see that in classical RD systems universal non mean-field behavior can only arise in the diffusion-limited regime as a consequence of spatial fluctuations in low dimensions. In the quantum RD dynamics, this is not the case: Universal behavior that departs from mean field is not only due to spatial fluctuations, but can also be present in the the reaction-limited regime where spatial fluctuations are smoothed out \cite{QRD20222}. We will explore this in the next section for general multi-body annihilation reactions.

\subsection{Quantum RD dynamics}
\label{subsec:quantum_RD}
Currently, very little is known about quantum RD dynamics. Equations \eqref{eq:master_equation} and \eqref{eq:dissipator} cannot be analytically solved since the Lindbladian is not quadratic due to the structure of the reaction jump operators \eqref{eq:2annihilation}-\eqref{eq:4annihilation}. Likewise, numerical simulations are hard due to the exponential scaling of the Hilbert space dimension with the system size: in contrast to the classical case where one can generate trajectories of configurations, a quantum trajectory unfolding of 
\eqref{eq:master_equation} requires the propagation of whole state, making large-scale numerics unfeasible. 

The diffusion-limited regime, with $\Gamma/\Omega=1$, for binary annihilation $2A\to \emptyset$ [cf. Eqs.~\eqref{eq:2annihilation} and \eqref{eq:free_fermion_Hamiltonian}] has been addressed numerically via exact diagonalization up to $L=22$ sites in Ref.~\cite{RDHorssen}. The system has been therein initialized in product state $\ket{\bullet \bullet  \dots \bullet \bullet}$ with unitary filling. The density is found to decay algebraically in time $\braket{n(t)} \sim t^{-b}$, with $1/2<b<1$. The decay exponent is different from the mean-field prediction, which for quantum systems is analogous to the classical one \eqref{eq:MF_rate_equation}, showing that the dynamics is controlled in one-dimension by spatial fluctuations. Quantum effects seem in this case to affect the decay exponent by making it presumably larger than the corresponding classical value --- $1/2$ --- in Eq.~\eqref{eq:diffusion_limited_2ann}, though extrapolation of the finite-size results to the thermodynamic limit is not trivial. 

The reaction-limited regime $\Gamma/\Omega \ll 1$ has been only recently studied in Ref.~\cite{QRD20222}. Using an analytical approach, this study could show that for binary reactions, such as $2A\to \emptyset$, the quantum reaction-limited dynamics is not always described by the mean-field approximation \eqref{eq:MF_rate_equation}. This turns out to be the case whenever quantum coherences in the initial state are present. The analysis of Ref.~\cite{QRD20222} is based on the time-dependent generalized Gibbs ensemble method (TGGE) introduced in  Refs.~\cite{tGGE1,tGGE2,tGGE3,tGGE4}. In the following, we briefly recall the main aspects of the TGGE, as this method will be employed in the next sections.

The TGGE approach is based on a separation of timescales between  reactions and hopping dynamics, which is possible in the limit $\Gamma/\Omega\ll 1$. Here, reactions are slow and long time intervals, on average, elapse between consecutive reaction events. The much faster hopping dynamics \eqref{eq:free_fermion_Hamiltonian} can thus be integrated out by considering the state $\rho(t)$ in between consecutive reactions as being relaxed with respect to the Hamiltonian $H$: $[H,\rho(t)]=0$. The time dependence of the state $\rho(t)$ accounts for the remaining slow evolution, taking place on the timescale $\Gamma^{-1}$, due to the reactions. The TGGE method then moves forward by making an ansatz for the state $\rho(t)=\rho_{\mathrm{GGE}}(t)$ of the form of a generalized Gibbs ensemble, see, e.g., the reviews \cite{GGErev1,GGErev2}.  In the case of the Hamiltonian \eqref{eq:free_fermion_Hamiltonian}, the GGE takes the form
\begin{equation}
\rho_{\mathrm{GGE}}(t)= \frac{1}{\mathcal{Z}(t)}\mbox{exp}\left(-\sum_{k}\lambda_k(t)\hat{n}_k\right),
\label{eq:tGGE_free_fermions}
\end{equation}
where $\mathcal{Z}(t)=\prod_{k}[1+e^{-\lambda_k(t)}]$. In the previous equation, $\lambda_{k}(t)$ are dubbed Lagrange mulipliers (or generalized inverse temperatures), $k\in (-\pi,\pi)$ is the quasimomentum, and $\hat{n}_k=\hat{c}_k^{\dagger} \hat{c}_k$ is the number operator in Fourier space, with $\hat{c}_k$ ($\hat{c}_k^{\dagger}$) fermionic desctruction (creation) operators (see Appendix \ref{app:3ann}). We note that the GGE state \eqref{eq:tGGE_free_fermions} gives direct access to the RD dynamics in the thermodynamic limit, as it describes the local relaxation of the system, i.e., the expectation $\braket{\dots}_{\mathrm{GGE}}(t)$ provides the exact average behavior of local observables in the thermodynamic limit. The state \eqref{eq:tGGE_free_fermions} is Gaussian and diagonal in momentum space. Its dynamics is therefore fully characterized by the momentum occupation functions $\braket{\hat{n}_q}_{\mathrm{GGE}}(t)=C_q(t)=1/[\mbox{exp}(\lambda_q(t))+1]$, which obey the equations \cite{lossth1,lossth2,lossth3,lossth6,QRD20222} 
 
\begin{equation}
\frac{\mbox{d} C_q(t)}{\mbox{d}t} = \sum_{j} \braket{{L_j^{\nu}}^{\dagger}[\hat{n}_q,L_j^{\nu}]}_{\mathrm{GGE}}(t),\,\quad  \forall q.
\label{eq:tgge_rate_equation_general}
\end{equation}
We remark that the previous equation provides a large reduction of complexity for the characterization of the many-body dynamics as it encodes the description of the dynamics in terms of the single function $C_q(t)$, rather than into the whole density matrix $\rho$ of the original Lindblad equation \eqref{eq:master_equation} and \eqref{eq:dissipator}, which is not tractable for a many-body system. Equation \eqref{eq:tgge_rate_equation_general} therefore represents an effective equation describing the large-scale, long times and large distances, physics of the model. It is also worth noting that the GGE-Gaussian form \eqref{eq:tGGE_free_fermions} holds equally for noninteracting bosons, i.e., Eq.~\eqref{eq:free_fermion_Hamiltonian} with the replacement $c_j \to b_j$ and $b_j$ a bosonic destruction operator. In this case, cf. Ref.~\cite{lossth3}, $k$-body losses $L_j=\sqrt{\Gamma_{k\alpha}}\,b_j^k$ yield the law of mass action asymptotic exponent \eqref{eq:MF_rate_equation}. 

The structure of Eq.~\eqref{eq:tgge_rate_equation_general} clearly also shows that $C_q(\tau)$ is a function of the rescaled time $\tau=\Gamma_{\nu} t$ according to the reaction rate, as in Eq.~\eqref{eq:MF_rate_equation} for the classical reaction-limited regime. This aspect further shows that the TGGE method is naturally apt to describe the reaction-limited dynamics.

\section{Quantum reaction-limited annihilation dynamics} 
\label{sec:results_annihilation_234}
In this Section, we present our results for the quantum annihilation process in the reaction-limited regime. In order to set the stage, we first discuss in Subsec.~\ref{sec:2ann} the binary annihilation $2A\to\emptyset$ case of Eq.~\eqref{eq:2annihilation}. We then consider in Subsec.~\ref{sec:3ann}, triplet annihilation $3A\to\emptyset$ in Eq.~\eqref{eq:3annihilation}. In Subsec.~\ref{sec:4ann}, quadruplet annihilation $4A\to \emptyset$ of Eq.~\eqref{eq:4annihilation} is eventually discussed.  

In all the three cases, we solve Eq.~\eqref{eq:tgge_rate_equation_general} considering two different classes of initial states. First, the Fermi sea (FS), i.e., the ground state of the Hamiltonian \eqref{eq:free_fermion_Hamiltonian} with an initial filling $n_0$, which is uniquely identified by the occupation functions 
\begin{equation}
C_q(t=0)=    \left\{
  \begin{array}{lr}
    1 \; \; \; \; \mbox{if} \, \, q\in [-\pi n_0,\pi n_0], \; \;  \\
    0 \; \; \; \;  \mbox{otherwise}.
  \end{array}
\right.
\label{eq:FS_initial_state}
\end{equation}
This initial state displays quantum coherences in real space, i.e., the associated density matrix possesses off-diagonal elements in the Fock-space basis $\prod_{j\in\Lambda}c_j^\dagger \ket{\circ \circ \circ \dots \circ}$, with $\Lambda$ an arbitrary set of lattice sites. In the case $n_0=1$, $\Lambda=\{1,2,\dots L\}$ and the state corresponds to the simple product state $\ket{\bullet \bullet \bullet \dots \bullet}$, where every lattice site is filled. Thus, as long as $n_0\neq1$ the FS initial state is, instead, quantum coherent. 

Second, we consider also initial incoherent states of the form $\rho_0 =\mbox{exp}(-\lambda N)/\mathcal{Z}_0$. This initial state, differently from the FS \eqref{eq:FS_initial_state}, is  diagonal in the (classical) basis introduced above and it is associated with a momentum-independent occupation function equal to the initial filling $n_0$:
\begin{equation}
C_q(0)=n_0.
\label{eq:initial_flat_incoherent}
\end{equation}
In both the cases \eqref{eq:FS_initial_state} and \eqref{eq:initial_flat_incoherent}, the density of particles $\braket{n(\tau)}_{\mathrm{GGE}}$ is computed from the momentum occupation function $C_q(\tau)$ as
\begin{equation}
\braket{n(\tau)}_{\mathrm{GGE}}=\frac{1}{L}\sum_{q}C_q(\tau).
\end{equation}
The decay exponent of the particle density as a function of time is quantified by computing the effective exponent $\delta_{\mathrm{eff}}(\tau)$, which is defined as \cite{hinrichsen2000non}
\begin{equation}
\delta_{\mathrm{eff}}(\tau)=-\frac{\log\left(\braket{n(b\tau)}_{\mathrm{GGE}}/\braket{n(\tau)}_{\mathrm{GGE}}\right)}{\log b},  \label{eq:effective_exponent}  
\end{equation}
with $b$ a scaling parameter. In the case of an asymptotic in time power-law behavior $\braket{n(\tau)}_{\mathrm{GGE}} \sim a \tau^{-\delta}$, the effective exponent $\delta_{\mathrm{eff}}(\tau)$ converges at long times to the exponent $\delta$ of the power law. In all the calculations of $\delta_{\mathrm{eff}}(\tau)$ in this paper, we use $b=2$.  

\subsection{Binary annihilation}
\label{sec:2ann}
The application of Eq.~\eqref{eq:tgge_rate_equation_general} to binary annihilation $2A\to \emptyset$ \eqref{eq:2annihilation} leads to the evolution equation \cite{QRD20222}
\begin{equation}
\frac{\mbox{d}C_q(\tau)}{\mbox{d}\tau}=-\frac{1}{L}\sum_k f_{2}(k,q) C_k(\tau) C_q(\tau),
\label{eq:ann_rate_equation} 
\end{equation}
with $\tau=\Gamma_{2\alpha}t$ and the function $f_2(k,q)$ given by
\begin{equation}
f_{2}(k,q)= 2[1-\cos(k-q)].
\label{eq:f2ann}
\end{equation}
In Fig.~\ref{fig:2ann}(a), we plot the density of particles as a function of time for the two different initial conditions \eqref{eq:FS_initial_state} and \eqref{eq:initial_flat_incoherent}. In the case of the incoherent initial state \eqref{eq:initial_flat_incoherent} (red-dashed line), we observe that the momentum distribution function $C_q(\tau)$ remains flat in $q$ at all times so that Eq.~\eqref{eq:ann_rate_equation} exactly reduces to the classical reaction-limited equation \eqref{eq:MF_rate_equation} (with $k=2$ therein). The asymptotic decay exponent is therefore the mean-field one $\braket{n(\tau)}_{\mathrm{GGE}} \sim \tau^{-1}$. The coherences in the initial state \eqref{eq:FS_initial_state}, on the other hand, strongly affect the asymptotic decay (blue-solid line), whose exponent 
\begin{equation}
\braket{n(\tau)}_{\mathrm{GGE}}\sim \tau^{-1/2},
\label{eq:2ann_GGE_FS}
\end{equation}
differs from the mean-field one. The algebraic decay exponent $1/2$ is in agreement with previous results concerning fermionic gases subject to two-body losses in a lattice \cite{lossth1,QRD20222} and in continuum space \cite{lossth2}. In the latter references, the $1/2$ decay exponent \eqref{eq:2ann_GGE_FS} represents an analytical result following from analytical-asymptotic calculations starting from the TGGE rate equation. In Fig.~\ref{fig:2ann}(b), one sees that $\delta_{\mathrm{eff}}(\tau)$ \eqref{eq:effective_exponent} neatly converges to $0.5$. This non mean-field asymptotic decay goes beyond the classical reaction-limited description, captured by the mean-field equation \eqref{eq:MF_rate_equation}, and it is determined by quantum coherences in the initial state. Coherences in the initial state are, indeed, fundamental as they allow for an inhomogeneous in $q$ initial occupation function $C_q(0)$. The latter, indeed, implies that the two-point fermionic correlation matrix $\braket{c_x^{\dagger} c_y}_{\mathrm{GGE}}$ the GGE \eqref{eq:tGGE_free_fermions} are not diagonal in real space. These states therefore possess quantum coherences in real space. For these kinds of initial conditions, the second term in Eq.~\eqref{eq:f2ann} for $f_2(k,q)$ does not vanish in the sum over $k$ in Eq.~\eqref{eq:ann_rate_equation} and it is responsible for the deviation of the power-law exponent from the mean-field description. The nontrivial function $f_2(k,q)$ is determined by the fermionic statistics, which hinders double (or higher) occupancy of lattice sites and it therefore determines the form of the considered jump operators \eqref{eq:2annihilation}. The joint effect of fermionic statistics and quantum coherences in the initial state is thus responsible for the emergence of universal behavior in the quantum RD dynamics beyond mean field. We remark that, because of this reason, the asymptotic decay \eqref{eq:2ann_GGE_FS} holds generically both for mixed and pure initial states, as long as the associated momentum occupation function $C_q(0)$ is not flat in $q$, as shown in Ref.~\cite{QRD20222} (cf. Fig.~S1 in the Supplemental Material therein). In the next Sections, we show that this behavior is even richer in the case of three and four-body annihilation reactions. 

\subsection{Three-body annihilation}
\label{sec:3ann}
In the case of three-body annihilation, $3A \to \emptyset$ \eqref{eq:3annihilation}, the rate equation \eqref{eq:tgge_rate_equation_general} takes the form 
\begin{equation}
\frac{\mbox{d}C_q(\tau)}{\mbox{d}\tau}=-\frac{C_q(\tau)}{L^2}\sum_{k,k'} f_3(k,k',q) C_k(\tau)C_{k'}(\tau),
\label{eq:3ann_rate_equation}
\end{equation}
with the rescaled time $\tau=\Gamma_{3\alpha}t$ and the function $f_3(k,k',q)$ given by 
\begin{equation}
f_3(k,k',q)=2\left[\sin(k-q)-\sin(k-k')+\sin(q-k') \right]^2.
\label{eq:f3_function}
\end{equation}
Equations \eqref{eq:3ann_rate_equation} and \eqref{eq:f3_function} are obtained by first expressing the fermionic operators $c_j$ in $L_j^{3\alpha}$ \eqref{eq:3annihilation} in terms of the Fourier-space operators $\hat{c}_k$ and then computing the right-hand side of Eq.~\eqref{eq:tgge_rate_equation_general} via Wick's theorem, exploiting the fact that the TGGE state \eqref{eq:tGGE_free_fermions} is Gaussian and diagonal in momentum space. The derivation is reported in Appendix \ref{app:3ann}.

\begin{figure}[t]
    \centering
\includegraphics[width=1\columnwidth]{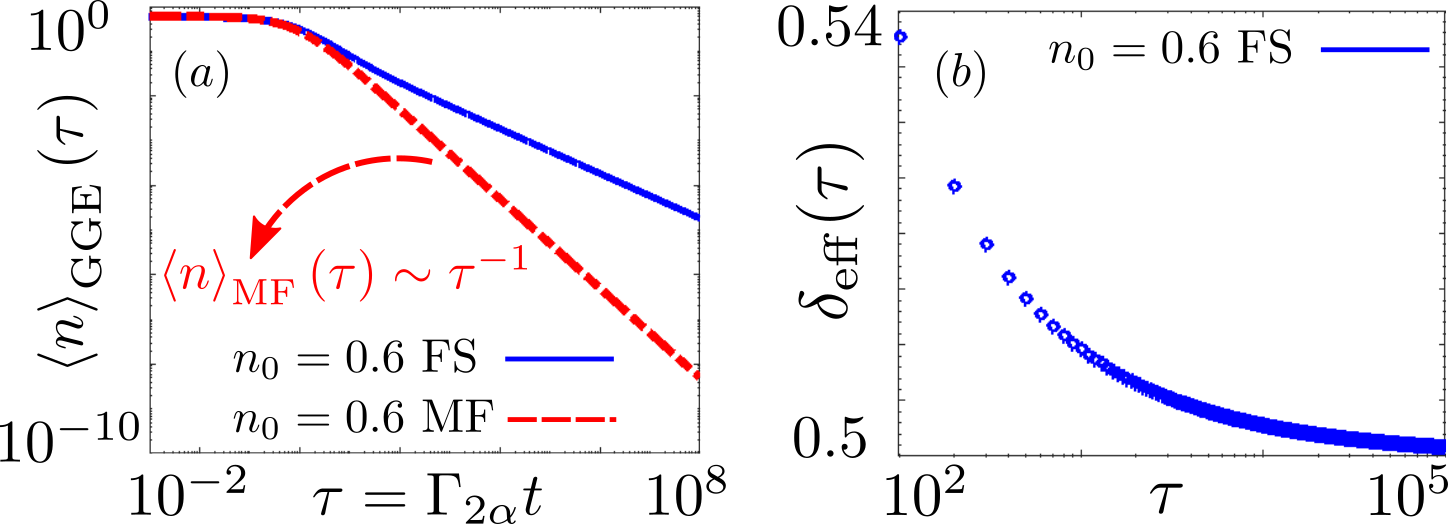}
    \caption{\textbf{Quantum binary annihilation dynamics in the reaction limited regime.} (a) Log-log plot of the density $\braket{n(\tau)}_{\mathrm{GGE}}$ as a function of the rescaled time $\tau=\Gamma_{2\alpha}t$. The blue solid line refers to the initial FS state \eqref{eq:FS_initial_state} at filling $n_0=0.6$, while the red-dashed line corresponds to the initial incoherent state \eqref{eq:initial_flat_incoherent} at the same filling $n_0=0.6$. In the latter case, the TGGE rate equation reduces to the law of mass action \eqref{eq:MF_rate_equation} and the density decays asymptotically as $\braket{n(\tau)}_{\mathrm{GGE}}=\braket{n}_{\mathrm{MF}}(\tau)\sim \tau^{-1}$. In the case of the FS state, on the contrary, the power-law exponent changes $\braket{n(\tau)}_{\mathrm{GGE}}\sim \tau^{-1/2}$. (b) Plot of the effective exponent $\delta_{\mathrm{eff}}(\tau)$ \eqref{eq:effective_exponent} as a function of $\tau$ (log scale only on the horizontal axis) for the FS initial state with $n_0=0.6$. The effective exponent converges to the value $0.5$.}
    
    \label{fig:2ann}
\end{figure} 

\begin{figure}[t]
    \centering
    \includegraphics[width=1\columnwidth]{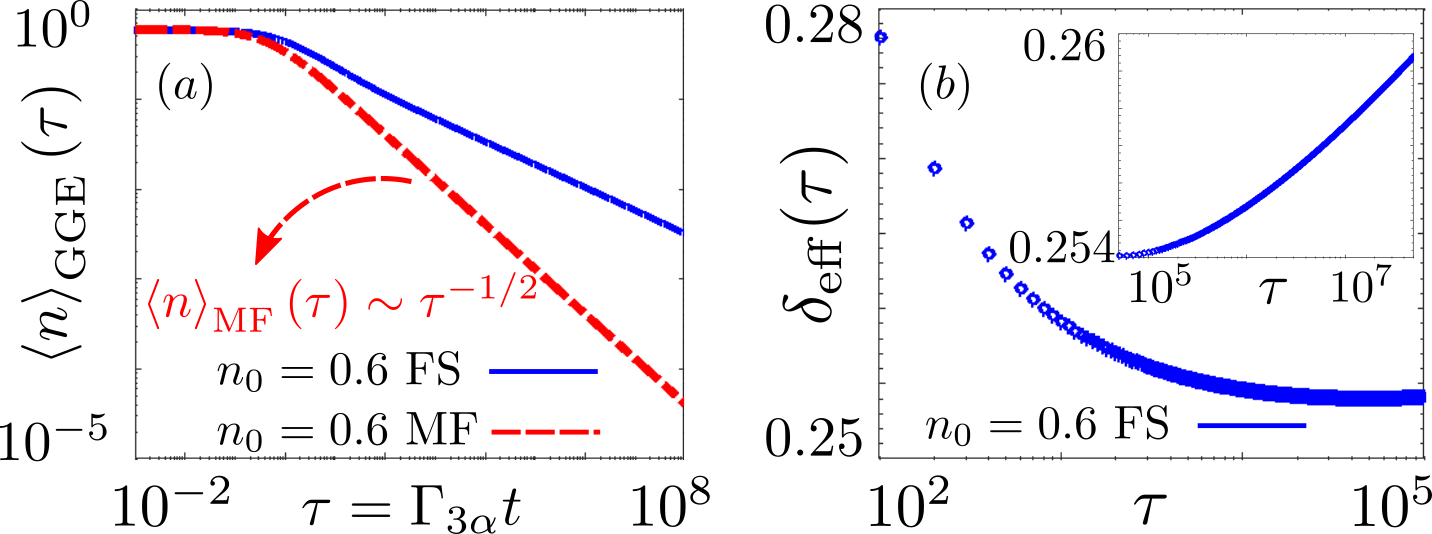}
    \caption{\textbf{Quantum three-body annihilation dynamics in the reaction limited regime.} (a) Log-log plot of the density $\braket{n(\tau)}_{\mathrm{GGE}}$ as a function of the rescaled time $\tau=\Gamma_{3\alpha}t$. The initial states chosen are identical to those considered in Fig.~\ref{fig:2ann} for $2A\to \emptyset$. In particular, the blue solid line refers to the initial FS state \eqref{eq:FS_initial_state}, while the red-dashed line corresponds to the initial incoherent state \eqref{eq:initial_flat_incoherent}. In both the cases the same initial filling value $n_0=0.6$ is taken. In the case of the initial state \eqref{eq:initial_flat_incoherent}, the TGGE rate equation reduces to the law of mass action \eqref{eq:MF_rate_equation} and \eqref{eq:GGE_MF_3ann} and the density decays asymptotically as $\braket{n}_{\mathrm{GGE}}=\braket{n}_{\mathrm{MF}}\sim \tau^{-1/2}$. In the case of the FS state, power-law decay is valid only up to $\tau \lesssim 10^5$ with a non mean-field exponent $\braket{n}_{\mathrm{GGE}}\sim \tau^{-0.25}$. (b) Plot of the effective exponent $\delta_{\mathrm{eff}}(\tau)$ \eqref{eq:effective_exponent} as a function of $\tau$ (log scale only on the horizontal axis). The effective exponent converges only for times $\tau\lesssim 10^5$. For longer times, plotted in the inset, $\delta_{\mathrm{eff}}(\tau)$ slowly drifts in time pointing out that the decay acquires a nonalgebraic correction. This correction is slower than the logarithmic correction \eqref{eq:diffusion_limited_3ann} observed for the classical analogue of the dynamics in $d=1$ in the different diffusion-limited regime.} 
    \label{fig:3ann}
\end{figure}

In Fig.~\ref{fig:3ann}(a), we show the density $\braket{n(\tau)}_{\mathrm{GGE}}$ of particles as a function of time $\tau$ both for the FS \eqref{eq:FS_initial_state} and the incoherent \eqref{eq:initial_flat_incoherent} initial state. In the latter case, similarly as in the case of $2A \to \emptyset$, elementary manipulations on the function $f_3(k,k',q)$ show that Eq.~\eqref{eq:3ann_rate_equation} reduces to 
\begin{equation}
\frac{\mbox{d}C_q(\tau)}{\mbox{d}\tau}=-3 C_q(\tau) \braket{n}^{2}_{\mathrm{GGE}}(\tau).
\label{eq:GGE_MF_3ann}
\end{equation}
Taking the sum over the quasimomenta $q$ on both sides of the previous equation, the law of mass action \eqref{eq:MF_rate_equation} with $k=3$ is retrieved. In this case the quantum reaction-limited regime therefore exactly coincides with its classical analogue and the asymptotic decay is $\braket{n(\tau)}_{\mathrm{GGE}}=\braket{n}_{\mathrm{MF}}(\tau) \sim \tau^{-1/2}$. 

The case of the FS initial state \eqref{eq:FS_initial_state} yields instead a much richer physics with a density dynamics which is markedly different from the mean-field prediction. At first inspection, from Fig.~\ref{fig:3ann}(a), the density $\braket{n(\tau)}_{\mathrm{GGE}}$ seems to follow a power law in time $\tau$ with an exponent different from the mean-field value, similarly as in the case of $2A\to \emptyset$ in Fig.~\ref{fig:2ann}. The numerical calculation of the effective exponent $\delta_{\mathrm{eff}}(\tau)$, in the main panel of Fig.~\ref{fig:3ann}(b), shows that the density follows the power-law behavior for times of the order $\tau \lesssim 10^5$ 
\begin{equation}
\braket{n(\tau)}_{\mathrm{GGE}} \sim \tau^{-0.25}, \,\, \mbox{for} \,\, \tau \lesssim 10^5,
\label{eq:3ann_intermediate}
\end{equation}
as anticipated in Eq.~\eqref{eq:q_react_lim_3} of the Introduction. Note that the decay exponent $0.25$ is an approximate value obtained from the numerical calculation of $\delta_{\mathrm{eff}}(\tau)$. It is in this sense different from the $1/2$ exponent for $2A\to \emptyset$ in \eqref{eq:2ann_GGE_FS}, which is an analytical result, as explained previously. Remarkably, in the quantum reaction-limited regime, the power-law decay exponent is different from the mean-field prediction ($1/2$) even in $d=d_c=1$. This contrasts the classical RD dynamics, where the power-law decay is always determined by the mean-field exponent $1/2$, both in the reaction-limited regime \eqref{eq:MF_rate_equation} and in the diffusion-limited one \eqref{eq:diffusion_limited_3ann}. In the latter case \eqref{eq:diffusion_limited_3ann}, deviations from the mean field are solely given by the  logarithmic correction $\sim (\ln \Omega t)^{1/2}$, the exponent of the algebraic decay $(\Omega t)^{-1/2}$ still being equal to $1/2$. The appearance of the non mean-field exponent at $d=d_c$ is only possible in the quantum RD dynamics where quantum effects, not determined by space dimensionality, induce correlations beyond mean field. In order to observe the decay dynamics \eqref{eq:3ann_intermediate}, one, indeed, needs the simultaneous presence of quantum coherences in the initial state and the fermionic statistics. The latter forces the annihilation reaction to take place only among adjacent particles \eqref{eq:3annihilation}, which is reflected in the nontrivial function $f_3(k,k',q)$ in Eq.~\eqref{eq:3ann_rate_equation}. Ultimately, this function renders the dynamics different from the law of mass action prediction \eqref{eq:GGE_MF_3ann} for quantum coherent initial conditions.  

A closer inspection of the effective exponent $\delta_{\mathrm{eff}}(\tau)$, in the inset of Fig.~\ref{fig:3ann}(b) for $\tau>10^5$, shows, however, that the power law \eqref{eq:3ann_intermediate} observed for triplet annihilation, $3A\to \emptyset$, remains valid only up to times $\tau \lesssim 10^5$. For longer times, the effective exponent $\delta_{\mathrm{eff}}(\tau)$ slowly increases as a function of time, indicating that the behavior in 
Eq.~\eqref{eq:3ann_intermediate} acquires a nontrivial non power-law correction. It is then natural to attempt to link this behavior to the logarithmic correction in Eq.~\eqref{eq:diffusion_limited_3ann} for the classical triplet annihilation in the diffusion-limited regime. Our analysis, however, shows that a multiplicative logarithmic correction to the power law \eqref{eq:3ann_intermediate} as $\braket{n(\tau)}_{\mathrm{GGE}}\sim b\,[\log(\tau)/\tau]^{\gamma}$, with $\gamma$ and $b$ being  fitting parameters, is not compatible with the long-time behavior of $\delta_{\mathrm{eff}}(\tau)$ in Fig.~\ref{fig:3ann}(b). In order to best capture also the intermediate-time algebraic decay \eqref{eq:3ann_intermediate}, we have also attempted to include the logarithmic correction in an additive form $\braket{n(\tau)}_{\mathrm{GGE}}\sim a/\tau^{0.25}+ b\,[\log(\tau)/\tau]^{\gamma}$, with $a$ being the amplitude of the algebraic decay. This additive logarithmic correction does not, however, improve the agreement with the late-time asymptotics of $\delta_{\mathrm{eff}}(\tau)$, since we numerically find that $\gamma<0.25$. In particular, we find that the asymptotic decay is slower than the one predicted by both the multiplicative and the additive logarithmic corrections.

The rich behavior in the $3A\to\emptyset$ decay from the FS initial condition \eqref{eq:FS_initial_state} is not qualitatively affected by changing the initial density parameter $n_0$ (cf. Fig.~\ref{fig:3ann_appendix} in Appendix \ref{app:3ann}). For all values of $n_0\neq 1$, we, indeed, observe that the effective exponent $\delta_{\mathrm{eff}}(\tau)$ converges to a value approximately equal to $0.25$ for $\tau \lesssim 10^5$. For longer times, we also observe that $\delta_{\mathrm{eff}}(\tau)$ increases in time according to a slow nonlogarithmic correction. Furthermore the same behavior in time is expected to apply for pure or mixed initial states featuring quantum coherences in real space, as already discussed after Eq.~\eqref{eq:2ann_GGE_FS} for $2A\to \emptyset$. The necessary requirement for the emergent collective behavior of Fig.~\ref{fig:3ann} is, indeed, solely that the initial occupation function $C_q(0)$ is not flat in $q$.

This highly nontrivial non-power-law behavior makes the dynamics of three-body annihilation particularly distinct from their classical counterpart: in the classical reaction-limited regime, indeed, only power-law behavior \eqref{eq:MF_rate_equation} is possible since spatial fluctuations are absent due to the rapid mixing through coherent hopping. In the diffusion-limited regime, instead, nonalgebraic asymptotic is possible only in the logarithmic form of Eq.~\eqref{eq:diffusion_limited_3ann} at $d=d_c$. The latter is, in turn, determined by the spatial fluctuations induced by the small diffusive mixing, as recalled in Subsec.~\ref{subsec:classical_RD}. The nonalgebraic behavior in Fig.~\ref{fig:3ann} is non logarithmic and it has a different origin determined not only by space dimensionality, but also by the interplay of the latter with quantum effects due to coherences in the initial states.

\subsection{Four-body annihilation} 
\label{sec:4ann}
In the case of quadruplet annihilation $4A\to\emptyset$ \eqref{eq:4annihilation}, Equation \eqref{eq:tgge_rate_equation_general} reads as 
\begin{equation}
\frac{\mbox{d}C_q(\tau)}{\mbox{d}\tau}\!=\!-\frac{C_q(\tau)}{L^3}\sum_{k,k',k''} \!\!\!f_4(k,k',k'',q) C_k(\tau)C_{k'}(\tau)C_{k''}(\tau),
\label{eq:4ann_rate_equation_GGE} 
\end{equation}
with the rescaled time $\tau=\Gamma_{4\alpha}t$. The function $f_4(k,k',k'',q)$ takes a rather cumbersome form. It results from a long calculation involving the application of Wick's theorem onto the eight-point function of the fermionic operators $\hat{c}_k$ deriving from the Fourier expression of the quadruplet annihilation jump operator \eqref{eq:4annihilation}. The expression for $f_4$ is therefore reported in Eq.~\eqref{eq:f4_function_4ann_complete} of Appendix \ref{app:4ann}, where the calculations leading to \eqref{eq:4ann_rate_equation_GGE} are summarized.   

\begin{figure}[t]
    \centering    \includegraphics[width=1\columnwidth]{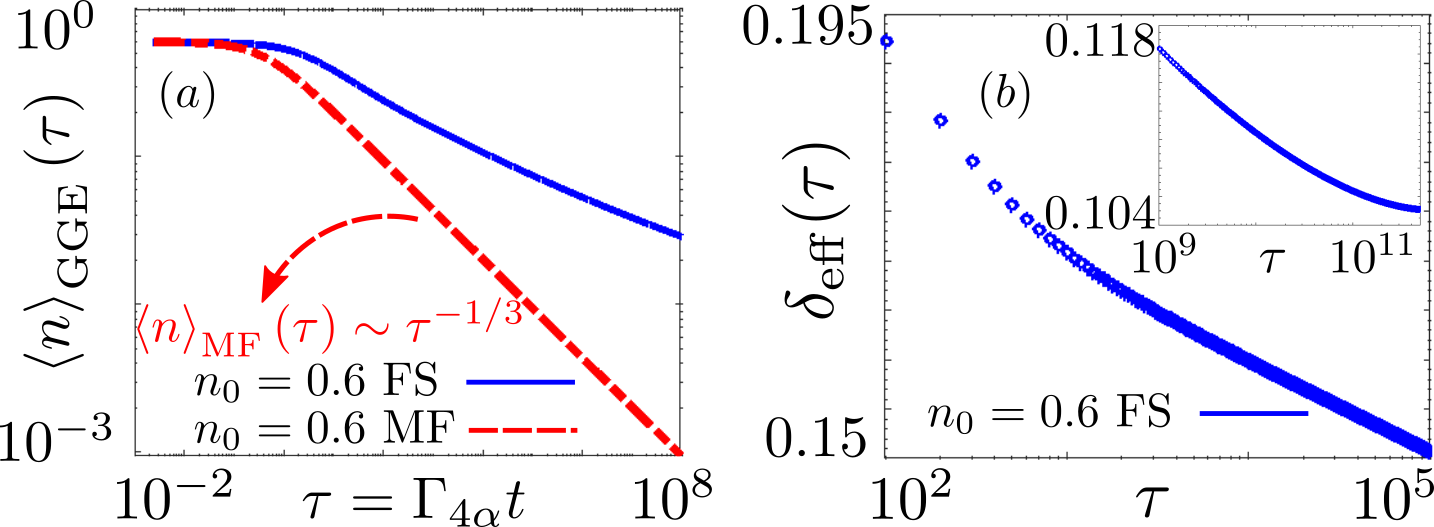}
     \caption{\textbf{Quantum four-body annihilation dynamics in the reaction limited regime.} (a) Log-log plot of the density $\braket{n(\tau)}_{\mathrm{GGE}}$ as a function of the rescaled time $\tau=\Gamma_{4\alpha}t$. The blue solid line represents the dynamics from the FS state \eqref{eq:FS_initial_state}, while the red dashed one the dynamics from the state \eqref{eq:initial_flat_incoherent}. The initial filling value is $n_0=0.6$, as in the case of Figs.~\ref{fig:2ann} and \ref{fig:3ann}. For the incoherent initial state \eqref{eq:initial_flat_incoherent}, the quantum-reaction limited dynamics exactly coincides with the mean-field law of mass action prediction \eqref{eq:GGE_MF_4ann}. The associated asymptotic density decay is $\braket{n}_{\mathrm{GGE}}=\braket{n}_{\mathrm{MF}}=(\tau)\sim \tau^{-1/3}$. For the coherent Fermi-sea initial state, the decay is slower and it is given by the power-law $\braket{n(\tau)}_{\mathrm{GGE}} \sim \tau^{-0.1}$. (b) Plot of the effective exponent $\delta_{\mathrm{eff}}(\tau)$ \eqref{eq:effective_exponent} as a function of $\tau$ (log scale only on the horizontal axis). The effective exponent converges in a slower way than in the case of $2A\to \emptyset$ of Fig.~\ref{fig:2ann}. In particular, observes that $\delta_{\mathrm{eff}}(\tau) \simeq 0.1$ at long times $\tau \gtrsim 10^{11}$, as shown in the inset. } 
    \label{fig:4ann}
\end{figure}

In Fig.~\ref{fig:4ann}(a), the density $\braket{n(\tau)}_{\mathrm{GGE}}$ is shown as a function of $\tau$ for the FS \eqref{eq:FS_initial_state} initial condition as well as for an initial incoherent state \eqref{eq:initial_flat_incoherent}. The dynamics from the incoherent initial state reduces, as in the case of binary $2A\to\emptyset$ and triplet $3A \to \emptyset$ annihilation, to the law of mass action \eqref{eq:MF_rate_equation} with $k=4$,
\begin{equation}
\frac{\mbox{d}C_q(\tau)}{\mbox{d}\tau}=-4 C_q(\tau) \braket{n(\tau)}^{3}_{\mathrm{GGE}}.
\label{eq:GGE_MF_4ann} 
\end{equation}
The decay of the density is therefore ruled by $\braket{n}_{\mathrm{GGE}}=\braket{n}_{\mathrm{MF}} \sim \tau^{-1/3}$. This is power-law decay is plotted with the dashed-red line in Fig.~\ref{fig:4ann}(a). We emphasize, as recalled in Subsec.~\ref{subsec:classical_RD}, that the decay \eqref{eq:GGE_MF_4ann} applies for classical RD both in the diffusion-limited and in the reaction-limited regime as for this process spatial fluctuations are irrelevant already in $d=1>d_c$. 

In the case of the FS initial state, however, the decay of the density, depicted with the blue-solid line of Fig.~\ref{fig:4ann}(a), $\braket{n(\tau)}_{\mathrm{GGE}}$ does not follow the mean-field prediction. In particular, we observe the algebraic decay \begin{equation}
\braket{n(\tau)}_{\mathrm{GGE}} \sim \tau^{-0.1},
\label{eq:FS_decay_4_ann}
\end{equation}
obtained by computing the effective exponent $\delta_{\mathrm{eff}}(\tau)$ \eqref{eq:effective_exponent}, which is plotted in Fig.~\ref{fig:4ann}(b). The $0.1$ decay exponent is therefore a numerically approximate value [in the same way as the decay exponent \eqref{eq:3ann_intermediate} for $3A\to\emptyset$]. We observe that the convergence of $\delta_{\mathrm{eff}}(\tau)$ to the asymptotic value is slower compared to the case of binary annihilation $2A\to \emptyset$ in Fig.~\ref{fig:2ann}. In the latter case $\delta_{\mathrm{eff}}(\tau)\simeq 0.5$ for $\tau \gtrsim 10^5$, while, in Fig.~\ref{fig:4ann}(b), $\delta_{\mathrm{eff}} \simeq 0.1$ for $\tau \gtrsim 10^{11}$. In all the cases discussed, $2A\to\emptyset$ \eqref{eq:2ann_GGE_FS}, $3A \to \emptyset$ \eqref{eq:3ann_intermediate}, and $4A\to\emptyset$ \eqref{eq:FS_decay_4_ann}, we observe that quantum coherences slow down the density decay, which is reflected in an exponent that is smaller than the one of the corresponding classical reaction-limited process \eqref{eq:MF_rate_equation}. In Fig.~\ref{fig:4ann_appendix} of Appendix \ref{app:4ann}, we study the $4A\to\emptyset$ dynamics for different fillings $n_0$ of the FS initial state. Similarly to the cases of $2A \to\emptyset$ and $3A\to \emptyset$, we observe no qualitative change upon varying $n_0$ as long as $n_0\neq 1$. The effective exponent is in all the cases monotonically decreasing towards a value approximately equal to $0.1$: $\delta_{\mathrm{eff}}(\tau)\simeq 0.1$. The very same asymtptotic exponent is also expected to apply for other coherent initial states (in real space) characterized by a momentum occupation function $C_q(0)$ not constant in the quasimomentum $q$.

The non mean-field algebraic decay \eqref{eq:FS_decay_4_ann} is valid in $d=1$ and, therefore, above the upper-critical dimension $d_c$ of the quadruplet annihilation reaction $4A\to\emptyset$. This might look surprising as one expects for $d>d_c$ the system dynamics to be captured by the mean-field approximation, as reported in Sec.~\ref{sec:reaction_lim}. One should note that the upper critical dimension $d_c$ characterizes diffusion-limited RD dynamics. In the diffusion-limited regime, correlations leading to non mean-field behavior are caused by spatial fluctuations in the density profile due to the diffusive motion of particles far apart from each other. Spatial fluctuations are relevant in low dimensions, while in higher dimensions diffusive mixing is effective in filling the whole space. This explains the emergence of an upper critical dimension $d_c$, beyond which critical exponents are given by the mean-field prediction. In the reaction-limited regime, where Eq.~\eqref{eq:FS_decay_4_ann} applies, spatial fluctuations in the density are by construction absent as one assumes the system to be relaxed to the homogeneous GGE state. Fluctuations beyond mean field are of strict quantum origin due quantum coherences in the initial state. These effects are valid in any space dimensionality. 
Consequently one has non mean-field behavior even for $d>d_c$. This is the hallmark of the different origin of universal dynamical behavior in quantum reaction-limited RD systems as compared to their classical counterparts. 

On the basis of these results we expect also 
that in quantum RD systems with 
five, $5A\to\emptyset$, and higher-body annihilation reactions will have algebraic decay with a non mean-field exponent due to to the same quantum effects [though possibly with a numerical value different from the one in Eq.~\eqref{eq:FS_decay_4_ann} for $4A\to\emptyset$]. The calculation of the decay exponents for such higher order systems is similar to that presented here (and in Appendix \ref{app:4ann}), only more cumbersome due to combinatorics, since higher fermionic correlation functions are involved (for instance a ten-point function for $5A\to \emptyset$). In any case, such behavior is again expected to be distinct to that of the corresponding classical RD dynamics, where for the reaction-limited regime all processes are described by the mean-field law of mass action, cf.\ \eqref{eq:MF_rate_equation}.           

\section{Discussion and Conclusions}
\label{sec:conclusion}
In this paper, we analyzed the quantum RD dynamics of fermionic quantum gases on a one-dimensional lattice subject to annihilation processes, formulated in terms of a Lindblad master equation \eqref{eq:master_equation} and \eqref{eq:dissipator}. Here classical diffusive motion is replaced by quantum coherent hopping \eqref{eq:free_fermion_Hamiltonian}, while annihilation reactions are irreversible and they are encoded into the jump operators \eqref{eq:2annihilation}-\eqref{eq:4annihilation}. We considered binary annihilation $2A\to\emptyset$ of pairs of neighbouring particles, triplet annihilation $3A\to\emptyset$, and quadruplet $4A\to \emptyset$ annihilation. We studied the dynamics from both coherent initial states, such as the FS \eqref{eq:FS_initial_state}, and from incoherent initial states as \eqref{eq:initial_flat_incoherent}. Quantum effects enter through the coherent Hamiltonian hopping and, possibly, through coherences within the initial state. 

We solved the problem analytically in the thermodynamic limit by exploiting the approximation of the dynamics obtained via the time-dependent generalized Gibbs ensemble method \eqref{eq:tGGE_free_fermions} \cite{tGGE1,tGGE2,tGGE3,tGGE4}. This method describes the reaction-limited, or weak-dissipation, regime $\Gamma_{\nu}/\Omega\ll 1$, where the irreversible reaction rate is much smaller than the coherent hopping rate.

In all the three cases considered, reactions $2A\to \emptyset$, $3A\to\emptyset$ and $4A\to \emptyset$, we observed that for incoherent initial states the quantum reaction-limited dynamics reduces to classical mean-field description in terms of the law of mass action \eqref{eq:MF_rate_equation}. Quantum coherences in the initial state, which amounts to considering an initial occupation function $C_q(0)$ not flat in the quasi-momentum $q$,  are therefore necessary in order to have a non mean-field universal decay of the particle density at long times. We remark that an inhomogeneous initial occupation function $C_q(0)$ can characterize both pure, such as the FS \eqref{eq:FS_initial_state}, and mixed states. Beyond mean-field quantum reaction-limited decay is therefore a robust feature of the dynamics which does not necessarily require considering pure states. The fermionic statistics is also important as it determines the structure \eqref{eq:2annihilation}-\eqref{eq:4annihilation} of the jump operators, which, consequently, determines the nontrivial functions $f_2$ [\eqref{eq:f2ann} in Eq.~\eqref{eq:ann_rate_equation}], $f_3$ [\eqref{eq:f3_function} in Eq.~\eqref{eq:3ann_rate_equation}] and $f_4$ [\eqref{eq:f4_function_4ann_complete} in Eq.~\eqref{eq:4ann_rate_equation_GGE}]. These functions eventually render the TGGE rate equations different from the law of mass action for quantum coherent initial conditions. In particular, in the case of binary annihilation $2A\to \emptyset$, the density decays algebraically in time as in Eq.~\eqref{eq:2ann_GGE_FS} with exponent $1/2$ (cf. Fig.~\ref{fig:2ann}). For triplet $3A\to \emptyset$ and quadruplet $4A\to\emptyset$ annihilation, the impact of the simultaneous presence of quantum coherences and fermionic statistics onto the asymptotic decay is even richer. In particular, for $3A\to\emptyset$, we find algebraic decay as in Eq.~\eqref{eq:3ann_intermediate} only in an intermediate time regime $\tau \lesssim 10^5$ with exponent approximately $0.25$. For later times, this decay acquires, however, a non power-law correction. In the classical RD dynamics nonalgebraic corrections are only possible in the diffusion-limited regime with a logarithmic form \eqref{eq:diffusion_limited_3ann}, which comes from the fact that the upper-critical dimension of the process is $d_c=d=1$ \cite{tauber2002dynamic,tauber2005applications,tauber2014critical}. Spatial density fluctuations are therefore marginal in one dimension for the reaction $3A\to \emptyset$. The decay we observe in the quantum reaction-limited regime in Fig.~\ref{fig:3ann} is, instead, slower than that predicted by a logarithmic correction. For quadruplet annihilation $4A\to\emptyset$, we find the algebraic decay in Eq.~\eqref{eq:FS_decay_4_ann} with exponent approximately equal to $0.1$ shown in Fig.~\ref{fig:4ann}. This result is in contrast with the classical description of the process which is always, both in the diffusion and in the reaction-limited regime, in agreement with the mean-field approximation \eqref{eq:diffusion_limited_4ann}. This is a consequence of the fact that for $4A\to \emptyset$ spatial fluctuations are irrelevant in any physical dimension since $d_c<1$. 

Our results for $3A\to \emptyset$ and $4A\to \emptyset$ show that in quantum RD dynamics correlations beyond mean field are not only determined by spatial fluctuations, but also by inherently quantum effects. These effects are present in any space dimensionality and even in the absence of spatial fluctuations of the density profile and therefore universal behavior, without any classical correspondence, is possible even at $d \geq d_c$. The mechanism behind critical behavior in quantum nonequilibrium RD models is therefore fundamentally different to that describing the emergence of universality in their classical counterpart.   

As a future direction, one may extend the present analysis to bosonic systems. In this case, we expect the absence of the exclusion principle to lead, in the reaction-limited regime, to mean-field results. In the case of annihilation channels exhibiting interference effects (see, e.g., \cite{QRD20222,lossth7,lossexp0,diehl2008quantum}), one may, however, still see a universal non mean-field decay. At the same time, it would be interesting to investigate the RD dynamics of fermionic and bosonic gases in the continuum. The time-dependent GGE description of the reaction-limited regime can be carried out in analogy to the lattice gases discussed here. 
The extension of the results here presented to spatial dimensions larger than one is important, as well. In particular, since the universal behavior in the quantum reaction-limited regime is dictated by quantum coherent effects, we expect the exponent for the algebraic decay to be nontrivial even for $d>d_c$. However, one expects the impact of these effects to depend on space dimensionality $d$ itself, and it would thus be interesting to understand the dependence of the decay exponent on $d$.  

\acknowledgments
G.P.~acknowledges support from the Alexander von Humboldt Foundation through a Humboldt research fellowship for postdoctoral researchers. We are grateful for financing from the Baden-W\"urttemberg Stiftung through Project No.~BWST\_ISF2019-23. We also acknowledge funding from the Deutsche Forschungsgemeinschaft (DFG, German Research Foundation) under Project No. 435696605, through the Research Unit FOR 5413/1, Grant No. 465199066 and through the Research Unit FOR 5522/1, Grant No. 499180199. F.C.~is indebted to the Baden-W\"urttemberg Stiftung for the financial support by the Eliteprogramme for Postdocs. 
We also acknowledge financial support from EPSRC Grant no.\ EP/V031201/1.

\onecolumngrid

\appendix

\section{Three-body annihilation dynamics}
\label{app:3ann}
In this Appendix, we provide details regarding the derivation of Eqs.~\eqref{eq:3ann_rate_equation} and \eqref{eq:f3_function} for triplet annihilation $3A\to \emptyset$. We consider the case of periodic boundary conditions $c_{j+L}=c_j$ for the Hamiltonian \eqref{eq:free_fermion_Hamiltonian}. As the analysis based on the TGGE method applies in the thermodynamic limit $L\to\infty$, the choice of boundary conditions does not impact on the final result in Eqs.~\eqref{eq:3ann_rate_equation} and \eqref{eq:f3_function}. We first introduce the Fourier transform $\hat{c}_k$ of the lattice fermionic operators $c_j$ as 
\begin{equation}
\hat{c}_{k_n}=\frac{1}{\sqrt{L}}\sum_{j=1}^{L} e^{-i k_n j} c_{j}, \qquad \mbox{with inverse} \qquad c_j=\frac{1}{\sqrt{L}}\sum_{k_n}e^{i k_n j} \hat{c}_{k_n},
\label{supeq:Fourier_transform}    
\end{equation}
where $k_n=2 \pi n/L$ are the quasimomenta and they are parametrized on the lattice in terms of the integer number $n=1,2\dots L$. In the previous summation the sum $\sum_{k_n}$ denotes a summation over the integer number $n$. In the main text and in the rest of the Supplemental material we use for brevity the shorter notation $\sum_{k_n} \to \sum_{k}$. When multiple summations over the quasimomenta are present, e.g., $\sum_{k_1}\sum_{k_2}\dots\sum_{k_n}$, we also use the compact notation $\sum_{k_1,k_2 \dots k_n}$ [see Eqs.~\eqref{eq:3ann_rate_equation} and \eqref{eq:4ann_rate_equation_GGE}]. We write the triplet annihilation jump operator $L_j^{3\alpha}$ in Fourier space as 
\begin{equation}
L_j^{3\alpha}=\sqrt{\Gamma_{3\alpha}}\,c_j c_{j+1}c_{j+2}= \frac{\sqrt{\Gamma_{3\alpha}}}{L^{3/2}}\sum_{k,k',k''} e^{ij(k+k'+k'')}g(k',k'') \hat{c}_{k}\hat{c}_{k'}\hat{c}_{k''},    
\end{equation}
with 
\begin{equation}
g(k',k'')=\mbox{exp}(i(k'+2k'')).
\label{eq:g_function_3ann}
\end{equation}
In order to simplify the commutator appearing in Eq.~\eqref{eq:tgge_rate_equation_general}, we use the identities 
\begin{equation}
[\hat{n}_q, \hat{c}_{k}\hat{c}_{k'}\hat{c}_{k''}]=-\hat{c}_{k}\hat{c}_{k'}\hat{c}_q \delta_{k'',q}-\hat{c}_{k}\hat{c}_q\hat{c}_{k''}\delta_{k',q}-\hat{c}_q\hat{c}_{k'}\hat{c}_{k''}\delta_{k,q}, \quad \mbox{from} \quad [\hat{n}_q,\hat{c}_k]=-\delta_{k,q}\hat{c}_q.
\label{eq:commutation_relations_3_ann}
\end{equation}
Using Eq.~\eqref{eq:commutation_relations_3_ann}, we then get for $[\hat{n}_q, L_j^{3\alpha}]$
\begin{align}
[\hat{n}_q,L_j^{3\alpha}]=-\frac{\sqrt{\Gamma_{3\alpha}}}{L^{3/2}}\left[\sum_{k,k'}\hat{c}_{k}\hat{c}_{k'} \hat{c}_q e^{i(k+k'+q) j}g(k',q)+\sum_{k,k'} \hat{c}_{k} \hat{c}_{q} \hat{c}_{k'}e^{i(k+k'+q)}g(q,k')+\sum_{k,k'} \hat{c}_q \hat{c}_{k} \hat{c}_{k'} e^{i(k+k'+q)j}g(k,k') \right].
\label{supeq:intermediate_commutator_nq_3ann}
\end{align}
Upon inserting Eq.~\eqref{supeq:intermediate_commutator_nq_3ann} into Eq.~\eqref{eq:tgge_rate_equation_general} one obtains three terms 
\begin{align}
\sum_j (L_j^{3\alpha})^{\dagger}[\hat{n}_q,L_j^{3\alpha}] &=-\frac{\Gamma_{3\alpha} }{L^{3}}\sum_j\sum_{k_1,k_2,k_3}e^{-ij(k_1+k_2+k_3) }\hat{c}^{\dagger}_{k_3} \hat{c}^{\dagger}_{k_2}\hat{c}^{\dagger}_{k_1}g^{\ast}(k_2,k_3)\left[ \sum_{k,k'}g(k',q)\hat{c}_k \hat{c}_{k'}\hat{c}_{q} e^{i j(k+k'+q)} \right. \nonumber \\
&\left.+\sum_{k,k'}g(q,k') \hat{c}_k \hat{c}_{q} \hat{c}_{k'}e^{ij(k+k'+q)} +\sum_{k,k'} g(k,k')\hat{c}_q \hat{c}_k \hat{c}_{k'}e^{ij(k+k'+q)} \right]\nonumber \\
&=-\frac{\Gamma_{3\alpha}}{L^2}\left[\sum_{k_1,k_2,k_3,k'} g^{\ast}(k_2,k_3)g(k',q)\hat{c}_{k_3}^{\dagger}\hat{c}_{k_2}^{\dagger}\hat{c}_{k_1}^{\dagger}\hat{c}_{k_1+k_2+k_3-k'-q}  \hat{c}_{k'} \hat{c}_q \right. \nonumber \\
&\quad \quad\quad\,\,+\sum_{k_1,k_2,k_3,k'}\left. g^{\ast}(k_2,k_3)g(q,k')\hat{c}_{k_3}^{\dagger} \hat{c}_{k_2}^{\dagger}\hat{c}_{k_1}^{\dagger}\hat{c}_{k_1+k_2+k_3-k'-q} \hat{c}_{q}\hat{c}_{k'} \right. \nonumber \\
&\quad \quad\quad\,\,\left.+\sum_{k_1,k_2,k_3,k'}g^{\ast}(k_2,k_3)g(k_1,k') \hat{c}_{k_3}^{\dagger}\hat{c}_{k_2}^{\dagger}\hat{c}_{k_1+k'+q-k_2-k_3}^{\dagger}\hat{c}_q \hat{c}_{k_1} \hat{c}_{k'} \right],
\label{eq:intermediate_six_3ann}
\end{align}
where in the last equality the Fourier representation of the Kronecker delta has been used. From the previous equation, one sees that the Lindblad dynamics of the two point function $\braket{\hat{n}_q}$ for the triplet annihilation $3A\to \emptyset$ is coupled to the dynamics of six-point functions. The evolution equation for $\braket{\hat{n}_q}$ is therefore not closed and one has a hierarchy of equations coupling the dynamics of correlation functions to higher-order correlation functions. In order to break this hierarchy, the TGGE assumption \eqref{eq:tGGE_free_fermions} is fundamental. The time-dependent GGE state \eqref{eq:tGGE_free_fermions} describes the dynamics in the reaction-limited, or, equivalently, weak-dissipation, limit $\Gamma_{3\alpha}/\Omega\ll 1$ and it amounts to replace $\braket{\hat{n}_q} \to \braket{\hat{n}_q}_{\mathrm{GGE}}$ (and analogously for other expectation values in the previous equation). The GGE state is Gaussian for the free-fermionic Hamiltonian \eqref{eq:free_fermion_Hamiltonian} and diagonal in momentum space. Higher-point correlation functions of fermionic operators $\hat{c}_k$ can therefore be computed in the GGE solely on the basis of the two-point function $\braket{\hat{c}^{\dagger}_q \hat{c}_k}_{\mathrm{GGE}}\equiv C_q \delta_{k,q}$ via Wick's theorem. In the case of the six-point function appearing  on the right-hand side of the first line of the second equality in Eq.~\eqref{eq:intermediate_six_3ann}, one has, applying Wick's theorem,
\begin{align}
\braket{\hat{c}_{k_3}^{\dagger}\hat{c}_{k_2}^{\dagger}\hat{c}_{k_1}^{\dagger}\hat{c}_{k_1+k_2+k_3-k'-q} \hat{c}_{k'} \hat{c}_{q}}_{\mathrm{GGE}}(\tau)= &-C_{k'}C_q C_{k_3}\delta_{k_2,k'}\delta_{q,k_1}+C_{k_3}C_{q}C_{k'}\delta_{k_1,k'}\delta_{k_2,q}+C_{k'}C_{k_2}C_{k_1}\delta_{k_3,k'}\delta_{k_1,q}  \nonumber \\ &-C_{k'}C_{k_2}C _{k_1}\delta_{k_3,k'}\delta_{k_2,q}+C_{k_3}C_{k'}C_{k_1}\delta_{k_3,q}\delta_{k_2,k'}-C_{k_3} C_{k_2}C_{k_1}\delta_{k_3,q}\delta_{k_1,k'}.
\label{eq:wick_six_points_3ann}
\end{align}
\begin{figure}[t]
    \centering   \includegraphics[width=1\columnwidth]{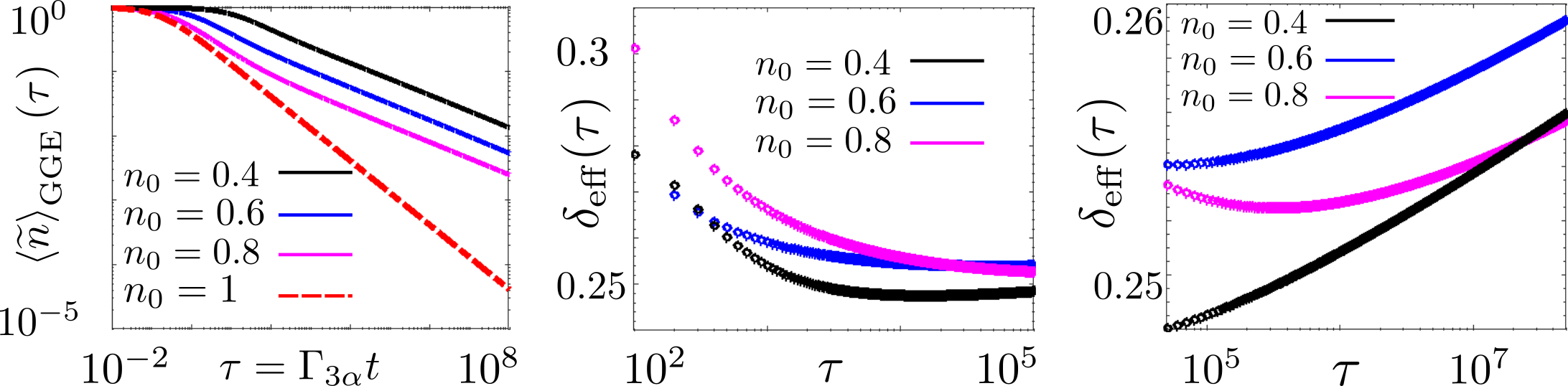}
    \caption{\textbf{Quantum three-body annihilation dynamics for different initial fillings of the FS initial state}. (a) Log-log plot, from the numerical solution of Eq.~\eqref{eq:3ann_rate_equation} for $3A\to\emptyset$, of the rescaled density $\braket{\widetilde{n}}_{\mathrm{GGE}}(\tau)=\braket{n}_{\mathrm{GGE}}(\tau)/n_0$ as a function of the rescaled time $\tau=\Gamma_{3\alpha}t$. Four different values $n_0=0.4,0.6,0.8$ and $1$ (from top to bottom) of the initial filling of the FS state \eqref{eq:FS_initial_state} are reported. The blue curve for $n_0=0.6$ is identical to that discussed in the main text in Fig.~\ref{fig:3ann}(a). For all the values $n_0 \neq 1$, the obtained curve differs from the mean-field prediction obtained with the associated initial density $n_0$ since the initial state displays quantum coherences in real space. In the case $n_0=1$ only, the FS has no real-space coherences and the density decays, red-dashed line in the figure, is exactly reproduced by the law of mass action equation \eqref{eq:MF_rate_equation} with $\braket{\widetilde{n}}_{\mathrm{GGE}}(\tau)=\braket{\widetilde{n}}_{\mathrm{MF}}(\tau)\sim \tau^{-1/2}$. (b) Plot of the effective exponent $\delta_{\mathrm{eff}}(\tau)$ \eqref{eq:effective_exponent} as a function of $\tau$ (log scale on the horizontal axis only) for the FS initial state and initial fillings $n_0=0.4,0.6$ and $0.8$. The blue curve for $n_0=0.6$ is, also in this case, equal to that of Fig.~\ref{fig:3ann}(b) of the main text. For the three cases the effective exponent is observed to converge to a value $\delta_{\mathrm{eff}}(\tau) \simeq 0.25$ for times $\tau \lesssim 10^5$. (c) Plot of the effective exponent for the very same values of the initial filling $n_0$ as in (b) but for longer times $\tau \gtrsim 10^5$. The effective exponent is for $\tau \gtrsim 10^5$ nonmonotonic. The density decay accordingly follows a nonalgebraic asymptotics.}   
    \label{fig:3ann_appendix}
\end{figure}%
\noindent The other two six-point correlation functions in Eq.~\eqref{eq:intermediate_six_3ann} can treated in an analogous way. In the previous equation, all momentum occupation functions $C_k(\tau)$ are function of the rescaled time $\tau=\Gamma_{3\alpha}t$ (not reported explicitly for brevity). We also note that $[\hat{n}_q,H]=0$ and therefore the right-hand side of the evolution equation for $\braket{n}_q$ is only given by Eq.~\eqref{eq:intermediate_six_3ann}. Upon using therein the TGGE assumption and the expression in Eq.~\eqref{eq:wick_six_points_3ann}, one obtains Eqs.~\eqref{eq:3ann_rate_equation} and \eqref{eq:f3_function} of the main text (after some lengthy but straightforward algebraic manipulation). In the numerical solution of the equation, plotted in Fig.~\ref{fig:3ann} of the main text, we take a value of $L$ large and we check the stability of the obtained curves upon further increasing $L$. In Fig.~\ref{fig:3ann}, we used $L=150000$ since for longer times, $\tau \simeq 10^7,10^8$ as in the inset of Fig.~\ref{fig:3ann}(b), larger values of $L$ are needed to get stable numbers.   

We remark that Eqs.~\eqref{eq:3ann_rate_equation} and \eqref{eq:f3_function} can be written in the equivalent form 
\begin{align}
\frac{\mbox{d}C_q(\tau)}{\mbox{d}\tau}=-3 C_q\braket{n}^{2}_{\mathrm{GGE}}(\tau) +\frac{C_q}{L^2}\sum_{k,k'}C_k C_{k'}&\left[ \cos(2k+2k')+2\cos(k-k')+2\cos(2k-2q)-2\cos(k+k'-2q)\right. \nonumber \\
&\left.+4\cos(k-q) -4\cos(2k-k'-q) \right].
\label{eq:triann_rate_simplified_GGE}
\end{align}
This form of the equation makes more transparent the connection with the classical reaction-limited dynamics ruled by the law of mass action \eqref{eq:MF_rate_equation}. In particular, the first term on the right-hand side of \eqref{eq:triann_rate_simplified_GGE} corresponds to the law of mass action for $3A \to \emptyset$, upon summing on both sides over all the possible values of $q$. The second term on the right-hand side of the previous equation, instead, couples the quasimomentum $q$ to all the other quasimomenta through a three-body term $C_q C_k C_{k'}$. The weight of this term is, however, non trivial and it is determined by the fermionic nature of the particles and the exclusion principle. Remarkably, this term contributes to the dynamics only if the initial momentum occupation function $C_q(0)$ is not flat in $q$. This amounts to considering initial states of the GGE form \eqref{eq:tGGE_free_fermions} with $\lambda_q(0)$ not flat in $q$ and, therefore, initial GGE states with a non diagonal, in real space, two-point fermionic correlation matrix. These kind of states display coherences in real space. The simulataneous presence of fermionic statistics and quantum coherences in the initial state makes the second term of \eqref{eq:triann_rate_simplified_GGE} relevant and it causes the non mean-field universal behavior of Fig.~\ref{fig:3ann} discussed in the main text.  

In Fig.~\ref{fig:3ann_appendix}, we further corroborate this finding by discussing the quantum reaction-limited dynamics ensuing from the FS initial state \eqref{eq:FS_initial_state} for different values of the initial filling $n_0$. In particular, in Fig.~\ref{fig:3ann_appendix}(a), universal behavior beyond mean field for the rescaled density $\braket{\widetilde{n}}_{\mathrm{GGE}}(\tau)=\braket{n}_{\mathrm{GGE}}(\tau)/n_0$ is observed for all values of $n_0\neq 1$. In the latter cases, indeed, $C_q(0)$ is not flat in $q$. In order to quantify such quantum universal decay, we compute the effective exponent $\delta_{\mathrm{eff}}(\tau)$, which is unexpectedly nonmonotonic in $\tau$. Indeed, in Fig.~\ref{fig:3ann_appendix}(b), $\delta_{\mathrm{eff}}(\tau)$ first decreases and saturates to a constant value, approximately equal to $\braket{\widetilde{n}}_{\mathrm{GGE}}(\tau) \sim \tau^{-0.25}$ for $\tau \lesssim 10^5$. For longer times $\tau \gtrsim 10^5$, in Fig.~\ref{fig:3ann_appendix}(c), $\delta_{\mathrm{eff}}(\tau)$ then increases signaling the onset of a nonalgebraic correction to the aforementioned algebraic scaling form. We remark that this behavior of $\delta_{\mathrm{eff}}$ does not qualitatively depend on the value of $n_0$, as long as $n_0\neq 1$. The value $\delta_{\mathrm{eff}}(\tau) \simeq 0.25$ to which the effective exponent converges, indeed, depends only weakly on the value of $n_0$, as shown in Fig.~\ref{fig:3ann_appendix}(b)-(c). At the same time, the onset of the nonalgebraic correction to the power-law asymptotic is also slightly shifted later in time as $n_0$ is increased (for $\tau \gtrsim 2 \cdot 10^5$ for $n_0=0.8$). The approximate algebraic decay $\braket{\widetilde{n}}_{\mathrm{GGE}}(\tau)\sim \tau^{-0.25}$ at intermediate times $\tau \lesssim 10^5$ and the subsequent nonalgebraic correction thereof are therefore robust features of the quantum reaction-limited $3A\to \emptyset$ decay for any value of the initial filling $n_0$ of the FS initial state. We similarly expect to observe a behavior qualitatively analogous to that of Fig.~\ref{fig:3ann_appendix} for other coherent initial states as long as the associated $C_q(0)$ is not flat in $q$. On the contrary, for $n_0=1$, one has $C_q(0)=1$ flat for all values of $q$, the initial state $\ket{\bullet \bullet \bullet \dots \bullet}$ has no real-space coherences and the law of mass action prediction $\braket{\widetilde{n}}_{\mathrm{MF}}(\tau) \sim \tau^{-1/2}$ is retrieved. We used $L=150000$, as for Fig.~\ref{fig:3ann} of the main text, to produce the numerical data in Fig.~\ref{fig:3ann_appendix}.

\section{Four-body annihilation dynamics}
\label{app:4ann}
In this Appendix, we report the main steps of the derivation of Eq.~\eqref{eq:4ann_rate_equation_GGE} for quadruplet annihilation $4A\to \emptyset$. We also give the expression of the function $f_4(k,k',k'',q)$. We write the jump operator $L_j^{4\alpha}$ in Eq.~\eqref{eq:4annihilation} in Fourier space \eqref{supeq:Fourier_transform} as
\begin{equation}
L_j^{4\alpha}=\sqrt{\Gamma_{4\alpha}}\,c_j c_{j+1}c_{j+2}c_{j+3}=\frac{\sqrt{\Gamma_{4\alpha}}}{L^2}\sum_{k_1,k_2,k_3,k_4}e^{ij(k_1+k_2+k_3+k_4)}g(k_2,k_3,k_4)\hat{c}_{k_1}\hat{c}_{k_2}\hat{c}_{k_3}\hat{c}_{k_4},
\label{eq:4ann_Fourier}
\end{equation}
with 
\begin{equation}
g(k_2,k_3,k_4)=\mbox{exp}[i(k_2+2k_3+3k_4)].    
\end{equation}
The commutator in Eq.~\eqref{eq:tgge_rate_equation_general} can be written by exploiting the chain rule for commutators and \eqref{eq:commutation_relations_3_ann} as 
\begin{align}
[\hat{n}_q,L_j^{4\alpha}]=-\frac{\sqrt{\Gamma_{4\alpha}}}{L^{2}}\Big[\sum_{k_2,k_3,k_4}e^{i(k_2+k_3+k_4+q) j}&[g(k_2,k_3,k_4) \hat{c}_q \hat{c}_{k_2}\hat{c}_{k_3} \hat{c}_{k_4} +g(q,k_3,k_4) \hat{c}_{k_2} \hat{c}_q \hat{c}_{k_3}\hat{c}_{k_4}\nonumber \\
&+g(k_2,q,k_4) \hat{c}_{k_3} \hat{c}_{k_2} \hat{c}_{q}\hat{c}_{k_4}+ g(k_2,k_3,q) \hat{c}_{k_4} \hat{c}_{k_2} \hat{c}_{k_{3}}\hat{c}_{q}]\Big].
\label{supeq:intermediate_commutator_nq_4ann}   
\end{align}
From Eq.~\eqref{supeq:intermediate_commutator_nq_4ann} into \eqref{eq:tgge_rate_equation_general} one has 
\begin{align}
\sum_j (L_j^{4\alpha})^{\dagger}[\hat{n}_q,L_j^{4\alpha}] =-\frac{\Gamma_{4\alpha} }{L^{4}}\sum_{j}\sum_{k,p,m,n} \sum_{k_2,k_3,k_4}&e^{ij(q+k_2+k_3+k_4-k-p-m-n)}g^{\ast}(p,m,n)\hat{c}^{\dagger}_n \hat{c}^{\dagger}_m  \hat{c}^{\dagger}_p \hat{c}^{\dagger}_k \nonumber \\   
&\qquad\qquad\qquad\qquad\Big[g(k_2,k_3,k_4) \hat{c}_q \hat{c}_{k_2}\hat{c}_{k_3} \hat{c}_{k_4} +g(q,k_3,k_4) \hat{c}_{k_2} \hat{c}_q \hat{c}_{k_3}\hat{c}_{k_4}\nonumber \\
&\qquad\qquad\qquad\qquad+g(k_2,q,k_4) \hat{c}_{k_3} \hat{c}_{k_2} \hat{c}_{q}\hat{c}_{k_4}+ g(k_2,k_3,q) \hat{c}_{k_4} \hat{c}_{k_2} \hat{c}_{k_{3}}\hat{c}_{q}\Big].
\label{supeq:intermediate_4ann_commutator_sum}
\end{align}
The previous equation can be split into four terms: 
\begin{align}
&\frac{1}{L^{4}}\sum_{j}\sum_{k,p,m,n} \sum_{k_2,k_3,k_4}e^{ij(q+k_2+k_3+k_4-k-p-m-n)}g^{\ast}(p,m,n) g(k_2,k_3,k_4) \hat{c}^{\dagger}_n \hat{c}^{\dagger}_m  \hat{c}^{\dagger}_p \hat{c}^{\dagger}_k  \hat{c}_q \hat{c}_{k_2}\hat{c}_{k_3} \hat{c}_{k_4} \nonumber \\
&=\frac{1}{L^{3}}\sum_{k,p,m} \sum_{k_2,k_3,k_4}g^{\ast}(p,m,q+k_2+k_3+k_4-k-p-m) g(k_2,k_3,k_4) \hat{c}^{\dagger}_{q+k_2+k_3+k_4-k-p-m} \hat{c}^{\dagger}_m  \hat{c}^{\dagger}_p \hat{c}^{\dagger}_k  \hat{c}_q \hat{c}_{k_2}\hat{c}_{k_3} \hat{c}_{k_4},
\label{supeq:term1_4ann}
\end{align}

\begin{align}
&\frac{1}{L^{4}}\sum_{j}\sum_{k,p,m,n} \sum_{k_2,k_3,k_4}e^{ij(q+k_2+k_3+k_4-k-p-m-n)}g^{\ast}(p,m,n) g(q,k_3,k_4) \hat{c}^{\dagger}_n \hat{c}^{\dagger}_m  \hat{c}^{\dagger}_p \hat{c}^{\dagger}_k  \hat{c}_{k_2} \hat{c}_{q}\hat{c}_{k_3} \hat{c}_{k_4} \nonumber \\
&=\frac{1}{L^{3}}\sum_{k,p,m} \sum_{k_2,k_3,k_4}g^{\ast}(p,m,q+k_2+k_3+k_4-k-p-m) g(q,k_3,k_4) \hat{c}^{\dagger}_{q+k_2+k_3+k_4-k-p-m} \hat{c}^{\dagger}_m  \hat{c}^{\dagger}_p \hat{c}^{\dagger}_k  \hat{c}_{k_2} \hat{c}_{q}\hat{c}_{k_3} \hat{c}_{k_4},
\label{supeq:term2_4ann}
\end{align}

\begin{align}
&\frac{1}{L^{4}}\sum_{j}\sum_{k,p,m,n} \sum_{k_2,k_3,k_4}e^{ij(q+k_2+k_3+k_4-k-p-m-n)}g^{\ast}(p,m,n) g(k_3,q,k_4) \hat{c}^{\dagger}_n \hat{c}^{\dagger}_m  \hat{c}^{\dagger}_p \hat{c}^{\dagger}_k  \hat{c}_{k_2} \hat{c}_{k_3}\hat{c}_{q} \hat{c}_{k_4} \nonumber \\
&=\frac{1}{L^{3}}\sum_{k,p,m} \sum_{k_2,k_3,k_4}g^{\ast}(p,m,q+k_2+k_3+k_4-k-p-m) g(k_3,q,k_4) \hat{c}^{\dagger}_{q+k_2+k_3+k_4-k-p-m} \hat{c}^{\dagger}_m  \hat{c}^{\dagger}_p \hat{c}^{\dagger}_k  \hat{c}_{k_2} \hat{c}_{k_3}\hat{c}_{q} \hat{c}_{k_4},
\label{supeq:term3_4ann}
\end{align}

\begin{align}
&\frac{1}{L^{4}}\sum_{j}\sum_{k,p,m,n} \sum_{k_2,k_3,k_4}e^{ij(q+k_2+k_3+k_4-k-p-m-n)}g^{\ast}(p,m,n) g(k_3,k_4,q) \hat{c}^{\dagger}_n \hat{c}^{\dagger}_m  \hat{c}^{\dagger}_p \hat{c}^{\dagger}_k  \hat{c}_{k_2} \hat{c}_{k_3}\hat{c}_{k_4} \hat{c}_{q} \nonumber \\
&=\frac{1}{L^{3}}\sum_{k,p,m} \sum_{k_2,k_3,k_4}g^{\ast}(p,m,q+k_2+k_3+k_4-k-p-m) g(k_3,k_4,q) \hat{c}^{\dagger}_{q+k_2+k_3+k_4-k-p-m} \hat{c}^{\dagger}_m  \hat{c}^{\dagger}_p \hat{c}^{\dagger}_k  \hat{c}_{k_2} \hat{c}_{k_3}\hat{c}_{k_4} \hat{c}_{q}.
\label{supeq:term4_4ann}
\end{align}
The four summations in Eqs.~\eqref{supeq:term1_4ann}-\eqref{supeq:term4_4ann} can be grouped into a single sum exploiting the fermionic anticommutation relations, so that Eq.~\eqref{supeq:intermediate_4ann_commutator_sum} reads as 
\begin{align}
\sum_j &(L_j^{4\alpha})^{\dagger}[\hat{n}_q,L_j^{4\alpha}]=\nonumber \\
&=\frac{1}{L^{3}}\sum_{k,p,m} \sum_{k_2,k_3,k_4}g^{\ast}(p,m,q+k_2+k_3+k_4-k-p-m) f(k_2,k_3,k_4,q) \hat{c}^{\dagger}_{q+k_2+k_3+k_4-k-p-m} \hat{c}^{\dagger}_m  \hat{c}^{\dagger}_p \hat{c}^{\dagger}_k  \hat{c}_{k_2} \hat{c}_{k_3}\hat{c}_{k_4} \hat{c}_{q},  
\label{supeq:4ann_intermediate_1sum}
\end{align}
with the function $f(k_2,k_3,k_4,q)$ given by
\begin{equation}
f(k_2,k_3,k_4,q)=g(k_2,k_3,k_4)-g(q,k_3,k_4)+g(k_3,q,k_4)-g(k_3,k_4,q).
\label{eq:function_4ann_intermediate}
\end{equation}
From the previous equation the calculation proceeds similarly as in Appendix \ref{app:3ann} for $3A\to \emptyset$. In particular, one resorts to the TGGE approximation in order to decompose the eight-point fermionic correlation function in Eq.~\eqref{supeq:4ann_intermediate_1sum} in terms of the two-point function $C_k(t)$ via Wick's theorem. A lengthy calculation leads after some algebraic manipulations to Eq.~\eqref{eq:4ann_rate_equation_GGE},  
\begin{equation}
\frac{\mbox{d}C_q(\tau)}{\mbox{d}\tau}\!=\!-\frac{C_q(\tau)}{L^3}\sum_{k,k',k''} \!\!\!f_4(k,k',k'',q) C_k(\tau)C_{k'}(\tau)C_{k''}(\tau),
\label{supeq:tGGE_4ann_final}
\end{equation}
with $f_4(k,k',k'',q)$ written as 
\begin{align}
f_4(k,k',k'',q)&=4-6\cos(k-k')-6\cos(k-q)-4\cos(2k-2k')-4\cos(2k-2q)-2\cos(3k-3k')-2\cos(3k-3q) \nonumber \\
&+2\cos(2k-k'-k'')+2\cos(k-2k'+k'')+8\cos(2k'-k-q)+4 \cos(k+k'-2q) \nonumber \\
&+4\cos(3k-2k'-k'')+4\cos(3k-2k'-q)+4\cos(3k-k'-2q)+4\cos(2k+k'-3q) \nonumber \\
&-6\cos(3k-k'-k''-q)-2\cos(k+k'+k''-3q)+4\cos(k+k'-k''-q)+2\cos(3k-3k'+k''-q) \nonumber \\
&+2\cos(3k+k'-k''-3q)-4\cos(2k+k'-2 k''-q)-4\cos(2k+k'-k''-2q)\nonumber \\
&-2\cos(3k-2k'-2k''+q)-4\cos(3k-2k'+k''-2q)-2\cos(2k-k'+2k''-3q)\nonumber \\
&+4\cos(2k-2k'+2k''-2q).
\label{eq:f4_function_4ann_complete}
\end{align}
Notice that in passing from Eqs.~\eqref{supeq:4ann_intermediate_1sum} and \eqref{eq:function_4ann_intermediate} to Eqs.~\eqref{supeq:tGGE_4ann_final} and \eqref{eq:f4_function_4ann_complete} we renamed the dummy summation variables $m,p$ with $k',k''$ in order to match the notation used in the main text in Eq.~\eqref{eq:4ann_rate_equation_GGE}. We also introduced the rescaled time $\tau=\Gamma_{4\alpha}t$. Equations \eqref{supeq:tGGE_4ann_final} and \eqref{eq:f4_function_4ann_complete} have been numerically solved in order to the get the results in Fig.~\ref{fig:4ann}. At the technical level, we remark that the calculation of the effective exponent $\delta_{\mathrm{eff}}(\tau)$ at long times, $\tau \simeq 10^{11}$ as in the inset in Fig.~\ref{fig:4ann}(b), requires a large value of $L$. We used $L=180000$ and we checked that the obtained numbers for $\delta_{\mathrm{eff}}(\tau)$ are stable upon further increasing $L$ to $L=240000$.

In the case of Eqs.~\eqref{supeq:tGGE_4ann_final} and \eqref{eq:f4_function_4ann_complete} considerations similar to those done in Appendix \ref{app:3ann} for Eq.~\eqref{eq:triann_rate_simplified_GGE} apply. In particular, the factor $4$ on the first line of the right-hand side of \eqref{eq:f4_function_4ann_complete} represents the law of mass action term \eqref{eq:GGE_MF_4ann}. All the other remaining terms couple the quasimomentum $q$ to all the other quasi momenta through the nontrivial form of the function $f_4$ determined by the fermionic statistics. These terms contribute to the dynamics when the initial occupation function $C_q(0)$ is not flat in $q$ and the initial state therefore displays quantum coherences. This determines the universal behavior of Fig.~\ref{fig:4ann}. We remark that this collective non mean-field behavior is valid already in one dimension, above the upper-critical dimension of $4A\to \emptyset$, as it is originates from quantum effects, due to quantum coherences in the initial state, that are present in any spatial dimension.

\begin{figure}[t]
    \centering
    \includegraphics[width=1\columnwidth]{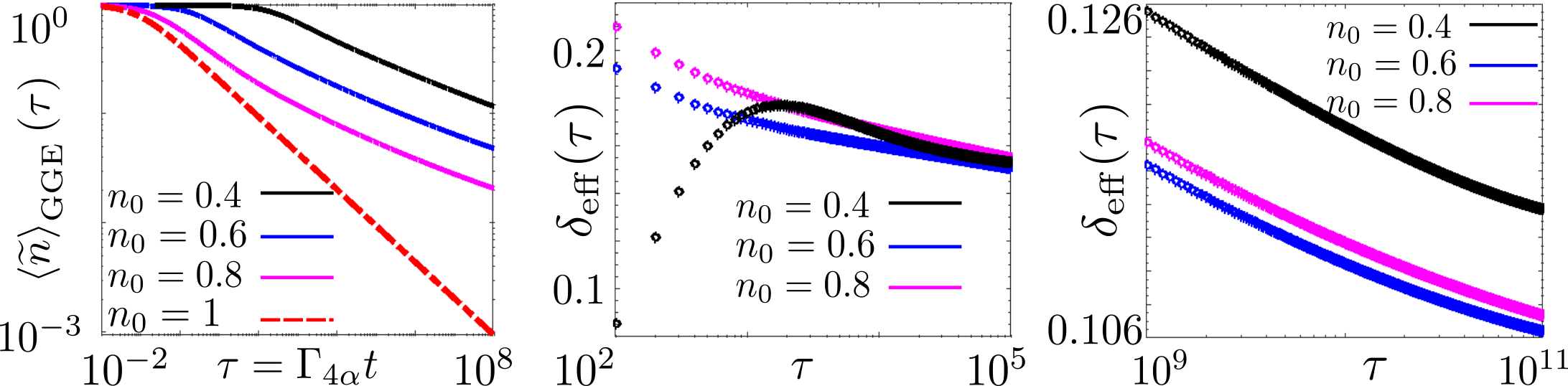}
    \caption{\textbf{Quantum four-body annihilation dynamics for different initial fillings of the FS initial state.}(a) Log-log plot, from the numerical solution of Eq.~\eqref{eq:4ann_rate_equation_GGE} for $4A\to\emptyset$, of the rescaled density $\braket{\widetilde{n}}_{\mathrm{GGE}}(\tau)=\braket{n}_{\mathrm{GGE}}(\tau)/n_0$ as a function of the rescaled time $\tau=\Gamma_{3\alpha}t$. Four different values $n_0=0.4,0.6,0.8$ and $1$ (from top to bottom) of the initial filling of the FS state \eqref{eq:FS_initial_state} are reported. The blue curve for $n_0=0.6$ is identical to that discussed in the main text in Fig.~\ref{fig:4ann}(a). Similarly to the case of Fig.~\ref{fig:3ann_appendix} for $3A\to\emptyset$, the density decay differs from the mean-field prediction obtained with the associated initial density $n_0$ since the initial state displays quantum coherences in real space. In the case $n_0=1$ only, the FS has no real-space coherences and the density decays, red-dashed line in the figure, is exactly reproduced by the law of mass action equation \eqref{eq:MF_rate_equation} with $\braket{\widetilde{n}}_{\mathrm{GGE}}(\tau)=\braket{\widetilde{n}}_{\mathrm{MF}}(\tau)\sim \tau^{-1/3}$. (b) Plot of the effective exponent $\delta_{\mathrm{eff}}(\tau)$ \eqref{eq:effective_exponent} as a function of $\tau$ (log scale on the horizontal axis only) for the FS initial state and initial fillings $n_0=0.4,0.6$ and $0.8$. The blue curve for $n_0=0.6$ is equal to that of Fig.~\ref{fig:4ann}(b) of the main text. After an initial transient, which is longer for $n_0=0.4$ ($\tau \lesssim 10^3$ ), the effective exponent monotonically decreases in time in all the three cases. (c) Plot of the effective exponent for the very same values of the initial filling $n_0$ as in (b) but for longer times $\tau \gtrsim 10^9$. The effective exponent is monotonically decreasing towards a value $\delta_{\mathrm{eff}}(\tau)\simeq 0.1$.}
    \label{fig:4ann_appendix}
\end{figure}

In Fig.~\ref{fig:4ann_appendix}, we report the quantum reaction-limited $4A\to \emptyset$ dynamics from the FS initial state for various values of the initial filling $n_0$. In Fig.~\ref{fig:4ann_appendix}(a), one can see that universal behavior beyond mean field is present for $n_0 \neq 1$. For $n_0=1$, instead, $C_q(0)$ is flat in $q$ and $\braket{\widetilde{n}}_{\mathrm{GGE}}(\tau)=\braket{n}_{\mathrm{GGE}}(\tau)/n_0=\braket{\widetilde{n}}_{\mathrm{MF}}\sim \tau^{-1/3}$. In Fig.~\ref{fig:4ann_appendix}(b), we show the effective exponent $\delta_{\mathrm{eff}}(\tau)$ for the same values of $n_0$ used in Fig.~\ref{fig:4ann_appendix}(a). For $n_0=0.4$, an initial increase of $\delta_{\mathrm{eff}}(\tau)$ for $\tau \lesssim 10^3$ is observed due to the early time nonuniversal and nonalgebraic decay of the density [cf. the topmost curve in Fig.~\ref{fig:4ann_appendix}(a)]. Such behavior is present also for $n_0=0.6$ and $0.8$ but it takes place for earlier times that $\tau\sim 10^2$ and it is therefore not visible in Fig.~\ref{fig:4ann_appendix}(b). Apart from this-early time nonuniversal regime, $\delta_{\mathrm{eff}}(\tau)$ is monotonically decreasing also for long times, reported in Fig.~\ref{fig:4ann_appendix}(c). This is in contrast with the case $3A\to\emptyset$ in Fig.~\ref{fig:3ann} of the main text and in Fig.~\ref{fig:3ann_appendix}. At long times, the effective exponent slowly converges to the value $\delta_{\mathrm{eff}}(\tau) \simeq 0.1$. Also in this case this asymptotic decay $\braket{\widetilde{n}}_{\mathrm{GGE}}(\tau)\sim \tau^{-0.1}$ holds for any $n_0 \neq 1$, i.e., whenever the FS initial state possesses quantum coherences in real space, i.e., $C_q(0)$ not flat in $q$. Here, $n_0$ solely determines the faster ($n_0=0.6$ and $0.8$) or slower ($n_0=0.4$) approach to power-law asymptotic with $\delta_{\mathrm{eff}}(\tau)\simeq 0.1$, as shown in Fig.~\ref{fig:4ann_appendix}(c). We also expect this behavior to apply generically to other coherent initial states, different from the FS, identified by a nonflat momentum occupation function $C_q(0)$. The numerical data in Fig.~\ref{fig:4ann_appendix} have been produced using $L=180000$ (as for Fig.~\ref{fig:4ann} of the main text).

\twocolumngrid
\bibliography{biblio}

\begin{thebibliography}{79}%
\makeatletter
\providecommand \@ifxundefined [1]{%
 \@ifx{#1\undefined}
}%
\providecommand \@ifnum [1]{%
 \ifnum #1\expandafter \@firstoftwo
 \else \expandafter \@secondoftwo
 \fi
}%
\providecommand \@ifx [1]{%
 \ifx #1\expandafter \@firstoftwo
 \else \expandafter \@secondoftwo
 \fi
}%
\providecommand \natexlab [1]{#1}%
\providecommand \enquote  [1]{``#1''}%
\providecommand \bibnamefont  [1]{#1}%
\providecommand \bibfnamefont [1]{#1}%
\providecommand \citenamefont [1]{#1}%
\providecommand \href@noop [0]{\@secondoftwo}%
\providecommand \href [0]{\begingroup \@sanitize@url \@href}%
\providecommand \@href[1]{\@@startlink{#1}\@@href}%
\providecommand \@@href[1]{\endgroup#1\@@endlink}%
\providecommand \@sanitize@url [0]{\catcode `\\12\catcode `\$12\catcode
  `\&12\catcode `\#12\catcode `\^12\catcode `\_12\catcode `\%12\relax}%
\providecommand \@@startlink[1]{}%
\providecommand \@@endlink[0]{}%
\providecommand \url  [0]{\begingroup\@sanitize@url \@url }%
\providecommand \@url [1]{\endgroup\@href {#1}{\urlprefix }}%
\providecommand \urlprefix  [0]{URL }%
\providecommand \Eprint [0]{\href }%
\providecommand \doibase [0]{https://doi.org/}%
\providecommand \selectlanguage [0]{\@gobble}%
\providecommand \bibinfo  [0]{\@secondoftwo}%
\providecommand \bibfield  [0]{\@secondoftwo}%
\providecommand \translation [1]{[#1]}%
\providecommand \BibitemOpen [0]{}%
\providecommand \bibitemStop [0]{}%
\providecommand \bibitemNoStop [0]{.\EOS\space}%
\providecommand \EOS [0]{\spacefactor3000\relax}%
\providecommand \BibitemShut  [1]{\csname bibitem#1\endcsname}%
\let\auto@bib@innerbib\@empty
\bibitem [{\citenamefont {Wilson}\ and\ \citenamefont
  {Kogut}(1974)}]{wilson1974renormalization}%
  \BibitemOpen
  \bibfield  {author} {\bibinfo {author} {\bibfnamefont {K.~G.}\ \bibnamefont
  {Wilson}}\ and\ \bibinfo {author} {\bibfnamefont {J.}~\bibnamefont {Kogut}},\
  }\bibfield  {title} {\bibinfo {title} {\emph{The renormalization group and
  the $\epsilon$ expansion}},\ }\href
  {https://doi.org/https://doi.org/10.1016/0370-1573(74)90023-4} {\bibfield
  {journal} {\bibinfo  {journal} {Phys. Rep.}\ }\textbf {\bibinfo {volume}
  {12}},\ \bibinfo {pages} {75} (\bibinfo {year} {1974})}\BibitemShut {NoStop}%
\bibitem [{\citenamefont {Kogut}(1979)}]{Kogut1979}%
  \BibitemOpen
  \bibfield  {author} {\bibinfo {author} {\bibfnamefont {J.~B.}\ \bibnamefont
  {Kogut}},\ }\bibfield  {title} {\bibinfo {title} {\emph{An introduction to
  lattice gauge theory and spin systems}},\ }\href
  {https://doi.org/10.1103/RevModPhys.51.659} {\bibfield  {journal} {\bibinfo
  {journal} {Rev. Mod. Phys.}\ }\textbf {\bibinfo {volume} {51}},\ \bibinfo
  {pages} {659} (\bibinfo {year} {1979})}\BibitemShut {NoStop}%
\bibitem [{\citenamefont {Henkel}\ \emph {et~al.}(2008)\citenamefont {Henkel},
  \citenamefont {Hinrichsen}, \citenamefont {L{\"u}beck},\ and\ \citenamefont
  {Pleimling}}]{henkel2008non}%
  \BibitemOpen
  \bibfield  {author} {\bibinfo {author} {\bibfnamefont {M.}~\bibnamefont
  {Henkel}}, \bibinfo {author} {\bibfnamefont {H.}~\bibnamefont {Hinrichsen}},
  \bibinfo {author} {\bibfnamefont {S.}~\bibnamefont {L{\"u}beck}},\ and\
  \bibinfo {author} {\bibfnamefont {M.}~\bibnamefont {Pleimling}},\ }\href
  {https://doi.org/https://doi.org/10.1007/978-1-4020-8765-3} {\emph {\bibinfo
  {title} {Non-equilibrium phase transitions}}},\ Vol.~\bibinfo {volume} {1}\
  (\bibinfo  {publisher} {Springer},\ \bibinfo {year} {2008})\BibitemShut
  {NoStop}%
\bibitem [{\citenamefont {Hinrichsen}(2000)}]{hinrichsen2000non}%
  \BibitemOpen
  \bibfield  {author} {\bibinfo {author} {\bibfnamefont {H.}~\bibnamefont
  {Hinrichsen}},\ }\bibfield  {title} {\bibinfo {title} {\emph{Non-equilibrium
  critical phenomena and phase transitions into absorbing states}},\ }\href
  {https://doi.org/https://doi.org/10.1080/00018730050198152} {\bibfield
  {journal} {\bibinfo  {journal} {Adv. Phys.}\ }\textbf {\bibinfo {volume}
  {49}},\ \bibinfo {pages} {815} (\bibinfo {year} {2000})}\BibitemShut
  {NoStop}%
\bibitem [{\citenamefont {Privman}(1997)}]{vladimir1997nonequilibrium}%
  \BibitemOpen
  \bibfield  {author} {\bibinfo {author} {\bibfnamefont {V.}~\bibnamefont
  {Privman}},\ }\href
  {https://doi.org/https://doi.org/10.1017/CBO9780511564284} {\emph {\bibinfo
  {title} {Nonequilibrium statistical mechanics in one dimension}}}\ (\bibinfo
  {publisher} {Cambridge University Press},\ \bibinfo {year}
  {1997})\BibitemShut {NoStop}%
\bibitem [{\citenamefont {T{\"a}uber}()}]{tauber2002dynamic}%
  \BibitemOpen
  \bibfield  {author} {\bibinfo {author} {\bibfnamefont {U.~C.}\ \bibnamefont
  {T{\"a}uber}},\ }\bibfield  {title} {\bibinfo {title} {\emph{Dynamic phase
  transitions in diffusion-limited reactions}},\ }\href
  {https://arxiv.org/abs/cond-mat/0205327} {\bibinfo  {journal}
  {arXiv:cond-mat/0205327 (2002)}\ }\BibitemShut {NoStop}%
\bibitem [{\citenamefont {T{\"a}uber}\ \emph {et~al.}(2005)\citenamefont
  {T{\"a}uber}, \citenamefont {Howard},\ and\ \citenamefont
  {Vollmayr-Lee}}]{tauber2005applications}%
  \BibitemOpen
\bibfield  {journal} {  }\bibfield  {author} {\bibinfo {author} {\bibfnamefont
  {U.~C.}\ \bibnamefont {T{\"a}uber}}, \bibinfo {author} {\bibfnamefont
  {M.}~\bibnamefont {Howard}},\ and\ \bibinfo {author} {\bibfnamefont {B.~P.}\
  \bibnamefont {Vollmayr-Lee}},\ }\bibfield  {title} {\bibinfo {title}
  {\emph{Applications of field-theoretic renormalization group methods to
  reaction--diffusion problems}},\ }\href
  {https://doi.org/https://doi.org/10.1088/0305-4470/38/17/R01} {\bibfield
  {journal} {\bibinfo  {journal} {J. Phys. A: Math. Gen.}\ }\textbf {\bibinfo
  {volume} {38}},\ \bibinfo {pages} {R79} (\bibinfo {year} {2005})}\BibitemShut
  {NoStop}%
\bibitem [{\citenamefont {T{\"a}uber}(2014)}]{tauber2014critical}%
  \BibitemOpen
  \bibfield  {author} {\bibinfo {author} {\bibfnamefont {U.~C.}\ \bibnamefont
  {T{\"a}uber}},\ }\href
  {https://doi.org/https://doi.org/10.1017/CBO9781139046213} {\emph {\bibinfo
  {title} {Critical dynamics: a field theory approach to equilibrium and
  non-equilibrium scaling behavior}}}\ (\bibinfo  {publisher} {Cambridge
  University Press},\ \bibinfo {year} {2014})\BibitemShut {NoStop}%
\bibitem [{\citenamefont {Krapivsky}\ \emph {et~al.}(2010)\citenamefont
  {Krapivsky}, \citenamefont {Redner},\ and\ \citenamefont
  {Ben-Naim}}]{krapivsky2010kinetic}%
  \BibitemOpen
  \bibfield  {author} {\bibinfo {author} {\bibfnamefont {P.~L.}\ \bibnamefont
  {Krapivsky}}, \bibinfo {author} {\bibfnamefont {S.}~\bibnamefont {Redner}},\
  and\ \bibinfo {author} {\bibfnamefont {E.}~\bibnamefont {Ben-Naim}},\ }\href
  {https://doi.org/https://doi.org/10.1017/CBO9780511780516} {\emph {\bibinfo
  {title} {A kinetic view of statistical physics}}}\ (\bibinfo  {publisher}
  {Cambridge University Press},\ \bibinfo {year} {2010})\BibitemShut {NoStop}%
\bibitem [{\citenamefont {Kang}\ and\ \citenamefont
  {Redner}(1985)}]{Redner1984}%
  \BibitemOpen
  \bibfield  {author} {\bibinfo {author} {\bibfnamefont {K.}~\bibnamefont
  {Kang}}\ and\ \bibinfo {author} {\bibfnamefont {S.}~\bibnamefont {Redner}},\
  }\bibfield  {title} {\bibinfo {title} {\emph{Fluctuation-dominated kinetics
  in diffusion-controlled reactions}},\ }\href
  {https://doi.org/10.1103/PhysRevA.32.435} {\bibfield  {journal} {\bibinfo
  {journal} {Phys. Rev. A}\ }\textbf {\bibinfo {volume} {32}},\ \bibinfo
  {pages} {435} (\bibinfo {year} {1985})}\BibitemShut {NoStop}%
\bibitem [{\citenamefont {Privman}\ and\ \citenamefont
  {Grynberg}(1992)}]{fastdiffusion1992}%
  \BibitemOpen
  \bibfield  {author} {\bibinfo {author} {\bibfnamefont {V.}~\bibnamefont
  {Privman}}\ and\ \bibinfo {author} {\bibfnamefont {M.~D.}\ \bibnamefont
  {Grynberg}},\ }\bibfield  {title} {\bibinfo {title} {\emph{Fast-diffusion
  mean-field theory for k-body reactions in one dimension}},\ }\href
  {https://doi.org/https://doi.org/10.1088/0305-4470/25/24/011} {\bibfield
  {journal} {\bibinfo  {journal} {J. Phys. A. Math. Gen.}\ }\textbf {\bibinfo
  {volume} {25}},\ \bibinfo {pages} {6567} (\bibinfo {year}
  {1992})}\BibitemShut {NoStop}%
\bibitem [{\citenamefont {Toussaint}\ and\ \citenamefont
  {Wilczek}(1983)}]{toussaint1983particle}%
  \BibitemOpen
  \bibfield  {author} {\bibinfo {author} {\bibfnamefont {D.}~\bibnamefont
  {Toussaint}}\ and\ \bibinfo {author} {\bibfnamefont {F.}~\bibnamefont
  {Wilczek}},\ }\bibfield  {title} {\bibinfo {title}
  {\emph{Particle--antiparticle annihilation in diffusive motion}},\ }\href
  {https://doi.org/https://doi.org/10.1063/1.445022} {\bibfield  {journal}
  {\bibinfo  {journal} {J. Chem. Phys.}\ }\textbf {\bibinfo {volume} {78}},\
  \bibinfo {pages} {2642} (\bibinfo {year} {1983})}\BibitemShut {NoStop}%
\bibitem [{\citenamefont {Spouge}(1988)}]{Spouge1988}%
  \BibitemOpen
  \bibfield  {author} {\bibinfo {author} {\bibfnamefont {J.~L.}\ \bibnamefont
  {Spouge}},\ }\bibfield  {title} {\bibinfo {title} {\emph{Exact solutions for
  a diffusion-reaction process in one dimension}},\ }\href
  {https://doi.org/10.1103/PhysRevLett.60.871} {\bibfield  {journal} {\bibinfo
  {journal} {Phys. Rev. Lett.}\ }\textbf {\bibinfo {volume} {60}},\ \bibinfo
  {pages} {871} (\bibinfo {year} {1988})}\BibitemShut {NoStop}%
\bibitem [{\citenamefont {Privman}(1994)}]{Privman1994}%
  \BibitemOpen
  \bibfield  {author} {\bibinfo {author} {\bibfnamefont {V.}~\bibnamefont
  {Privman}},\ }\bibfield  {title} {\bibinfo {title} {\emph{Exact results for
  diffusion-limited reactions with synchronous dynamics}},\ }\href
  {https://doi.org/10.1103/PhysRevE.50.50} {\bibfield  {journal} {\bibinfo
  {journal} {Phys. Rev. E}\ }\textbf {\bibinfo {volume} {50}},\ \bibinfo
  {pages} {50} (\bibinfo {year} {1994})}\BibitemShut {NoStop}%
\bibitem [{\citenamefont {Torney}\ and\ \citenamefont
  {McConnell}(1983)}]{torney1983diffusion}%
  \BibitemOpen
  \bibfield  {author} {\bibinfo {author} {\bibfnamefont {D.~C.}\ \bibnamefont
  {Torney}}\ and\ \bibinfo {author} {\bibfnamefont {H.~M.}\ \bibnamefont
  {McConnell}},\ }\bibfield  {title} {\bibinfo {title} {\emph{Diffusion-limited
  reactions in one dimension}},\ }\href
  {https://doi.org/https://doi.org/10.1021/j100234a023} {\bibfield  {journal}
  {\bibinfo  {journal} {J. Phys. Chem.}\ }\textbf {\bibinfo {volume} {87}},\
  \bibinfo {pages} {1941} (\bibinfo {year} {1983})}\BibitemShut {NoStop}%
\bibitem [{\citenamefont {Kang}\ and\ \citenamefont
  {Redner}(1984{\natexlab{a}})}]{fluctuationseffects}%
  \BibitemOpen
  \bibfield  {author} {\bibinfo {author} {\bibfnamefont {K.}~\bibnamefont
  {Kang}}\ and\ \bibinfo {author} {\bibfnamefont {S.}~\bibnamefont {Redner}},\
  }\bibfield  {title} {\bibinfo {title} {\emph{Fluctuation effects in
  Smoluchowski reaction kinetics}},\ }\href
  {https://doi.org/10.1103/PhysRevA.30.2833} {\bibfield  {journal} {\bibinfo
  {journal} {Phys. Rev. A}\ }\textbf {\bibinfo {volume} {30}},\ \bibinfo
  {pages} {2833} (\bibinfo {year} {1984}{\natexlab{a}})}\BibitemShut {NoStop}%
\bibitem [{\citenamefont {Kang}\ and\ \citenamefont
  {Redner}(1984{\natexlab{b}})}]{redner1984scaling}%
  \BibitemOpen
  \bibfield  {author} {\bibinfo {author} {\bibfnamefont {K.}~\bibnamefont
  {Kang}}\ and\ \bibinfo {author} {\bibfnamefont {S.}~\bibnamefont {Redner}},\
  }\bibfield  {title} {\bibinfo {title} {\emph{Scaling Approach for the
  Kinetics of Recombination Processes}},\ }\href
  {https://doi.org/10.1103/PhysRevLett.52.955} {\bibfield  {journal} {\bibinfo
  {journal} {Phys. Rev. Lett.}\ }\textbf {\bibinfo {volume} {52}},\ \bibinfo
  {pages} {955} (\bibinfo {year} {1984}{\natexlab{b}})}\BibitemShut {NoStop}%
\bibitem [{\citenamefont {Kang}\ \emph {et~al.}(1984)\citenamefont {Kang},
  \citenamefont {Meakin}, \citenamefont {Oh},\ and\ \citenamefont
  {Redner}}]{kang1984universal}%
  \BibitemOpen
  \bibfield  {author} {\bibinfo {author} {\bibfnamefont {K.}~\bibnamefont
  {Kang}}, \bibinfo {author} {\bibfnamefont {P.}~\bibnamefont {Meakin}},
  \bibinfo {author} {\bibfnamefont {J.}~\bibnamefont {Oh}},\ and\ \bibinfo
  {author} {\bibfnamefont {S.}~\bibnamefont {Redner}},\ }\bibfield  {title}
  {\bibinfo {title} {\emph{Universal behaviour of $N$-body decay processes}},\
  }\href {https://doi.org/10.1088/0305-4470/17/12/007} {\bibfield  {journal}
  {\bibinfo  {journal} {J. Phys. A: Math. Gen.}\ }\textbf {\bibinfo {volume}
  {17}},\ \bibinfo {pages} {L665} (\bibinfo {year} {1984})}\BibitemShut
  {NoStop}%
\bibitem [{\citenamefont {R\'acz}(1985)}]{Racz1985}%
  \BibitemOpen
  \bibfield  {author} {\bibinfo {author} {\bibfnamefont {Z.}~\bibnamefont
  {R\'acz}},\ }\bibfield  {title} {\bibinfo {title} {\emph{Diffusion-controlled
  annihilation in the presence of particle sources: Exact results in one
  dimension}},\ }\href {https://doi.org/10.1103/PhysRevLett.55.1707} {\bibfield
   {journal} {\bibinfo  {journal} {Phys. Rev. Lett.}\ }\textbf {\bibinfo
  {volume} {55}},\ \bibinfo {pages} {1707} (\bibinfo {year}
  {1985})}\BibitemShut {NoStop}%
\bibitem [{\citenamefont {Doi}(1976)}]{doi1976stochastic}%
  \BibitemOpen
  \bibfield  {author} {\bibinfo {author} {\bibfnamefont {M.}~\bibnamefont
  {Doi}},\ }\bibfield  {title} {\bibinfo {title} {\emph{Stochastic theory of
  diffusion-controlled reaction}},\ }\href
  {https://doi.org/https://doi.org/10.1088/0305-4470/9/9/009} {\bibfield
  {journal} {\bibinfo  {journal} {J. Phys. A: Math. Gen.}\ }\textbf {\bibinfo
  {volume} {9}},\ \bibinfo {pages} {1479} (\bibinfo {year} {1976})}\BibitemShut
  {NoStop}%
\bibitem [{\citenamefont {Peliti}(1985)}]{peliti1985path}%
  \BibitemOpen
  \bibfield  {author} {\bibinfo {author} {\bibfnamefont {L.}~\bibnamefont
  {Peliti}},\ }\bibfield  {title} {\bibinfo {title} {Path integral approach to
  birth-death processes on a lattice},\ }\href
  {https://doi.org/https://doi.org/10.1051/jphys:019850046090146900} {\bibfield
   {journal} {\bibinfo  {journal} {J. Phys. France}\ }\textbf {\bibinfo
  {volume} {46}},\ \bibinfo {pages} {1469} (\bibinfo {year}
  {1985})}\BibitemShut {NoStop}%
\bibitem [{\citenamefont {Peliti}(1986)}]{peliti1986renormalisation}%
  \BibitemOpen
  \bibfield  {author} {\bibinfo {author} {\bibfnamefont {L.}~\bibnamefont
  {Peliti}},\ }\bibfield  {title} {\bibinfo {title} {\emph{Renormalisation of
  fluctuation effects in the $A+ A \to A$ reaction}},\ }\href
  {https://doi.org/https://doi.org/10.1088/0305-4470/19/6/012} {\bibfield
  {journal} {\bibinfo  {journal} {J. Phys. A: Math. Gen.}\ }\textbf {\bibinfo
  {volume} {19}},\ \bibinfo {pages} {L365} (\bibinfo {year}
  {1986})}\BibitemShut {NoStop}%
\bibitem [{\citenamefont {Mattis}\ and\ \citenamefont
  {Glasser}(1998)}]{QFT_RD_1998}%
  \BibitemOpen
  \bibfield  {author} {\bibinfo {author} {\bibfnamefont {D.~C.}\ \bibnamefont
  {Mattis}}\ and\ \bibinfo {author} {\bibfnamefont {M.~L.}\ \bibnamefont
  {Glasser}},\ }\bibfield  {title} {\bibinfo {title} {\emph{The uses of quantum
  field theory in diffusion-limited reactions}},\ }\href
  {https://doi.org/10.1103/RevModPhys.70.979} {\bibfield  {journal} {\bibinfo
  {journal} {Rev. Mod. Phys.}\ }\textbf {\bibinfo {volume} {70}},\ \bibinfo
  {pages} {979} (\bibinfo {year} {1998})}\BibitemShut {NoStop}%
\bibitem [{\citenamefont {Syassen}\ \emph {et~al.}(2008)\citenamefont
  {Syassen}, \citenamefont {Bauer}, \citenamefont {Lettner}, \citenamefont
  {Volz}, \citenamefont {Dietze}, \citenamefont {García-Ripoll}, \citenamefont
  {Cirac}, \citenamefont {Rempe},\ and\ \citenamefont {Dürr}}]{lossexp0}%
  \BibitemOpen
  \bibfield  {author} {\bibinfo {author} {\bibfnamefont {N.}~\bibnamefont
  {Syassen}}, \bibinfo {author} {\bibfnamefont {D.~M.}\ \bibnamefont {Bauer}},
  \bibinfo {author} {\bibfnamefont {M.}~\bibnamefont {Lettner}}, \bibinfo
  {author} {\bibfnamefont {T.}~\bibnamefont {Volz}}, \bibinfo {author}
  {\bibfnamefont {D.}~\bibnamefont {Dietze}}, \bibinfo {author} {\bibfnamefont
  {J.~J.}\ \bibnamefont {García-Ripoll}}, \bibinfo {author} {\bibfnamefont
  {J.~I.}\ \bibnamefont {Cirac}}, \bibinfo {author} {\bibfnamefont
  {G.}~\bibnamefont {Rempe}},\ and\ \bibinfo {author} {\bibfnamefont
  {S.}~\bibnamefont {Dürr}},\ }\bibfield  {title} {\bibinfo {title}
  {\emph{Strong Dissipation Inhibits Losses and Induces Correlations in Cold
  Molecular Gases}},\ }\href {https://doi.org/10.1126/science.1155309}
  {\bibfield  {journal} {\bibinfo  {journal} {Science}\ }\textbf {\bibinfo
  {volume} {320}},\ \bibinfo {pages} {1329} (\bibinfo {year}
  {2008})}\BibitemShut {NoStop}%
\bibitem [{\citenamefont {Traverso}\ \emph {et~al.}(2009)\citenamefont
  {Traverso}, \citenamefont {Chakraborty}, \citenamefont {Martinez~de Escobar},
  \citenamefont {Mickelson}, \citenamefont {Nagel}, \citenamefont {Yan},\ and\
  \citenamefont {Killian}}]{lossexp3}%
  \BibitemOpen
  \bibfield  {author} {\bibinfo {author} {\bibfnamefont {A.}~\bibnamefont
  {Traverso}}, \bibinfo {author} {\bibfnamefont {R.}~\bibnamefont
  {Chakraborty}}, \bibinfo {author} {\bibfnamefont {Y.~N.}\ \bibnamefont
  {Martinez~de Escobar}}, \bibinfo {author} {\bibfnamefont {P.~G.}\
  \bibnamefont {Mickelson}}, \bibinfo {author} {\bibfnamefont {S.~B.}\
  \bibnamefont {Nagel}}, \bibinfo {author} {\bibfnamefont {M.}~\bibnamefont
  {Yan}},\ and\ \bibinfo {author} {\bibfnamefont {T.~C.}\ \bibnamefont
  {Killian}},\ }\bibfield  {title} {\bibinfo {title} {\emph{Inelastic and
  elastic collision rates for triplet states of ultracold strontium}},\ }\href
  {https://doi.org/10.1103/PhysRevA.79.060702} {\bibfield  {journal} {\bibinfo
  {journal} {Phys. Rev. A}\ }\textbf {\bibinfo {volume} {79}},\ \bibinfo
  {pages} {060702} (\bibinfo {year} {2009})}\BibitemShut {NoStop}%
\bibitem [{\citenamefont {Yamaguchi}\ \emph {et~al.}(2008)\citenamefont
  {Yamaguchi}, \citenamefont {Uetake}, \citenamefont {Hashimoto}, \citenamefont
  {Doyle},\ and\ \citenamefont {Takahashi}}]{lossexp4}%
  \BibitemOpen
  \bibfield  {author} {\bibinfo {author} {\bibfnamefont {A.}~\bibnamefont
  {Yamaguchi}}, \bibinfo {author} {\bibfnamefont {S.}~\bibnamefont {Uetake}},
  \bibinfo {author} {\bibfnamefont {D.}~\bibnamefont {Hashimoto}}, \bibinfo
  {author} {\bibfnamefont {J.~M.}\ \bibnamefont {Doyle}},\ and\ \bibinfo
  {author} {\bibfnamefont {Y.}~\bibnamefont {Takahashi}},\ }\bibfield  {title}
  {\bibinfo {title} {\emph{Inelastic Collisions in Optically Trapped Ultracold
  Metastable Ytterbium}},\ }\href
  {https://doi.org/10.1103/PhysRevLett.101.233002} {\bibfield  {journal}
  {\bibinfo  {journal} {Phys. Rev. Lett.}\ }\textbf {\bibinfo {volume} {101}},\
  \bibinfo {pages} {233002} (\bibinfo {year} {2008})}\BibitemShut {NoStop}%
\bibitem [{\citenamefont {Yan}\ \emph {et~al.}(2013)\citenamefont {Yan},
  \citenamefont {Moses}, \citenamefont {Gadway}, \citenamefont {Covey},
  \citenamefont {Hazzard}, \citenamefont {Rey}, \citenamefont {Jin},\ and\
  \citenamefont {Ye}}]{lossexpF1}%
  \BibitemOpen
  \bibfield  {author} {\bibinfo {author} {\bibfnamefont {B.}~\bibnamefont
  {Yan}}, \bibinfo {author} {\bibfnamefont {S.~A.}\ \bibnamefont {Moses}},
  \bibinfo {author} {\bibfnamefont {B.}~\bibnamefont {Gadway}}, \bibinfo
  {author} {\bibfnamefont {J.~P.}\ \bibnamefont {Covey}}, \bibinfo {author}
  {\bibfnamefont {K.~R.}\ \bibnamefont {Hazzard}}, \bibinfo {author}
  {\bibfnamefont {A.~M.}\ \bibnamefont {Rey}}, \bibinfo {author} {\bibfnamefont
  {D.~S.}\ \bibnamefont {Jin}},\ and\ \bibinfo {author} {\bibfnamefont
  {J.}~\bibnamefont {Ye}},\ }\bibfield  {title} {\bibinfo {title}
  {\emph{Observation of dipolar spin-exchange interactions with
  lattice-confined polar molecules}},\ }\href
  {https://doi.org/https://doi.org/10.1038/nature12483} {\bibfield  {journal}
  {\bibinfo  {journal} {Nature}\ }\textbf {\bibinfo {volume} {501}},\ \bibinfo
  {pages} {521} (\bibinfo {year} {2013})}\BibitemShut {NoStop}%
\bibitem [{\citenamefont {Zhu}\ \emph {et~al.}(2014)\citenamefont {Zhu},
  \citenamefont {Gadway}, \citenamefont {Foss-Feig}, \citenamefont
  {Schachenmayer}, \citenamefont {Wall}, \citenamefont {Hazzard}, \citenamefont
  {Yan}, \citenamefont {Moses}, \citenamefont {Covey}, \citenamefont {Jin},
  \citenamefont {Ye}, \citenamefont {Holland},\ and\ \citenamefont
  {Rey}}]{lossexpF2}%
  \BibitemOpen
  \bibfield  {author} {\bibinfo {author} {\bibfnamefont {B.}~\bibnamefont
  {Zhu}}, \bibinfo {author} {\bibfnamefont {B.}~\bibnamefont {Gadway}},
  \bibinfo {author} {\bibfnamefont {M.}~\bibnamefont {Foss-Feig}}, \bibinfo
  {author} {\bibfnamefont {J.}~\bibnamefont {Schachenmayer}}, \bibinfo {author}
  {\bibfnamefont {M.~L.}\ \bibnamefont {Wall}}, \bibinfo {author}
  {\bibfnamefont {K.~R.~A.}\ \bibnamefont {Hazzard}}, \bibinfo {author}
  {\bibfnamefont {B.}~\bibnamefont {Yan}}, \bibinfo {author} {\bibfnamefont
  {S.~A.}\ \bibnamefont {Moses}}, \bibinfo {author} {\bibfnamefont {J.~P.}\
  \bibnamefont {Covey}}, \bibinfo {author} {\bibfnamefont {D.~S.}\ \bibnamefont
  {Jin}}, \bibinfo {author} {\bibfnamefont {J.}~\bibnamefont {Ye}}, \bibinfo
  {author} {\bibfnamefont {M.}~\bibnamefont {Holland}},\ and\ \bibinfo {author}
  {\bibfnamefont {A.~M.}\ \bibnamefont {Rey}},\ }\bibfield  {title} {\bibinfo
  {title} {\emph{Suppressing the Loss of Ultracold Molecules Via the Continuous
  Quantum Zeno Effect}},\ }\href
  {https://doi.org/10.1103/PhysRevLett.112.070404} {\bibfield  {journal}
  {\bibinfo  {journal} {Phys. Rev. Lett.}\ }\textbf {\bibinfo {volume} {112}},\
  \bibinfo {pages} {070404} (\bibinfo {year} {2014})}\BibitemShut {NoStop}%
\bibitem [{\citenamefont {Sponselee}\ \emph {et~al.}(2018)\citenamefont
  {Sponselee}, \citenamefont {Freystatzky}, \citenamefont {Abeln},
  \citenamefont {Diem}, \citenamefont {Hundt}, \citenamefont {Kochanke},
  \citenamefont {Ponath}, \citenamefont {Santra}, \citenamefont {Mathey},
  \citenamefont {Sengstock} \emph {et~al.}}]{lossexpF3}%
  \BibitemOpen
  \bibfield  {author} {\bibinfo {author} {\bibfnamefont {K.}~\bibnamefont
  {Sponselee}}, \bibinfo {author} {\bibfnamefont {L.}~\bibnamefont
  {Freystatzky}}, \bibinfo {author} {\bibfnamefont {B.}~\bibnamefont {Abeln}},
  \bibinfo {author} {\bibfnamefont {M.}~\bibnamefont {Diem}}, \bibinfo {author}
  {\bibfnamefont {B.}~\bibnamefont {Hundt}}, \bibinfo {author} {\bibfnamefont
  {A.}~\bibnamefont {Kochanke}}, \bibinfo {author} {\bibfnamefont
  {T.}~\bibnamefont {Ponath}}, \bibinfo {author} {\bibfnamefont
  {B.}~\bibnamefont {Santra}}, \bibinfo {author} {\bibfnamefont
  {L.}~\bibnamefont {Mathey}}, \bibinfo {author} {\bibfnamefont
  {K.}~\bibnamefont {Sengstock}}, \emph {et~al.},\ }\bibfield  {title}
  {\bibinfo {title} {\emph{Dynamics of ultracold quantum gases in the
  dissipative Fermi--Hubbard model}},\ }\href
  {https://doi.org/10.1088/2058-9565/aadccd} {\bibfield  {journal} {\bibinfo
  {journal} {Quantum Sci. Technol.}\ }\textbf {\bibinfo {volume} {4}},\
  \bibinfo {pages} {014002} (\bibinfo {year} {2018})}\BibitemShut {NoStop}%
\bibitem [{\citenamefont {Kinoshita}\ \emph {et~al.}(2005)\citenamefont
  {Kinoshita}, \citenamefont {Wenger},\ and\ \citenamefont {Weiss}}]{lossexp5}%
  \BibitemOpen
  \bibfield  {author} {\bibinfo {author} {\bibfnamefont {T.}~\bibnamefont
  {Kinoshita}}, \bibinfo {author} {\bibfnamefont {T.}~\bibnamefont {Wenger}},\
  and\ \bibinfo {author} {\bibfnamefont {D.~S.}\ \bibnamefont {Weiss}},\
  }\bibfield  {title} {\bibinfo {title} {\emph{Local Pair Correlations in
  One-Dimensional Bose Gases}},\ }\href
  {https://doi.org/10.1103/PhysRevLett.95.190406} {\bibfield  {journal}
  {\bibinfo  {journal} {Phys. Rev. Lett.}\ }\textbf {\bibinfo {volume} {95}},\
  \bibinfo {pages} {190406} (\bibinfo {year} {2005})}\BibitemShut {NoStop}%
\bibitem [{\citenamefont {Tomita}\ \emph {et~al.}(2017)\citenamefont {Tomita},
  \citenamefont {Nakajima}, \citenamefont {Danshita}, \citenamefont {Takasu},\
  and\ \citenamefont {Takahashi}}]{lossexp5bis}%
  \BibitemOpen
  \bibfield  {author} {\bibinfo {author} {\bibfnamefont {T.}~\bibnamefont
  {Tomita}}, \bibinfo {author} {\bibfnamefont {S.}~\bibnamefont {Nakajima}},
  \bibinfo {author} {\bibfnamefont {I.}~\bibnamefont {Danshita}}, \bibinfo
  {author} {\bibfnamefont {Y.}~\bibnamefont {Takasu}},\ and\ \bibinfo {author}
  {\bibfnamefont {Y.}~\bibnamefont {Takahashi}},\ }\bibfield  {title} {\bibinfo
  {title} {\emph{Observation of the Mott insulator to superfluid crossover of a
  driven-dissipative Bose-Hubbard system}},\ }\href {https://doi.org/DOI:
  10.1126/sciadv.1701513} {\bibfield  {journal} {\bibinfo  {journal} {Sci.
  Adv.}\ }\textbf {\bibinfo {volume} {3}},\ \bibinfo {pages} {e1701513}
  (\bibinfo {year} {2017})}\BibitemShut {NoStop}%
\bibitem [{\citenamefont {Bouchoule}\ and\ \citenamefont
  {Schemmer}(2020)}]{lossexp2}%
  \BibitemOpen
  \bibfield  {author} {\bibinfo {author} {\bibfnamefont {I.}~\bibnamefont
  {Bouchoule}}\ and\ \bibinfo {author} {\bibfnamefont {M.}~\bibnamefont
  {Schemmer}},\ }\bibfield  {title} {\bibinfo {title} {\emph{Asymptotic
  temperature of a lossy condensate}},\ }\href
  {https://doi.org/10.21468/SciPostPhys.8.4.060} {\bibfield  {journal}
  {\bibinfo  {journal} {SciPost Phys.}\ }\textbf {\bibinfo {volume} {8}},\
  \bibinfo {pages} {60} (\bibinfo {year} {2020})}\BibitemShut {NoStop}%
\bibitem [{\citenamefont {S{\"o}ding}\ \emph {et~al.}(1999)\citenamefont
  {S{\"o}ding}, \citenamefont {Gu{\'e}ry-Odelin}, \citenamefont {Desbiolles},
  \citenamefont {Chevy}, \citenamefont {Inamori},\ and\ \citenamefont
  {Dalibard}}]{lossexp6}%
  \BibitemOpen
  \bibfield  {author} {\bibinfo {author} {\bibfnamefont {J.}~\bibnamefont
  {S{\"o}ding}}, \bibinfo {author} {\bibfnamefont {D.}~\bibnamefont
  {Gu{\'e}ry-Odelin}}, \bibinfo {author} {\bibfnamefont {P.}~\bibnamefont
  {Desbiolles}}, \bibinfo {author} {\bibfnamefont {F.}~\bibnamefont {Chevy}},
  \bibinfo {author} {\bibfnamefont {H.}~\bibnamefont {Inamori}},\ and\ \bibinfo
  {author} {\bibfnamefont {J.}~\bibnamefont {Dalibard}},\ }\bibfield  {title}
  {\bibinfo {title} {\emph{Three-body decay of a rubidium Bose--Einstein
  condensate}},\ }\href {https://doi.org/https://doi.org/10.1007/s003400050805}
  {\bibfield  {journal} {\bibinfo  {journal} {Appl. Phys. B}\ }\textbf
  {\bibinfo {volume} {69}},\ \bibinfo {pages} {257} (\bibinfo {year}
  {1999})}\BibitemShut {NoStop}%
\bibitem [{\citenamefont {Tolra}\ \emph {et~al.}(2004)\citenamefont {Tolra},
  \citenamefont {O'Hara}, \citenamefont {Huckans}, \citenamefont {Phillips},
  \citenamefont {Rolston},\ and\ \citenamefont {Porto}}]{lossexp7}%
  \BibitemOpen
  \bibfield  {author} {\bibinfo {author} {\bibfnamefont {B.~L.}\ \bibnamefont
  {Tolra}}, \bibinfo {author} {\bibfnamefont {K.~M.}\ \bibnamefont {O'Hara}},
  \bibinfo {author} {\bibfnamefont {J.~H.}\ \bibnamefont {Huckans}}, \bibinfo
  {author} {\bibfnamefont {W.~D.}\ \bibnamefont {Phillips}}, \bibinfo {author}
  {\bibfnamefont {S.~L.}\ \bibnamefont {Rolston}},\ and\ \bibinfo {author}
  {\bibfnamefont {J.~V.}\ \bibnamefont {Porto}},\ }\bibfield  {title} {\bibinfo
  {title} {\emph{Observation of Reduced Three-Body Recombination in a
  Correlated 1D Degenerate Bose Gas}},\ }\href
  {https://doi.org/10.1103/PhysRevLett.92.190401} {\bibfield  {journal}
  {\bibinfo  {journal} {Phys. Rev. Lett.}\ }\textbf {\bibinfo {volume} {92}},\
  \bibinfo {pages} {190401} (\bibinfo {year} {2004})}\BibitemShut {NoStop}%
\bibitem [{\citenamefont {Ferlaino}\ \emph {et~al.}(2009)\citenamefont
  {Ferlaino}, \citenamefont {Knoop}, \citenamefont {Berninger}, \citenamefont
  {Harm}, \citenamefont {D'Incao}, \citenamefont {N\"agerl},\ and\
  \citenamefont {Grimm}}]{lossexp8}%
  \BibitemOpen
  \bibfield  {author} {\bibinfo {author} {\bibfnamefont {F.}~\bibnamefont
  {Ferlaino}}, \bibinfo {author} {\bibfnamefont {S.}~\bibnamefont {Knoop}},
  \bibinfo {author} {\bibfnamefont {M.}~\bibnamefont {Berninger}}, \bibinfo
  {author} {\bibfnamefont {W.}~\bibnamefont {Harm}}, \bibinfo {author}
  {\bibfnamefont {J.~P.}\ \bibnamefont {D'Incao}}, \bibinfo {author}
  {\bibfnamefont {H.-C.}\ \bibnamefont {N\"agerl}},\ and\ \bibinfo {author}
  {\bibfnamefont {R.}~\bibnamefont {Grimm}},\ }\bibfield  {title} {\bibinfo
  {title} {\emph{Evidence for Universal Four-Body States Tied to an Efimov
  Trimer}},\ }\href {https://doi.org/10.1103/PhysRevLett.102.140401} {\bibfield
   {journal} {\bibinfo  {journal} {Phys. Rev. Lett.}\ }\textbf {\bibinfo
  {volume} {102}},\ \bibinfo {pages} {140401} (\bibinfo {year}
  {2009})}\BibitemShut {NoStop}%
\bibitem [{\citenamefont {Gurian}\ \emph {et~al.}(2012)\citenamefont {Gurian},
  \citenamefont {Cheinet}, \citenamefont {Huillery}, \citenamefont {Fioretti},
  \citenamefont {Zhao}, \citenamefont {Gould}, \citenamefont {Comparat},\ and\
  \citenamefont {Pillet}}]{lossexp9}%
  \BibitemOpen
  \bibfield  {author} {\bibinfo {author} {\bibfnamefont {J.~H.}\ \bibnamefont
  {Gurian}}, \bibinfo {author} {\bibfnamefont {P.}~\bibnamefont {Cheinet}},
  \bibinfo {author} {\bibfnamefont {P.}~\bibnamefont {Huillery}}, \bibinfo
  {author} {\bibfnamefont {A.}~\bibnamefont {Fioretti}}, \bibinfo {author}
  {\bibfnamefont {J.}~\bibnamefont {Zhao}}, \bibinfo {author} {\bibfnamefont
  {P.~L.}\ \bibnamefont {Gould}}, \bibinfo {author} {\bibfnamefont
  {D.}~\bibnamefont {Comparat}},\ and\ \bibinfo {author} {\bibfnamefont
  {P.}~\bibnamefont {Pillet}},\ }\bibfield  {title} {\bibinfo {title}
  {\emph{Observation of a Resonant Four-Body Interaction in Cold Cesium Rydberg
  Atoms}},\ }\href {https://doi.org/10.1103/PhysRevLett.108.023005} {\bibfield
  {journal} {\bibinfo  {journal} {Phys. Rev. Lett.}\ }\textbf {\bibinfo
  {volume} {108}},\ \bibinfo {pages} {023005} (\bibinfo {year}
  {2012})}\BibitemShut {NoStop}%
\bibitem [{\citenamefont {Diehl}\ \emph {et~al.}(2008)\citenamefont {Diehl},
  \citenamefont {Micheli}, \citenamefont {Kantian}, \citenamefont {Kraus},
  \citenamefont {B{\"u}chler},\ and\ \citenamefont
  {Zoller}}]{diehl2008quantum}%
  \BibitemOpen
  \bibfield  {author} {\bibinfo {author} {\bibfnamefont {S.}~\bibnamefont
  {Diehl}}, \bibinfo {author} {\bibfnamefont {A.}~\bibnamefont {Micheli}},
  \bibinfo {author} {\bibfnamefont {A.}~\bibnamefont {Kantian}}, \bibinfo
  {author} {\bibfnamefont {B.}~\bibnamefont {Kraus}}, \bibinfo {author}
  {\bibfnamefont {H.}~\bibnamefont {B{\"u}chler}},\ and\ \bibinfo {author}
  {\bibfnamefont {P.}~\bibnamefont {Zoller}},\ }\bibfield  {title} {\bibinfo
  {title} {\emph{Quantum states and phases in driven open quantum systems with
  cold atoms}},\ }\href {https://doi.org/https://doi.org/10.1038/nphys1073}
  {\bibfield  {journal} {\bibinfo  {journal} {Nat. Phys.}\ }\textbf {\bibinfo
  {volume} {4}},\ \bibinfo {pages} {878} (\bibinfo {year} {2008})}\BibitemShut
  {NoStop}%
\bibitem [{\citenamefont {Bouchoule}\ and\ \citenamefont
  {Dubail}(2021)}]{lossth4}%
  \BibitemOpen
  \bibfield  {author} {\bibinfo {author} {\bibfnamefont {I.}~\bibnamefont
  {Bouchoule}}\ and\ \bibinfo {author} {\bibfnamefont {J.}~\bibnamefont
  {Dubail}},\ }\bibfield  {title} {\bibinfo {title} {\emph{Breakdown of Tan's
  Relation in Lossy One-Dimensional Bose Gases}},\ }\href
  {https://doi.org/10.1103/PhysRevLett.126.160603} {\bibfield  {journal}
  {\bibinfo  {journal} {Phys. Rev. Lett.}\ }\textbf {\bibinfo {volume} {126}},\
  \bibinfo {pages} {160603} (\bibinfo {year} {2021})}\BibitemShut {NoStop}%
\bibitem [{\citenamefont {Bouchoule}\ and\ \citenamefont
  {Dubail}(2022)}]{lossth5}%
  \BibitemOpen
  \bibfield  {author} {\bibinfo {author} {\bibfnamefont {I.}~\bibnamefont
  {Bouchoule}}\ and\ \bibinfo {author} {\bibfnamefont {J.}~\bibnamefont
  {Dubail}},\ }\bibfield  {title} {\bibinfo {title} {\emph{Generalized
  hydrodynamics in the one-dimensional Bose gas: theory and experiments}},\
  }\href {https://doi.org/https://doi.org/10.1088/1742-5468/ac3659} {\bibfield
  {journal} {\bibinfo  {journal} {J. Stat. Mech.: Theory Exp.}\ }\textbf
  {\bibinfo {volume} {2022}}\bibinfo  {number} { (1)},\ \bibinfo {pages}
  {014003}}\BibitemShut {NoStop}%
\bibitem [{\citenamefont {Garc{\'\i}a-Ripoll}\ \emph
  {et~al.}(2009)\citenamefont {Garc{\'\i}a-Ripoll}, \citenamefont {D{\"u}rr},
  \citenamefont {Syassen}, \citenamefont {Bauer}, \citenamefont {Lettner},
  \citenamefont {Rempe},\ and\ \citenamefont {Cirac}}]{lossth7}%
  \BibitemOpen
\bibfield  {number} {  }\bibfield  {author} {\bibinfo {author} {\bibfnamefont
  {J.~J.}\ \bibnamefont {Garc{\'\i}a-Ripoll}}, \bibinfo {author} {\bibfnamefont
  {S.}~\bibnamefont {D{\"u}rr}}, \bibinfo {author} {\bibfnamefont
  {N.}~\bibnamefont {Syassen}}, \bibinfo {author} {\bibfnamefont {D.~M.}\
  \bibnamefont {Bauer}}, \bibinfo {author} {\bibfnamefont {M.}~\bibnamefont
  {Lettner}}, \bibinfo {author} {\bibfnamefont {G.}~\bibnamefont {Rempe}},\
  and\ \bibinfo {author} {\bibfnamefont {J.~I.}\ \bibnamefont {Cirac}},\
  }\bibfield  {title} {\bibinfo {title} {\emph{Dissipation-induced hard-core
  boson gas in an optical lattice}},\ }\href
  {https://doi.org/https://doi.org/10.1088/1367-2630/11/1/013053} {\bibfield
  {journal} {\bibinfo  {journal} {New J. Phys.}\ }\textbf {\bibinfo {volume}
  {11}},\ \bibinfo {pages} {013053} (\bibinfo {year} {2009})}\BibitemShut
  {NoStop}%
\bibitem [{\citenamefont {Ates}\ \emph {et~al.}(2012)\citenamefont {Ates},
  \citenamefont {Olmos}, \citenamefont {Li},\ and\ \citenamefont
  {Lesanovsky}}]{lossth8}%
  \BibitemOpen
  \bibfield  {author} {\bibinfo {author} {\bibfnamefont {C.}~\bibnamefont
  {Ates}}, \bibinfo {author} {\bibfnamefont {B.}~\bibnamefont {Olmos}},
  \bibinfo {author} {\bibfnamefont {W.}~\bibnamefont {Li}},\ and\ \bibinfo
  {author} {\bibfnamefont {I.}~\bibnamefont {Lesanovsky}},\ }\bibfield  {title}
  {\bibinfo {title} {\emph{Dissipative Binding of Lattice Bosons through
  Distance-Selective Pair Loss}},\ }\href
  {https://doi.org/10.1103/PhysRevLett.109.233003} {\bibfield  {journal}
  {\bibinfo  {journal} {Phys. Rev. Lett.}\ }\textbf {\bibinfo {volume} {109}},\
  \bibinfo {pages} {233003} (\bibinfo {year} {2012})}\BibitemShut {NoStop}%
\bibitem [{\citenamefont {Everest}\ \emph {et~al.}(2014)\citenamefont
  {Everest}, \citenamefont {Hush},\ and\ \citenamefont
  {Lesanovsky}}]{lossth8bis}%
  \BibitemOpen
  \bibfield  {author} {\bibinfo {author} {\bibfnamefont {B.}~\bibnamefont
  {Everest}}, \bibinfo {author} {\bibfnamefont {M.~R.}\ \bibnamefont {Hush}},\
  and\ \bibinfo {author} {\bibfnamefont {I.}~\bibnamefont {Lesanovsky}},\
  }\bibfield  {title} {\bibinfo {title} {\emph{Many-body out-of-equilibrium
  dynamics of hard-core lattice bosons with nonlocal loss}},\ }\href
  {https://doi.org/10.1103/PhysRevB.90.134306} {\bibfield  {journal} {\bibinfo
  {journal} {Phys. Rev. B}\ }\textbf {\bibinfo {volume} {90}},\ \bibinfo
  {pages} {134306} (\bibinfo {year} {2014})}\BibitemShut {NoStop}%
\bibitem [{\citenamefont {Bouchoule}\ \emph {et~al.}(2021)\citenamefont
  {Bouchoule}, \citenamefont {Dubois},\ and\ \citenamefont
  {Barbier}}]{lossth9}%
  \BibitemOpen
  \bibfield  {author} {\bibinfo {author} {\bibfnamefont {I.}~\bibnamefont
  {Bouchoule}}, \bibinfo {author} {\bibfnamefont {L.}~\bibnamefont {Dubois}},\
  and\ \bibinfo {author} {\bibfnamefont {L.-P.}\ \bibnamefont {Barbier}},\
  }\bibfield  {title} {\bibinfo {title} {\emph{Losses in interacting quantum
  gases: Ultraviolet divergence and its regularization}},\ }\href
  {https://doi.org/10.1103/PhysRevA.104.L031304} {\bibfield  {journal}
  {\bibinfo  {journal} {Phys. Rev. A}\ }\textbf {\bibinfo {volume} {104}},\
  \bibinfo {pages} {L031304} (\bibinfo {year} {2021})}\BibitemShut {NoStop}%
\bibitem [{\citenamefont {Rossini}\ \emph {et~al.}(2021)\citenamefont
  {Rossini}, \citenamefont {Ghermaoui}, \citenamefont {Aguilera}, \citenamefont
  {Vatr\'e}, \citenamefont {Bouganne}, \citenamefont {Beugnon}, \citenamefont
  {Gerbier},\ and\ \citenamefont {Mazza}}]{lossth1}%
  \BibitemOpen
  \bibfield  {author} {\bibinfo {author} {\bibfnamefont {D.}~\bibnamefont
  {Rossini}}, \bibinfo {author} {\bibfnamefont {A.}~\bibnamefont {Ghermaoui}},
  \bibinfo {author} {\bibfnamefont {M.~B.}\ \bibnamefont {Aguilera}}, \bibinfo
  {author} {\bibfnamefont {R.}~\bibnamefont {Vatr\'e}}, \bibinfo {author}
  {\bibfnamefont {R.}~\bibnamefont {Bouganne}}, \bibinfo {author}
  {\bibfnamefont {J.}~\bibnamefont {Beugnon}}, \bibinfo {author} {\bibfnamefont
  {F.}~\bibnamefont {Gerbier}},\ and\ \bibinfo {author} {\bibfnamefont
  {L.}~\bibnamefont {Mazza}},\ }\bibfield  {title} {\bibinfo {title}
  {\emph{Strong correlations in lossy one-dimensional quantum gases: From the
  quantum Zeno effect to the generalized Gibbs ensemble}},\ }\href
  {https://doi.org/10.1103/PhysRevA.103.L060201} {\bibfield  {journal}
  {\bibinfo  {journal} {Phys. Rev. A}\ }\textbf {\bibinfo {volume} {103}},\
  \bibinfo {pages} {L060201} (\bibinfo {year} {2021})}\BibitemShut {NoStop}%
\bibitem [{\citenamefont {Rosso}\ \emph {et~al.}(2022)\citenamefont {Rosso},
  \citenamefont {Biella},\ and\ \citenamefont {Mazza}}]{lossth2}%
  \BibitemOpen
  \bibfield  {author} {\bibinfo {author} {\bibfnamefont {L.}~\bibnamefont
  {Rosso}}, \bibinfo {author} {\bibfnamefont {A.}~\bibnamefont {Biella}},\ and\
  \bibinfo {author} {\bibfnamefont {L.}~\bibnamefont {Mazza}},\ }\bibfield
  {title} {\bibinfo {title} {\emph{The one-dimensional Bose gas with strong
  two-body losses: the effect of the harmonic confinement}},\ }\href
  {https://doi.org/10.21468/SciPostPhys.12.1.044} {\bibfield  {journal}
  {\bibinfo  {journal} {SciPost Phys.}\ }\textbf {\bibinfo {volume} {12}},\
  \bibinfo {pages} {44} (\bibinfo {year} {2022})}\BibitemShut {NoStop}%
\bibitem [{\citenamefont {Rosso}\ \emph {et~al.}(2023)\citenamefont {Rosso},
  \citenamefont {Biella}, \citenamefont {De~Nardis},\ and\ \citenamefont
  {Mazza}}]{lossth6}%
  \BibitemOpen
  \bibfield  {author} {\bibinfo {author} {\bibfnamefont {L.}~\bibnamefont
  {Rosso}}, \bibinfo {author} {\bibfnamefont {A.}~\bibnamefont {Biella}},
  \bibinfo {author} {\bibfnamefont {J.}~\bibnamefont {De~Nardis}},\ and\
  \bibinfo {author} {\bibfnamefont {L.}~\bibnamefont {Mazza}},\ }\bibfield
  {title} {\bibinfo {title} {\emph{Dynamical theory for one-dimensional
  fermions with strong two-body losses: Universal non-Hermitian Zeno physics
  and spin-charge separation}},\ }\href
  {https://doi.org/10.1103/PhysRevA.107.013303} {\bibfield  {journal} {\bibinfo
   {journal} {Phys. Rev. A}\ }\textbf {\bibinfo {volume} {107}},\ \bibinfo
  {pages} {013303} (\bibinfo {year} {2023})}\BibitemShut {NoStop}%
\bibitem [{\citenamefont {Bouchoule}\ \emph {et~al.}(2020)\citenamefont
  {Bouchoule}, \citenamefont {Doyon},\ and\ \citenamefont {Dubail}}]{lossth3}%
  \BibitemOpen
  \bibfield  {author} {\bibinfo {author} {\bibfnamefont {I.}~\bibnamefont
  {Bouchoule}}, \bibinfo {author} {\bibfnamefont {B.}~\bibnamefont {Doyon}},\
  and\ \bibinfo {author} {\bibfnamefont {J.}~\bibnamefont {Dubail}},\
  }\bibfield  {title} {\bibinfo {title} {\emph{The effect of atom losses on the
  distribution of rapidities in the one-dimensional Bose gas}},\ }\href
  {https://doi.org/10.21468/SciPostPhys.9.4.044} {\bibfield  {journal}
  {\bibinfo  {journal} {SciPost Phys.}\ }\textbf {\bibinfo {volume} {9}},\
  \bibinfo {pages} {44} (\bibinfo {year} {2020})}\BibitemShut {NoStop}%
\bibitem [{\citenamefont {Lesanovsky}\ and\ \citenamefont
  {Garrahan}(2013)}]{lesanovsky2013}%
  \BibitemOpen
  \bibfield  {author} {\bibinfo {author} {\bibfnamefont {I.}~\bibnamefont
  {Lesanovsky}}\ and\ \bibinfo {author} {\bibfnamefont {J.~P.}\ \bibnamefont
  {Garrahan}},\ }\bibfield  {title} {\bibinfo {title} {\emph{Kinetic
  Constraints, Hierarchical Relaxation, and Onset of Glassiness in Strongly
  Interacting and Dissipative Rydberg Gases}},\ }\href
  {https://doi.org/10.1103/PhysRevLett.111.215305} {\bibfield  {journal}
  {\bibinfo  {journal} {Phys. Rev. Lett.}\ }\textbf {\bibinfo {volume} {111}},\
  \bibinfo {pages} {215305} (\bibinfo {year} {2013})}\BibitemShut {NoStop}%
\bibitem [{\citenamefont {Olmos}\ \emph {et~al.}(2014)\citenamefont {Olmos},
  \citenamefont {Lesanovsky},\ and\ \citenamefont {Garrahan}}]{olmos2014}%
  \BibitemOpen
  \bibfield  {author} {\bibinfo {author} {\bibfnamefont {B.}~\bibnamefont
  {Olmos}}, \bibinfo {author} {\bibfnamefont {I.}~\bibnamefont {Lesanovsky}},\
  and\ \bibinfo {author} {\bibfnamefont {J.~P.}\ \bibnamefont {Garrahan}},\
  }\bibfield  {title} {\bibinfo {title} {\emph{Out-of-equilibrium evolution of
  kinetically constrained many-body quantum systems under purely dissipative
  dynamics}},\ }\href {https://doi.org/10.1103/PhysRevE.90.042147} {\bibfield
  {journal} {\bibinfo  {journal} {Phys. Rev. E}\ }\textbf {\bibinfo {volume}
  {90}},\ \bibinfo {pages} {042147} (\bibinfo {year} {2014})}\BibitemShut
  {NoStop}%
\bibitem [{\citenamefont {Everest}\ \emph {et~al.}(2016)\citenamefont
  {Everest}, \citenamefont {Marcuzzi}, \citenamefont {Garrahan},\ and\
  \citenamefont {Lesanovsky}}]{everest2016}%
  \BibitemOpen
  \bibfield  {author} {\bibinfo {author} {\bibfnamefont {B.}~\bibnamefont
  {Everest}}, \bibinfo {author} {\bibfnamefont {M.}~\bibnamefont {Marcuzzi}},
  \bibinfo {author} {\bibfnamefont {J.~P.}\ \bibnamefont {Garrahan}},\ and\
  \bibinfo {author} {\bibfnamefont {I.}~\bibnamefont {Lesanovsky}},\ }\bibfield
   {title} {\bibinfo {title} {\emph{Emergent kinetic constraints, ergodicity
  breaking, and cooperative dynamics in noisy quantum systems}},\ }\href
  {https://doi.org/10.1103/PhysRevE.94.052108} {\bibfield  {journal} {\bibinfo
  {journal} {Phys. Rev. E}\ }\textbf {\bibinfo {volume} {94}},\ \bibinfo
  {pages} {052108} (\bibinfo {year} {2016})}\BibitemShut {NoStop}%
\bibitem [{\citenamefont {Marcuzzi}\ \emph {et~al.}(2016)\citenamefont
  {Marcuzzi}, \citenamefont {Buchhold}, \citenamefont {Diehl},\ and\
  \citenamefont {Lesanovsky}}]{marcuzzi2016}%
  \BibitemOpen
  \bibfield  {author} {\bibinfo {author} {\bibfnamefont {M.}~\bibnamefont
  {Marcuzzi}}, \bibinfo {author} {\bibfnamefont {M.}~\bibnamefont {Buchhold}},
  \bibinfo {author} {\bibfnamefont {S.}~\bibnamefont {Diehl}},\ and\ \bibinfo
  {author} {\bibfnamefont {I.}~\bibnamefont {Lesanovsky}},\ }\bibfield  {title}
  {\bibinfo {title} {\emph{Absorbing State Phase Transition with Competing
  Quantum and Classical Fluctuations}},\ }\href
  {https://doi.org/10.1103/PhysRevLett.116.245701} {\bibfield  {journal}
  {\bibinfo  {journal} {Phys. Rev. Lett.}\ }\textbf {\bibinfo {volume} {116}},\
  \bibinfo {pages} {245701} (\bibinfo {year} {2016})}\BibitemShut {NoStop}%
\bibitem [{\citenamefont {Buchhold}\ \emph {et~al.}(2017)\citenamefont
  {Buchhold}, \citenamefont {Everest}, \citenamefont {Marcuzzi}, \citenamefont
  {Lesanovsky},\ and\ \citenamefont {Diehl}}]{buchhold2017}%
  \BibitemOpen
  \bibfield  {author} {\bibinfo {author} {\bibfnamefont {M.}~\bibnamefont
  {Buchhold}}, \bibinfo {author} {\bibfnamefont {B.}~\bibnamefont {Everest}},
  \bibinfo {author} {\bibfnamefont {M.}~\bibnamefont {Marcuzzi}}, \bibinfo
  {author} {\bibfnamefont {I.}~\bibnamefont {Lesanovsky}},\ and\ \bibinfo
  {author} {\bibfnamefont {S.}~\bibnamefont {Diehl}},\ }\bibfield  {title}
  {\bibinfo {title} {\emph{Nonequilibrium effective field theory for absorbing
  state phase transitions in driven open quantum spin systems}},\ }\href
  {https://doi.org/10.1103/PhysRevB.95.014308} {\bibfield  {journal} {\bibinfo
  {journal} {Phys. Rev. B}\ }\textbf {\bibinfo {volume} {95}},\ \bibinfo
  {pages} {014308} (\bibinfo {year} {2017})}\BibitemShut {NoStop}%
\bibitem [{\citenamefont {Guti\'errez}\ \emph {et~al.}(2017)\citenamefont
  {Guti\'errez}, \citenamefont {Simonelli}, \citenamefont {Archimi},
  \citenamefont {Castellucci}, \citenamefont {Arimondo}, \citenamefont
  {Ciampini}, \citenamefont {Marcuzzi}, \citenamefont {Lesanovsky},\ and\
  \citenamefont {Morsch}}]{gutierrez2017}%
  \BibitemOpen
  \bibfield  {author} {\bibinfo {author} {\bibfnamefont {R.}~\bibnamefont
  {Guti\'errez}}, \bibinfo {author} {\bibfnamefont {C.}~\bibnamefont
  {Simonelli}}, \bibinfo {author} {\bibfnamefont {M.}~\bibnamefont {Archimi}},
  \bibinfo {author} {\bibfnamefont {F.}~\bibnamefont {Castellucci}}, \bibinfo
  {author} {\bibfnamefont {E.}~\bibnamefont {Arimondo}}, \bibinfo {author}
  {\bibfnamefont {D.}~\bibnamefont {Ciampini}}, \bibinfo {author}
  {\bibfnamefont {M.}~\bibnamefont {Marcuzzi}}, \bibinfo {author}
  {\bibfnamefont {I.}~\bibnamefont {Lesanovsky}},\ and\ \bibinfo {author}
  {\bibfnamefont {O.}~\bibnamefont {Morsch}},\ }\bibfield  {title} {\bibinfo
  {title} {\emph{Experimental signatures of an absorbing-state phase transition
  in an open driven many-body quantum system}},\ }\href
  {https://doi.org/10.1103/PhysRevA.96.041602} {\bibfield  {journal} {\bibinfo
  {journal} {Phys. Rev. A}\ }\textbf {\bibinfo {volume} {96}},\ \bibinfo
  {pages} {041602} (\bibinfo {year} {2017})}\BibitemShut {NoStop}%
\bibitem [{\citenamefont {Roscher}\ \emph {et~al.}(2018)\citenamefont
  {Roscher}, \citenamefont {Diehl},\ and\ \citenamefont
  {Buchhold}}]{roscher2018}%
  \BibitemOpen
  \bibfield  {author} {\bibinfo {author} {\bibfnamefont {D.}~\bibnamefont
  {Roscher}}, \bibinfo {author} {\bibfnamefont {S.}~\bibnamefont {Diehl}},\
  and\ \bibinfo {author} {\bibfnamefont {M.}~\bibnamefont {Buchhold}},\
  }\bibfield  {title} {\bibinfo {title} {\emph{Phenomenology of first-order
  dark-state phase transitions}},\ }\href
  {https://doi.org/10.1103/PhysRevA.98.062117} {\bibfield  {journal} {\bibinfo
  {journal} {Phys. Rev. A}\ }\textbf {\bibinfo {volume} {98}},\ \bibinfo
  {pages} {062117} (\bibinfo {year} {2018})}\BibitemShut {NoStop}%
\bibitem [{\citenamefont {Wintermantel}\ \emph {et~al.}(2020)\citenamefont
  {Wintermantel}, \citenamefont {Wang}, \citenamefont {Lochead}, \citenamefont
  {Shevate}, \citenamefont {Brennen},\ and\ \citenamefont
  {Whitlock}}]{wintermantel2020}%
  \BibitemOpen
  \bibfield  {author} {\bibinfo {author} {\bibfnamefont {T.~M.}\ \bibnamefont
  {Wintermantel}}, \bibinfo {author} {\bibfnamefont {Y.}~\bibnamefont {Wang}},
  \bibinfo {author} {\bibfnamefont {G.}~\bibnamefont {Lochead}}, \bibinfo
  {author} {\bibfnamefont {S.}~\bibnamefont {Shevate}}, \bibinfo {author}
  {\bibfnamefont {G.~K.}\ \bibnamefont {Brennen}},\ and\ \bibinfo {author}
  {\bibfnamefont {S.}~\bibnamefont {Whitlock}},\ }\bibfield  {title} {\bibinfo
  {title} {\emph{Unitary and Nonunitary Quantum Cellular Automata with Rydberg
  Arrays}},\ }\href {https://doi.org/10.1103/PhysRevLett.124.070503} {\bibfield
   {journal} {\bibinfo  {journal} {Phys. Rev. Lett.}\ }\textbf {\bibinfo
  {volume} {124}},\ \bibinfo {pages} {070503} (\bibinfo {year}
  {2020})}\BibitemShut {NoStop}%
\bibitem [{\citenamefont {Helmrich}\ \emph {et~al.}(2020)\citenamefont
  {Helmrich}, \citenamefont {Arias}, \citenamefont {Lochead}, \citenamefont
  {Wintermantel}, \citenamefont {Buchhold}, \citenamefont {Diehl},\ and\
  \citenamefont {Whitlock}}]{helmrich2020}%
  \BibitemOpen
  \bibfield  {author} {\bibinfo {author} {\bibfnamefont {S.}~\bibnamefont
  {Helmrich}}, \bibinfo {author} {\bibfnamefont {A.}~\bibnamefont {Arias}},
  \bibinfo {author} {\bibfnamefont {G.}~\bibnamefont {Lochead}}, \bibinfo
  {author} {\bibfnamefont {T.~M.}\ \bibnamefont {Wintermantel}}, \bibinfo
  {author} {\bibfnamefont {M.}~\bibnamefont {Buchhold}}, \bibinfo {author}
  {\bibfnamefont {S.}~\bibnamefont {Diehl}},\ and\ \bibinfo {author}
  {\bibfnamefont {S.}~\bibnamefont {Whitlock}},\ }\bibfield  {title} {\bibinfo
  {title} {\emph{Signatures of self-organized criticality in an ultracold
  atomic gas}},\ }\href {https://doi.org/10.1038/s41586-019-1908-6} {\bibfield
  {journal} {\bibinfo  {journal} {Nature}\ }\textbf {\bibinfo {volume} {577}},\
  \bibinfo {pages} {481} (\bibinfo {year} {2020})}\BibitemShut {NoStop}%
\bibitem [{\citenamefont {Nigmatullin}\ \emph {et~al.}(2021)\citenamefont
  {Nigmatullin}, \citenamefont {Wagner},\ and\ \citenamefont
  {Brennen}}]{nigmatullin2021}%
  \BibitemOpen
  \bibfield  {author} {\bibinfo {author} {\bibfnamefont {R.}~\bibnamefont
  {Nigmatullin}}, \bibinfo {author} {\bibfnamefont {E.}~\bibnamefont
  {Wagner}},\ and\ \bibinfo {author} {\bibfnamefont {G.~K.}\ \bibnamefont
  {Brennen}},\ }\bibfield  {title} {\bibinfo {title} {\emph{Directed
  percolation in nonunitary quantum cellular automata}},\ }\href
  {https://doi.org/10.1103/PhysRevResearch.3.043167} {\bibfield  {journal}
  {\bibinfo  {journal} {Phys. Rev. Research}\ }\textbf {\bibinfo {volume}
  {3}},\ \bibinfo {pages} {043167} (\bibinfo {year} {2021})}\BibitemShut
  {NoStop}%
\bibitem [{\citenamefont {Kazemi}\ and\ \citenamefont
  {Weimer}(2021)}]{kazemi2021}%
  \BibitemOpen
  \bibfield  {author} {\bibinfo {author} {\bibfnamefont {J.}~\bibnamefont
  {Kazemi}}\ and\ \bibinfo {author} {\bibfnamefont {H.}~\bibnamefont
  {Weimer}},\ }\bibfield  {title} {\bibinfo {title} {\emph{Genuine Bistability
  in Open Quantum Many-Body Systems}},\ }\href
  {https://arxiv.org/abs/2111.05352} {\bibfield  {journal} {\bibinfo  {journal}
  {arXiv:211.05352}\ } (\bibinfo {year} {2021})}\BibitemShut {NoStop}%
\bibitem [{\citenamefont {Carollo}\ \emph {et~al.}(2022)\citenamefont
  {Carollo}, \citenamefont {Gnann}, \citenamefont {Perfetto},\ and\
  \citenamefont {Lesanovsky}}]{carollo2022quantum}%
  \BibitemOpen
  \bibfield  {author} {\bibinfo {author} {\bibfnamefont {F.}~\bibnamefont
  {Carollo}}, \bibinfo {author} {\bibfnamefont {M.}~\bibnamefont {Gnann}},
  \bibinfo {author} {\bibfnamefont {G.}~\bibnamefont {Perfetto}},\ and\
  \bibinfo {author} {\bibfnamefont {I.}~\bibnamefont {Lesanovsky}},\ }\bibfield
   {title} {\bibinfo {title} {\emph{Signatures of a quantum stabilized
  fluctuating phase and critical dynamics in a kinetically constrained open
  many-body system with two absorbing states}},\ }\href
  {https://doi.org/10.1103/PhysRevB.106.094315} {\bibfield  {journal} {\bibinfo
   {journal} {Phys. Rev. B}\ }\textbf {\bibinfo {volume} {106}},\ \bibinfo
  {pages} {094315} (\bibinfo {year} {2022})}\BibitemShut {NoStop}%
\bibitem [{\citenamefont {van Horssen}\ and\ \citenamefont
  {Garrahan}(2015)}]{RDHorssen}%
  \BibitemOpen
  \bibfield  {author} {\bibinfo {author} {\bibfnamefont {M.}~\bibnamefont {van
  Horssen}}\ and\ \bibinfo {author} {\bibfnamefont {J.~P.}\ \bibnamefont
  {Garrahan}},\ }\bibfield  {title} {\bibinfo {title} {\emph{Open quantum
  reaction-diffusion dynamics: Absorbing states and relaxation}},\ }\href
  {https://doi.org/10.1103/PhysRevE.91.032132} {\bibfield  {journal} {\bibinfo
  {journal} {Phys. Rev. E}\ }\textbf {\bibinfo {volume} {91}},\ \bibinfo
  {pages} {032132} (\bibinfo {year} {2015})}\BibitemShut {NoStop}%
\bibitem [{\citenamefont {Carollo}\ \emph {et~al.}(2019)\citenamefont
  {Carollo}, \citenamefont {Gillman}, \citenamefont {Weimer},\ and\
  \citenamefont {Lesanovsky}}]{carollo2019}%
  \BibitemOpen
  \bibfield  {author} {\bibinfo {author} {\bibfnamefont {F.}~\bibnamefont
  {Carollo}}, \bibinfo {author} {\bibfnamefont {E.}~\bibnamefont {Gillman}},
  \bibinfo {author} {\bibfnamefont {H.}~\bibnamefont {Weimer}},\ and\ \bibinfo
  {author} {\bibfnamefont {I.}~\bibnamefont {Lesanovsky}},\ }\bibfield  {title}
  {\bibinfo {title} {\emph{Critical Behavior of the Quantum Contact Process in
  One Dimension}},\ }\href {https://doi.org/10.1103/PhysRevLett.123.100604}
  {\bibfield  {journal} {\bibinfo  {journal} {Phys. Rev. Lett.}\ }\textbf
  {\bibinfo {volume} {123}},\ \bibinfo {pages} {100604} (\bibinfo {year}
  {2019})}\BibitemShut {NoStop}%
\bibitem [{\citenamefont {Gillman}\ \emph {et~al.}(2019)\citenamefont
  {Gillman}, \citenamefont {Carollo},\ and\ \citenamefont
  {Lesanovsky}}]{gillman2019}%
  \BibitemOpen
  \bibfield  {author} {\bibinfo {author} {\bibfnamefont {E.}~\bibnamefont
  {Gillman}}, \bibinfo {author} {\bibfnamefont {F.}~\bibnamefont {Carollo}},\
  and\ \bibinfo {author} {\bibfnamefont {I.}~\bibnamefont {Lesanovsky}},\
  }\bibfield  {title} {\bibinfo {title} {\emph{Numerical simulation of critical
  dissipative non-equilibrium quantum systems with an absorbing state}},\
  }\href {https://doi.org/10.1088/1367-2630/ab43b0} {\bibfield  {journal}
  {\bibinfo  {journal} {New J. Phys.}\ }\textbf {\bibinfo {volume} {21}},\
  \bibinfo {pages} {093064} (\bibinfo {year} {2019})}\BibitemShut {NoStop}%
\bibitem [{\citenamefont {Gillman}\ \emph {et~al.}(2020)\citenamefont
  {Gillman}, \citenamefont {Carollo},\ and\ \citenamefont
  {Lesanovsky}}]{gillman2020}%
  \BibitemOpen
  \bibfield  {author} {\bibinfo {author} {\bibfnamefont {E.}~\bibnamefont
  {Gillman}}, \bibinfo {author} {\bibfnamefont {F.}~\bibnamefont {Carollo}},\
  and\ \bibinfo {author} {\bibfnamefont {I.}~\bibnamefont {Lesanovsky}},\
  }\bibfield  {title} {\bibinfo {title} {\emph{Nonequilibrium Phase Transitions
  in ($1+1$)-Dimensional Quantum Cellular Automata with Controllable Quantum
  Correlations}},\ }\href {https://doi.org/10.1103/PhysRevLett.125.100403}
  {\bibfield  {journal} {\bibinfo  {journal} {Phys. Rev. Lett.}\ }\textbf
  {\bibinfo {volume} {125}},\ \bibinfo {pages} {100403} (\bibinfo {year}
  {2020})}\BibitemShut {NoStop}%
\bibitem [{\citenamefont {Jo}\ \emph {et~al.}(2021)\citenamefont {Jo},
  \citenamefont {Lee}, \citenamefont {Choi},\ and\ \citenamefont
  {Kahng}}]{jo2021}%
  \BibitemOpen
  \bibfield  {author} {\bibinfo {author} {\bibfnamefont {M.}~\bibnamefont
  {Jo}}, \bibinfo {author} {\bibfnamefont {J.}~\bibnamefont {Lee}}, \bibinfo
  {author} {\bibfnamefont {K.}~\bibnamefont {Choi}},\ and\ \bibinfo {author}
  {\bibfnamefont {B.}~\bibnamefont {Kahng}},\ }\bibfield  {title} {\bibinfo
  {title} {\emph{Absorbing phase transition with a continuously varying
  exponent in a quantum contact process: A neural network approach}},\ }\href
  {https://doi.org/10.1103/PhysRevResearch.3.013238} {\bibfield  {journal}
  {\bibinfo  {journal} {Phys. Rev. Research}\ }\textbf {\bibinfo {volume}
  {3}},\ \bibinfo {pages} {013238} (\bibinfo {year} {2021})}\BibitemShut
  {NoStop}%
\bibitem [{\citenamefont {Zeiher}\ \emph {et~al.}(2016)\citenamefont {Zeiher},
  \citenamefont {Van~Bijnen}, \citenamefont {Schau{\ss}}, \citenamefont {Hild},
  \citenamefont {Choi}, \citenamefont {Pohl}, \citenamefont {Bloch},\ and\
  \citenamefont {Gross}}]{quantum_simulator_1}%
  \BibitemOpen
  \bibfield  {author} {\bibinfo {author} {\bibfnamefont {J.}~\bibnamefont
  {Zeiher}}, \bibinfo {author} {\bibfnamefont {R.}~\bibnamefont {Van~Bijnen}},
  \bibinfo {author} {\bibfnamefont {P.}~\bibnamefont {Schau{\ss}}}, \bibinfo
  {author} {\bibfnamefont {S.}~\bibnamefont {Hild}}, \bibinfo {author}
  {\bibfnamefont {J.-y.}\ \bibnamefont {Choi}}, \bibinfo {author}
  {\bibfnamefont {T.}~\bibnamefont {Pohl}}, \bibinfo {author} {\bibfnamefont
  {I.}~\bibnamefont {Bloch}},\ and\ \bibinfo {author} {\bibfnamefont
  {C.}~\bibnamefont {Gross}},\ }\bibfield  {title} {\bibinfo {title}
  {\emph{Many-body interferometry of a Rydberg-dressed spin lattice}},\ }\href
  {https://doi.org/https://doi.org/10.1038/nphys3835} {\bibfield  {journal}
  {\bibinfo  {journal} {Nat. Phys.}\ }\textbf {\bibinfo {volume} {12}},\
  \bibinfo {pages} {1095} (\bibinfo {year} {2016})}\BibitemShut {NoStop}%
\bibitem [{\citenamefont {Kim}\ \emph {et~al.}(2018)\citenamefont {Kim},
  \citenamefont {Park}, \citenamefont {Kim}, \citenamefont {Sim},\ and\
  \citenamefont {Ahn}}]{quantum_simulator_2}%
  \BibitemOpen
  \bibfield  {author} {\bibinfo {author} {\bibfnamefont {H.}~\bibnamefont
  {Kim}}, \bibinfo {author} {\bibfnamefont {Y.}~\bibnamefont {Park}}, \bibinfo
  {author} {\bibfnamefont {K.}~\bibnamefont {Kim}}, \bibinfo {author}
  {\bibfnamefont {H.-S.}\ \bibnamefont {Sim}},\ and\ \bibinfo {author}
  {\bibfnamefont {J.}~\bibnamefont {Ahn}},\ }\bibfield  {title} {\bibinfo
  {title} {\emph{Detailed Balance of Thermalization Dynamics in Rydberg-Atom
  Quantum Simulators}},\ }\href
  {https://doi.org/10.1103/PhysRevLett.120.180502} {\bibfield  {journal}
  {\bibinfo  {journal} {Phys. Rev. Lett.}\ }\textbf {\bibinfo {volume} {120}},\
  \bibinfo {pages} {180502} (\bibinfo {year} {2018})}\BibitemShut {NoStop}%
\bibitem [{\citenamefont {Ebadi}\ \emph {et~al.}(2021)\citenamefont {Ebadi},
  \citenamefont {Wang}, \citenamefont {Levine}, \citenamefont {Keesling},
  \citenamefont {Semeghini}, \citenamefont {Omran}, \citenamefont {Bluvstein},
  \citenamefont {Samajdar}, \citenamefont {Pichler}, \citenamefont {Ho} \emph
  {et~al.}}]{quantum_simulator_3}%
  \BibitemOpen
  \bibfield  {author} {\bibinfo {author} {\bibfnamefont {S.}~\bibnamefont
  {Ebadi}}, \bibinfo {author} {\bibfnamefont {T.~T.}\ \bibnamefont {Wang}},
  \bibinfo {author} {\bibfnamefont {H.}~\bibnamefont {Levine}}, \bibinfo
  {author} {\bibfnamefont {A.}~\bibnamefont {Keesling}}, \bibinfo {author}
  {\bibfnamefont {G.}~\bibnamefont {Semeghini}}, \bibinfo {author}
  {\bibfnamefont {A.}~\bibnamefont {Omran}}, \bibinfo {author} {\bibfnamefont
  {D.}~\bibnamefont {Bluvstein}}, \bibinfo {author} {\bibfnamefont
  {R.}~\bibnamefont {Samajdar}}, \bibinfo {author} {\bibfnamefont
  {H.}~\bibnamefont {Pichler}}, \bibinfo {author} {\bibfnamefont {W.~W.}\
  \bibnamefont {Ho}}, \emph {et~al.},\ }\bibfield  {title} {\bibinfo {title}
  {\emph{Quantum phases of matter on a 256-atom programmable quantum
  simulator}},\ }\href
  {https://doi.org/https://doi.org/10.1038/s41586-021-03582-4} {\bibfield
  {journal} {\bibinfo  {journal} {Nature}\ }\textbf {\bibinfo {volume} {595}},\
  \bibinfo {pages} {227} (\bibinfo {year} {2021})}\BibitemShut {NoStop}%
\bibitem [{\citenamefont {Jo}\ and\ \citenamefont
  {Kim}(2022)}]{quantum_simulator_4}%
  \BibitemOpen
  \bibfield  {author} {\bibinfo {author} {\bibfnamefont {M.}~\bibnamefont
  {Jo}}\ and\ \bibinfo {author} {\bibfnamefont {M.}~\bibnamefont {Kim}},\
  }\bibfield  {title} {\bibinfo {title} {\emph{Simulating open quantum
  many-body systems using optimised circuits in digital quantum simulation}},\
  }\href {https://arxiv.org/abs/2203.14295} {\bibfield  {journal} {\bibinfo
  {journal} {arXiv:2203.14295}\ } (\bibinfo {year} {2022})}\BibitemShut
  {NoStop}%
\bibitem [{\citenamefont {Perfetto}\ \emph {et~al.}(2023)\citenamefont
  {Perfetto}, \citenamefont {Carollo}, \citenamefont {Garrahan},\ and\
  \citenamefont {Lesanovsky}}]{QRD20222}%
  \BibitemOpen
  \bibfield  {author} {\bibinfo {author} {\bibfnamefont {G.}~\bibnamefont
  {Perfetto}}, \bibinfo {author} {\bibfnamefont {F.}~\bibnamefont {Carollo}},
  \bibinfo {author} {\bibfnamefont {J.~P.}\ \bibnamefont {Garrahan}},\ and\
  \bibinfo {author} {\bibfnamefont {I.}~\bibnamefont {Lesanovsky}},\ }\bibfield
   {title} {\bibinfo {title} {\emph{Reaction-Limited Quantum Reaction-Diffusion
  Dynamics}},\ }\href {https://doi.org/10.1103/PhysRevLett.130.210402}
  {\bibfield  {journal} {\bibinfo  {journal} {Phys. Rev. Lett.}\ }\textbf
  {\bibinfo {volume} {130}},\ \bibinfo {pages} {210402} (\bibinfo {year}
  {2023})}\BibitemShut {NoStop}%
\bibitem [{\citenamefont {Gorini}\ \emph {et~al.}(1976)\citenamefont {Gorini},
  \citenamefont {Kossakowski},\ and\ \citenamefont {Sudarshan}}]{gorini1976}%
  \BibitemOpen
  \bibfield  {author} {\bibinfo {author} {\bibfnamefont {V.}~\bibnamefont
  {Gorini}}, \bibinfo {author} {\bibfnamefont {A.}~\bibnamefont
  {Kossakowski}},\ and\ \bibinfo {author} {\bibfnamefont {E.~C.~G.}\
  \bibnamefont {Sudarshan}},\ }\bibfield  {title} {\bibinfo {title}
  {\emph{Completely positive dynamical semigroups of N‐level systems}},\
  }\href {https://doi.org/10.1063/1.522979} {\bibfield  {journal} {\bibinfo
  {journal} {J. Math. Phys.}\ }\textbf {\bibinfo {volume} {17}},\ \bibinfo
  {pages} {821} (\bibinfo {year} {1976})}\BibitemShut {NoStop}%
\bibitem [{\citenamefont {Lindblad}(1976)}]{lindblad1976}%
  \BibitemOpen
  \bibfield  {author} {\bibinfo {author} {\bibfnamefont {G.}~\bibnamefont
  {Lindblad}},\ }\bibfield  {title} {\bibinfo {title} {\emph{On the generators
  of quantum dynamical semigroups}},\ }\href
  {https://doi.org/10.1007/BF01608499} {\bibfield  {journal} {\bibinfo
  {journal} {Comm. Math. Phys.}\ }\textbf {\bibinfo {volume} {48}},\ \bibinfo
  {pages} {119} (\bibinfo {year} {1976})}\BibitemShut {NoStop}%
\bibitem [{\citenamefont {Breuer}\ and\ \citenamefont
  {Petruccione}(2002)}]{breuer2002}%
  \BibitemOpen
  \bibfield  {author} {\bibinfo {author} {\bibfnamefont {H.-P.}\ \bibnamefont
  {Breuer}}\ and\ \bibinfo {author} {\bibfnamefont {F.}~\bibnamefont
  {Petruccione}},\ }\href
  {https://doi.org/DOI:10.1093/acprof:oso/9780199213900.001.0001} {\emph
  {\bibinfo {title} {The theory of open quantum systems}}}\ (\bibinfo
  {publisher} {Oxford University Press on Demand},\ \bibinfo {year}
  {2002})\BibitemShut {NoStop}%
\bibitem [{\citenamefont {Lange}\ \emph {et~al.}(2018)\citenamefont {Lange},
  \citenamefont {Lenar\ifmmode \check{c}\else
  \v{c}\fi{}i\ifmmode~\check{c}\else \v{c}\fi{}},\ and\ \citenamefont
  {Rosch}}]{tGGE1}%
  \BibitemOpen
  \bibfield  {author} {\bibinfo {author} {\bibfnamefont {F.}~\bibnamefont
  {Lange}}, \bibinfo {author} {\bibfnamefont {Z.}~\bibnamefont {Lenar\ifmmode
  \check{c}\else \v{c}\fi{}i\ifmmode~\check{c}\else \v{c}\fi{}}},\ and\
  \bibinfo {author} {\bibfnamefont {A.}~\bibnamefont {Rosch}},\ }\bibfield
  {title} {\bibinfo {title} {\emph{Time-dependent generalized Gibbs ensembles
  in open quantum systems}},\ }\href
  {https://doi.org/10.1103/PhysRevB.97.165138} {\bibfield  {journal} {\bibinfo
  {journal} {Phys. Rev. B}\ }\textbf {\bibinfo {volume} {97}},\ \bibinfo
  {pages} {165138} (\bibinfo {year} {2018})}\BibitemShut {NoStop}%
\bibitem [{\citenamefont {Mallayya}\ \emph {et~al.}(2019)\citenamefont
  {Mallayya}, \citenamefont {Rigol},\ and\ \citenamefont {De~Roeck}}]{tGGE2}%
  \BibitemOpen
  \bibfield  {author} {\bibinfo {author} {\bibfnamefont {K.}~\bibnamefont
  {Mallayya}}, \bibinfo {author} {\bibfnamefont {M.}~\bibnamefont {Rigol}},\
  and\ \bibinfo {author} {\bibfnamefont {W.}~\bibnamefont {De~Roeck}},\
  }\bibfield  {title} {\bibinfo {title} {\emph{Prethermalization and
  Thermalization in Isolated Quantum Systems}},\ }\href
  {https://doi.org/10.1103/PhysRevX.9.021027} {\bibfield  {journal} {\bibinfo
  {journal} {Phys. Rev. X}\ }\textbf {\bibinfo {volume} {9}},\ \bibinfo {pages}
  {021027} (\bibinfo {year} {2019})}\BibitemShut {NoStop}%
\bibitem [{\citenamefont {Lange}\ \emph {et~al.}(2017)\citenamefont {Lange},
  \citenamefont {Lenar{\v{c}}i{\v{c}}},\ and\ \citenamefont {Rosch}}]{tGGE3}%
  \BibitemOpen
  \bibfield  {author} {\bibinfo {author} {\bibfnamefont {F.}~\bibnamefont
  {Lange}}, \bibinfo {author} {\bibfnamefont {Z.}~\bibnamefont
  {Lenar{\v{c}}i{\v{c}}}},\ and\ \bibinfo {author} {\bibfnamefont
  {A.}~\bibnamefont {Rosch}},\ }\bibfield  {title} {\bibinfo {title}
  {\emph{Pumping approximately integrable systems}},\ }\href
  {https://doi.org/https://doi.org/10.1038/ncomms15767} {\bibfield  {journal}
  {\bibinfo  {journal} {Nat. Commun.}\ }\textbf {\bibinfo {volume} {8}},\
  \bibinfo {pages} {1} (\bibinfo {year} {2017})}\BibitemShut {NoStop}%
\bibitem [{\citenamefont {Lenar\ifmmode \check{c}\else
  \v{c}\fi{}i\ifmmode~\check{c}\else \v{c}\fi{}}\ \emph
  {et~al.}(2018)\citenamefont {Lenar\ifmmode \check{c}\else
  \v{c}\fi{}i\ifmmode~\check{c}\else \v{c}\fi{}}, \citenamefont {Lange},\ and\
  \citenamefont {Rosch}}]{tGGE4}%
  \BibitemOpen
  \bibfield  {author} {\bibinfo {author} {\bibfnamefont {Z.}~\bibnamefont
  {Lenar\ifmmode \check{c}\else \v{c}\fi{}i\ifmmode~\check{c}\else
  \v{c}\fi{}}}, \bibinfo {author} {\bibfnamefont {F.}~\bibnamefont {Lange}},\
  and\ \bibinfo {author} {\bibfnamefont {A.}~\bibnamefont {Rosch}},\ }\bibfield
   {title} {\bibinfo {title} {\emph{Perturbative approach to weakly driven
  many-particle systems in the presence of approximate conservation laws}},\
  }\href {https://doi.org/10.1103/PhysRevB.97.024302} {\bibfield  {journal}
  {\bibinfo  {journal} {Phys. Rev. B}\ }\textbf {\bibinfo {volume} {97}},\
  \bibinfo {pages} {024302} (\bibinfo {year} {2018})}\BibitemShut {NoStop}%
\bibitem [{\citenamefont {Feller}(1968)}]{feller1957introduction}%
  \BibitemOpen
  \bibfield  {author} {\bibinfo {author} {\bibfnamefont {W.}~\bibnamefont
  {Feller}},\ }\href
  {https://www.wiley.com/en-us/An+Introduction+to+Probability+Theory+and+Its+Applications%2C+Volume+1%2C+3rd+Edition-p-9780471257080}
  {\emph {\bibinfo {title} {An introduction to probability theory and its
  applications}}}\ (\bibinfo  {publisher} {Wiley},\ \bibinfo {year}
  {1968})\BibitemShut {NoStop}%
\bibitem [{\citenamefont {Essler}\ and\ \citenamefont
  {Fagotti}(2016)}]{GGErev1}%
  \BibitemOpen
  \bibfield  {author} {\bibinfo {author} {\bibfnamefont {F.~H.}\ \bibnamefont
  {Essler}}\ and\ \bibinfo {author} {\bibfnamefont {M.}~\bibnamefont
  {Fagotti}},\ }\bibfield  {title} {\bibinfo {title} {\emph{Quench dynamics and
  relaxation in isolated integrable quantum spin chains}},\ }\href
  {https://doi.org/https://doi.org/10.1088/1742-5468/2016/06/064002} {\bibfield
   {journal} {\bibinfo  {journal} {J. Stat. Mech.: Theory Exp.}\ }\textbf
  {\bibinfo {volume} {2016}}\bibinfo  {number} { (6)},\ \bibinfo {pages}
  {064002}}\BibitemShut {NoStop}%
\bibitem [{\citenamefont {Vidmar}\ and\ \citenamefont {Rigol}(2016)}]{GGErev2}%
  \BibitemOpen
\bibfield  {number} {  }\bibfield  {author} {\bibinfo {author} {\bibfnamefont
  {L.}~\bibnamefont {Vidmar}}\ and\ \bibinfo {author} {\bibfnamefont
  {M.}~\bibnamefont {Rigol}},\ }\bibfield  {title} {\bibinfo {title}
  {\emph{Generalized Gibbs ensemble in integrable lattice models}},\ }\href
  {https://doi.org/https://doi.org/10.1088/1742-5468/2016/06/064007} {\bibfield
   {journal} {\bibinfo  {journal} {J. Stat. Mech.: Theory Exp.}\ }\textbf
  {\bibinfo {volume} {2016}}\bibinfo  {number} { (6)},\ \bibinfo {pages}
  {064007}}\BibitemShut {NoStop}%
\end{thebibliography}%

\end{document}